\documentclass[final,1p,times]{elsarticle}

\usepackage{lineno}
\usepackage{graphicx}
\usepackage{mathtools}
\usepackage{natbib}
\usepackage{xstring}
\usepackage{placeins}
\usepackage{amssymb}
\usepackage{amsmath}
\usepackage{xcolor}
\usepackage{xcolor}

\usepackage{color}      
\usepackage[colorlinks=true,
            linkcolor=BrickRed,
            urlcolor=blue,
            citecolor=blue]{hyperref}
            
\definecolor{revone}{RGB}{160,30,120}     
\definecolor{revthree}{RGB}{0,120,50}     
\definecolor{openred}{RGB}{200,0,0}

\definecolor{revboth}{RGB}{200,110,0}    

\newif\ifshownotes  \shownotestrue
\newif\ifshowfixes  \showfixestrue

\ifshownotes
  \usepackage[colorinlistoftodos,textsize=footnotesize]{todonotes}
\else
  \ifshowfixes
    \usepackage[colorinlistoftodos,textsize=footnotesize]{todonotes}
  \fi
\fi

\ifshowfixes
  \newcommand{\rfixone}[3]{\todo[inline,color=revone!8,bordercolor=revone!70,%
    caption={#1}]{\textbf{\textcolor{revone}{REVIEWER 1, COMMENT #1.}}\\
    \emph{Asked:} #2\\ \emph{Done:} #3}}
  \newcommand{\rfixthree}[3]{\todo[inline,color=revthree!8,bordercolor=revthree!70,%
    caption={#1}]{\textbf{\textcolor{revthree}{REVIEWER 3, COMMENT #1.}}\\
    \emph{Asked:} #2\\ \emph{Done:} #3}}
  
\else
  \newcommand{\rfixone}[3]{}
  \newcommand{\rfixthree}[3]{}
  
\fi

\ifshownotes
  \newcommand{\cjudge}[1]{\todo[inline,color=blue!10,bordercolor=blue!55,%
    caption={JUDGEMENT CALL}]{\textbf{JUDGEMENT CALL.} #1}}
  \newcommand{\cdisc}[1]{\todo[inline,color=red!10,bordercolor=red!60,%
    caption={DISCREPANCY}]{\textbf{DISCREPANCY vs.\ THE PUBLISHED PAPER.} #1}}
  \newcommand{\cconcern}[1]{\todo[inline,color=orange!13,bordercolor=orange!75,%
    caption={CONCERN}]{\textbf{CONCERN.} #1}}
  \newcommand{\cguide}[1]{\todo[inline,color=black!8,bordercolor=black!55,%
    caption={How to read these notes}]{#1}}
\else
  \newcommand{\cjudge}[1]{}
  \newcommand{\cdisc}[1]{}
  \newcommand{\cconcern}[1]{}
  \newcommand{\cguide}[1]{}
\fi

\ifshownotes
  
\else
  \ifshowfixes
    
  \else
    
  \fi
\fi

\usepackage{multicol}
\usepackage{lipsum}
\usepackage{amsthm}

\usepackage[utf8]{inputenc}
\usepackage[T1]{fontenc}
\usepackage{amsmath,amssymb,amsthm}
\usepackage{mathtools}
\usepackage{bm}
\usepackage{tikz}
\usepackage{pgfplots}
\usepackage{booktabs}
\usepackage{array}
\usepackage{geometry}
\usepackage{hyperref}
\usepackage{cleveref}
\usepackage{xcolor}
\usepackage{float}
\usepackage{enumitem}
\usepackage{caption}
\usepackage{subcaption}
\usepackage{graphicx}
\usepackage{placeins}
\usetikzlibrary{shapes.geometric, arrows.meta, positioning, calc, fit, backgrounds, decorations.pathreplacing, patterns, matrix}

\usepackage{algorithm}
\usepackage{algpseudocode}

\definecolor{bayesblue}{RGB}{70,130,180}
\definecolor{detgreen}{RGB}{60,150,60}
\definecolor{basisorange}{RGB}{230,140,50}
\definecolor{outputred}{RGB}{180,60,60}
\definecolor{lightblue}{RGB}{220,235,250}
\definecolor{lightgreen}{RGB}{220,245,220}
\definecolor{lightorange}{RGB}{255,240,220}
\definecolor{lightred}{RGB}{255,230,230}

\newcommand{\KL}{D_{\mathrm{KL}}}
\newcommand{\Normal}{\mathcal{N}}
\DeclareMathOperator*{\argmin}{arg\,min}
\newcommand{\bW}{\bm{W}}
\newcommand{\bx}{\bm{x}}

\newcommand{\corr}[2]{\underbrace{#1}_{\text{#2}}}
\usepackage{pgfplots}
\pgfplotsset{compat=1.18}
\biboptions{sort&compress}

\journal{} 

\begin{document}

\begin{frontmatter}

\title{Bayesian neural network correction of RANS turbulence models with uncertainty quantification in separated flows}

\author[mymainaddress]{Tyler Buchanan\fnref{equal}}

\author[mymainaddress]{Ali Eidi\fnref{equal}\corref{cor}}

\author[mymainaddress]{Richard P. Dwight\corref{cor}}

\fntext[equal]{These authors contributed equally to this work.}

\cortext[cor]{Corresponding authors: Ali Eidi (a.eidi@tudelft.nl) and Richard P. Dwight (r.p.dwight@tudelft.nl).}

\address[mymainaddress]{Department of Flow Physics and Technology, Faculty of Aerospace Engineering, Delft University of Technology, 2629 HS Delft, Netherlands}

\begin{abstract}

Data-driven correction of turbulence models offers a promising route for improving
Reynolds-averaged Navier--Stokes (RANS) predictions, but quantifying uncertainty in such
corrections and ensuring generalization across flows remain key challenges. This work
presents a Bayesian neural network (BNN) framework for uncertainty-aware correction of
RANS models. The BNN represents both epistemic and input-dependent aleatoric
uncertainty, interpreted here as irreducible scatter in the feature-to-correction
relationship, with the latter propagated as spatially correlated fields through the RANS
solver.
The framework is trained exclusively on a periodic-hill configuration and evaluated without retraining on five unseen separated-flow configurations.
The learned corrections improve the training-flow prediction, but the strong momentum-field benefit obtained from the anisotropy correction
does not transfer consistently to the unseen flows.
Out-of-distribution under-coverage is already present at the correction-field
level, indicating that the loss of calibration is already present in transfer of the
learned correction rather than being introduced primarily by CFD propagation.
Overall, the framework enables joint propagation of epistemic and aleatoric
uncertainty while exposing the present limitations of uncertainty calibration under
distribution shift.

\end{abstract}



\end{frontmatter}

\section{Introduction} \label{sec:Introduction}

Separated flows from curved surfaces are among the most challenging regimes to predict in computational aerodynamics. They appear in a wide variety of engineering contexts, from stalled wings and diffusers to turbomachinery passages and combustors, and they strongly influence performance, efficiency, and safety. Yet, despite their prevalence and importance, robust prediction of separation onset, reattachment, and associated unsteady phenomena remains elusive. This is primarily due to the fact that industry-standard Reynolds-averaged Navier–Stokes (RANS) methods, while computationally affordable, invoke structural assumptions that break down in separated regions. In particular, separated shear layers exhibit turbulence anisotropy, strong non-equilibrium, and coherent vortex dynamics that violate the linear stress–strain assumptions of most eddy-viscosity closures \cite{li2020aerodynamic}. As a result, RANS models typically underpredict shear stresses, misplace reattachment points, and fail to capture the unsteady dynamics that dominate separated flows \cite{xiao2019quantification}. Conversely, high-fidelity (HF) methods such as wall-resolved large-eddy simulation (LES) or direct numerical simulation (DNS) resolve separation physics faithfully but remain prohibitively expensive for design and optimization workflows. As a result, RANS continues to serve as the workhorse of engineering computational fluid dynamics for the foreseeable future, even though its credibility is limited by model-form uncertainties \cite{eidi2022data}.

The deficiencies of RANS models in separated flows stem largely from the assumptions built into linear eddy-viscosity closures. For example, widely used linear eddy-viscosity models (such as the $k$–$\omega$ SST model \cite{menter1993zonal}) assume a linear relationship between the Reynolds stress anisotropy and the mean strain rate. Such simplifications suppress anisotropy and lead to insufficient turbulent mixing in shear layers, causing large errors in predicting separation and reattachment \cite{wilcox1998turbulence, schmitt2007boussinesq}. Numerous studies have shown that these errors are most pronounced outside the boundary layer, in the separated shear layers where turbulence exhibits strong non-equilibrium dynamics and coherent structures \cite{rumsey2009exploring}.

Beyond specific models, these deficiencies reflect the broader uncertainty landscape in turbulence modeling. As emphasized in \cite{duraisamy2019turbulence}, RANS closures are affected by uncertainties at multiple levels: (i) irrecoverable information loss during ensemble averaging, (ii) assumptions in functional representations of Reynolds stress, (iii) structural choices of closure forms, and (iv) calibration of model coefficients. Both \emph{epistemic} uncertainties (arising from lack of knowledge or inadequate data) and \emph{aleatoric} uncertainties (due to intrinsic variability) degrade predictive reliability \cite{ferson1996different}. For RANS closures, the dominant source is epistemic: structural inadequacies in the closure ansatz cannot be resolved by parameter tuning alone, and remain the primary driver of predictive failure in separated flows. This recognition has led to two major research thrusts. First, uncertainty quantification (UQ) methods have been developed to represent RANS predictions probabilistically. Forward (data-free) UQ propagates prior distributions of closure coefficients or model discrepancies through RANS solvers, while backward (data-driven) UQ assimilates HF or experimental data via Bayesian inference to constrain model errors \cite{xiao2019quantification}. These approaches have provided valuable confidence intervals for engineering predictions, but they remain primarily diagnostic and do not by themselves improve the underlying closures.

Parallel to UQ, a complementary line of research has sought to improve RANS closures by learning model discrepancies from HF data. A seminal contribution in this direction was the field inversion and machine learning (FIML) framework of Parish and Duraisamy \cite{parish2016paradigm}, which combined model inversion with machine learning to infer and generalize corrective information for turbulence closures. Subsequent studies have developed different representations of such corrections, including symbolic regression and machine learning approaches.
Early methods applied symbolic regression to discover interpretable algebraic stress models. Weatheritt and Sandberg \cite{weatheritt2017development} developed genetic programming-based methods, while Schmelzer et al.\ \cite{schmelzer2020discovery} introduced SpaRTA (Sparse Regression of Turbulence Anisotropy), yielding algebraic correction terms based on Pope’s tensor basis \cite{pope2001turbulent}, thereby retaining interpretability while improving accuracy. However, symbolic regression approaches also face challenges. The expressions discovered by these methods are often highly non-unique and may overfit the limited set of training flows, yielding complex algebraic forms that lack robustness across different flow conditions. Moreover, ensuring numerical stability of the resulting corrections when embedded in a RANS solver is nontrivial, as small changes in the discovered formula can lead to large variations in predicted stresses. Consequently, the generalization of symbolic corrections beyond the training set remains difficult, as discrepancies vary significantly across geometries and flow conditions \cite{mandler2024generalization, jigjid2025discovery}. These methods have since evolved into Bayesian symbolic regression frameworks such as SBL-SpaRTA, which integrate regression with UQ \cite{cherroud2022sparse}. Nevertheless, the limitations of symbolic regression have underscored the trade-off between interpretability and flexibility, motivating the exploration of more expressive neural-network-based corrections.

Machine learning approaches have also introduced more flexible representations of Reynolds-stress corrections. Wu et al.\ \cite{wu2018physics} developed a physics-informed framework based on invariant mean flow features and a decomposition of the Reynolds stress designed to facilitate stable propagation of the learned corrections through RANS solvers. A milestone in neural network-based closure modeling was the tensor basis neural network (TBNN) of Ling et al. \cite{ling2016reynolds}, which embedded Galilean and rotational invariance into the architecture through a tensor-basis representation of Reynolds-stress anisotropy. Building on this formulation, Kaandorp and Dwight\ \cite{kaandorp2020data} proposed the Tensor Basis Random Forest (TBRF), retaining the invariant tensor basis representation while replacing the neural network mapping with a random forest model. These approaches demonstrated the potential of invariant machine learning models for Reynolds stress prediction and their subsequent use within RANS solvers, but remained deterministic.

An important milestone was the work of Geneva and Zabaras\ \cite{geneva2019quantifying}, who pioneered the use of Bayesian deep neural networks to quantify model-form uncertainty in RANS closures. Their framework applied invariant neural networks to correct Reynolds stress discrepancies and demonstrated that Bayesian neural networks (BNNs) can provide probabilistic bounds on RANS predictions by propagating epistemic uncertainty through stochastic RANS ensembles. However, their corrections were applied globally across the flow field and primarily addressed epistemic uncertainty, leaving aleatoric effects unmodeled. 
Building on these ideas, Tang et al.\ \cite{tang2023data} developed a Bayesian deep learning framework for Reynolds stress closure that combines invariant tensor bases with a decomposition of the stress tensor into linear and nonlinear components. The nonlinear contribution is learned using a BNN, enabling both epistemic and aleatoric UQ while improving generalization across flows with significantly different geometries and Reynolds numbers. Their results highlight the potential of Bayesian approaches not only for uncertainty-aware predictions but also for enhancing robustness and extrapolation capability in data-driven turbulence models. 

More recently, Pash et al.\ \cite{pash2025priori} extended this paradigm to turbulent combustion closures, introducing architectures that capture both epistemic and aleatoric uncertainty, thereby offering predictive distributions that separate data noise from model inadequacy. 
BNNs therefore offer a natural probabilistic extension of deterministic data-driven
closures by representing uncertainty in the learned correction rather than providing only
a point estimate. However, existing applications remain limited in how these uncertainties
are propagated to the resulting flow solution and in how corrections are localized to
regions where the baseline turbulence model is deficient.

Despite these advances, two gaps remain. First, predictive uncertainty in a learned
turbulence model correction should distinguish epistemic uncertainty in the inferred model
from input-dependent aleatoric variability in the correction conditioned on the available
features. In the present setting, the latter represents irreducible scatter in the
feature-to-correction relationship rather than repeated measurement noise. More
importantly, uncertainty quantified only at the correction level does not establish the
uncertainty of the resulting flow solution. Both contributions must therefore be
propagated through the RANS equations if they are to provide uncertainty bounds on the
predicted flow quantities.

Second, most data-driven correction strategies are formulated as global modifications of the turbulence closure. However, RANS models are often reasonably accurate in attached regions, while the dominant errors are concentrated in separated shear layers. Applying corrections uniformly across the domain may therefore degrade predictions in regions where the baseline model already performs well \cite{brunton2020machine}. This limitation has been highlighted in recent studies \cite{wu2025development}, which show that global corrections can compromise robustness in canonical boundary layer flows and distort near-wall behavior \cite{srivastava2024generalizably}. These observations point to the need for selective correction strategies that target only the regions where model deficiencies are significant.

The present study addresses these gaps through a zonal Bayesian framework for data-driven turbulence model correction. Separate Bayesian models are constructed for the Reynolds stress anisotropy correction and the turbulent kinetic energy production correction. Invariant tensor basis representations are used to preserve the required transformation properties of the learned corrections, while the probabilistic formulation provides both epistemic and input-dependent aleatoric uncertainty. Rather than reporting these uncertainties only at the correction level, samples from the resulting predictive distributions are propagated through the RANS solver to quantify their effect on the predicted flow quantities. The corrections are activated selectively in separated shear layers using a physics-based zonal classifier \cite{buchanan2025data}, leaving regions in which the baseline closure performs adequately unmodified.
The same trained models are then evaluated across separated-flow configurations and Reynolds numbers using case-independent normalization and invariant features. During
propagation, the aleatoric contribution is represented through spatially correlated random fields so that its spatial coherence is retained in the RANS solution.

The present framework builds on two closely related earlier contributions. The Relative Importance Term Analysis (RITA) methodology introduced in \cite{buchanan2025data} provides the physics-based classifier used here to localize the corrections to separated shear-layer regions. A preliminary version of the Bayesian correction framework was subsequently presented in \cite{eidi2025physicsguidedbayesianneuralnetworks}. The formulation considered in the present study has since been substantially revised, including the Bayesian model architecture and normalization procedure, and the uncertainty propagation is extended to account for both epistemic and aleatoric contributions. The numerical assessment is also broadened from the flow configurations considered in the preliminary study to six cases spanning different geometries and Reynolds numbers. The present work therefore focuses on the probabilistic formulation, propagation, and transferability of zonal data-driven turbulence model corrections, while adopting the RITA classifier and the underlying tensor basis and discrepancy correction concepts from the established methodologies described above.

The remainder of this paper is organized as follows.
Section~\ref{sec:methodology} presents the framework: frozen-RANS discrepancy extraction,
the invariant and tensor-basis representation, the Bayesian formulation with
heteroscedastic training and Monte Carlo inference, and the correlated sampling used to
propagate the corrections.
Section~\ref{sec:results} examines the two correction models and their propagation on the training flow, first the scalar correction alone and then the two combined, which isolates what each contributes.
Section~\ref{sec:generalization} applies the same models, without retraining, to five unseen configurations, and asks where the calibration holds and where it fails.
Section~\ref{sec:conclusion} summarizes the findings and their implications for
uncertainty-aware turbulence modeling.

\section{Methodology} \label{sec:methodology}

This section introduces the full framework used to construct, train, and propagate the data-driven turbulence corrections. The first two subsections establish the physical and mathematical foundations of the approach: model-form discrepancies are first extracted from the baseline SST model using a frozen-RANS procedure, and the resulting correction fields are then expressed in terms of invariant features and tensor basis representations. Building on this setup, the main methodological contribution of the present work begins in \S\ref{sec:correction_models}, where Bayesian correction models are introduced for both the scalar and anisotropy discrepancies. The subsequent subsections then present the BNN formulation, the heteroscedastic variational training objective, and the Monte Carlo inference strategy used to decompose uncertainty and propagate stochastic correction fields through the RANS solver.

\subsection{Frozen-RANS discrepancy extraction}
\label{sec:frozen_rans}

Model-form discrepancies of the baseline $k$--$\omega$ SST turbulence model \cite{menter1993zonal} are extracted using the frozen-RANS framework of \cite{schmelzer2020discovery}, in which HF flow fields are imposed while the RANS operators are left unchanged. This isolates turbulence closure error from numerical and discretization effects.
Under the Boussinesq approximation, the SST model represents the Reynolds stresses through an eddy viscosity-based anisotropy tensor. Using the imposed HF fields, the corresponding Reynolds stress anisotropy can be evaluated directly and compared with the baseline Boussinesq form. The difference defines an anisotropy correction, denoted $b_{ij}^{\Delta}$, which quantifies the part of the stress tensor that cannot be represented by the linear eddy viscosity assumption. The corrected anisotropy tensor is then written as
\begin{equation}
b_{ij} = b_{ij}^{\text{RANS}} + b_{ij}^{\Delta}.
\end{equation}

The primary role of $b_{ij}^{\Delta}$ is to correct the Reynolds stress tensor, which directly affects the momentum equations. Through this modification, the correction also enters the production terms of the $k$ and $\omega$ transport equations, as these depend on the Reynolds stresses. However, modifying the Reynolds stress alone does not generally recover the correct turbulent kinetic energy (TKE) balance observed in HF data. This is because the altered production terms are not accompanied by a consistent adjustment of dissipation and transport mechanisms within the turbulence model.

To address this imbalance, an additional scalar correction, denoted $k_{\text{deficit}}$, is introduced in the turbulence transport equations. This term is defined as the residual required to satisfy the $k$ and $\omega$-equations when the HF velocity and TKE fields are imposed under the frozen-RANS framework. In other words, $k_{\text{deficit}}$ represents the missing contribution needed to close the TKE budget once the anisotropy correction has been applied.
In the frozen inversion procedure, the mean velocity and TKE are fixed to their HF values, while the $\omega$ equation is solved iteratively using the augmented anisotropy tensor.

The resulting augmented $k$--$\omega$ SST equations are

\begin{equation}
\rho\frac{\partial k}{\partial t}
+
\rho\frac{\partial(U_jk)}{\partial x_j}
=
\corr{\left(\hat{P}_k + k_{\text{deficit}}\right)}{corrected production}
-
\beta^*\rho\omega k
+
\frac{\partial}{\partial x_j}
\left[
(\mu+\sigma_k\mu_t)
\frac{\partial k}{\partial x_j}
\right],
\end{equation}

\begin{equation}
\rho\frac{\partial\omega}{\partial t}
+
\rho\frac{\partial(U_j\omega)}{\partial x_j}
=
\gamma\frac{\nu_t}{k}
\corr{\left(\hat{P}_k + k_{\text{deficit}}\right)}{corrected production}
-
\beta\rho\omega^2
+
\frac{\partial}{\partial x_j}
\left[
(\mu+\sigma_\omega\mu_t)
\frac{\partial\omega}{\partial x_j}
\right]
+
2\rho(1-F_1)\sigma_{\omega2}
\frac{1}{\omega}
\frac{\partial k}{\partial x_j}
\frac{\partial\omega}{\partial x_j},
\end{equation}

with modified production term,

\begin{equation}
\hat{P}_k
=
\min
\left(
-2k\left(b_{ij}^{\text{RANS}} + b_{ij}^{\Delta}\right)
\frac{\partial U_i}{\partial x_j},
10\beta^*\omega k
\right).
\end{equation}

The frozen RANS procedure therefore produces two discrepancy fields: the anisotropy correction $b_{ij}^{\Delta}$ and the scalar correction $k_{\text{deficit}}$. These fields represent the model-form error of the baseline turbulence model and constitute the targets for the data-driven correction models.

The corrections are applied only in regions identified by the RITA classifier \cite{buchanan2025data}. Specifically, the classifier defines a binary mask $\sigma \in \{0,1\}$, where cells with $\sigma = 1$ are classified as belonging to the separated shear layer and receive the learned corrections, while cells with $\sigma = 0$ retain the baseline RANS formulation. This selective strategy ensures that modifications are introduced only where the baseline model exhibits systematic deficiencies, while preserving its performance in well-predicted regions such as attached boundary layers and free-stream flow. Because the classifier is evaluated from local term balances in the turbulence model equations, the identified correction region adapts to the underlying flow state in a physically consistent manner. Further details of the RITA framework and the choice of classification criteria are provided in \ref{app:rita}.

\subsection{Flow configurations}
\label{sec:configurations}
The framework is trained exclusively on the periodic hill (PH) configuration of Breuer et al.\ \cite{breuer2009flow} at $Re_H = 10{,}595$, where the Reynolds number is based on the hill height $H$ and the bulk velocity above the crest. The corresponding LES data are used as the HF reference for the inversion of the discrepancy fields. All quantities required during training, including the normalization parameters, classifier threshold, and correlation length introduced below, are determined from this configuration and retained unchanged in all subsequent applications. Generalization is assessed on five configurations that are excluded from training. These comprise the curved backward-facing step (CBFS) of Bentaleb et al.\ \cite{bentaleb2012large} at $Re_H = 13{,}700$; three parameterized periodic-hill (Parm-PH) configurations from Xiao et al.\ \cite{xiao2020flows}, with hill-width parameters $\alpha = 0.5$, $1.0$, and $1.5$, all at $Re_H = 5600$; and the NASA wall-mounted hump of Uzun and Malik \cite{uzun2017nasahump} at $Re_c = 9.36 \times 10^{5}$. The NASA hump is formulated in the $x$--$z$ plane and is the only configuration considered in dimensional units. No retraining or case-specific adjustment of the learned models is performed for any of these five cases. Further details of the test configurations and their corresponding results are provided in \S\ref{sec:generalization}.

\begin{table}[htbp]
\caption{Computational setup for the six flow configurations. For the periodic-hill cases and the CBFS, $H$ denotes the hill and step height, respectively, and the Reynolds number is based on $H$ and the corresponding bulk velocity. For the NASA hump, $c$ denotes the chord and $Re_c$ is based on the free-stream velocity; this case is formulated in dimensional units in the $x$--$z$ plane. The maximum wall slope is reported as $|\mathrm{d}y/\mathrm{d}x|$ expressed as an angle.}
\centering
\small
\setlength{\tabcolsep}{4pt}
\begin{tabular}{l l c c r l}
\toprule
Case & Domain ($x \times y$) & Steepest & $Re$ & Cells & Inflow \\
     &                       & wall slope &    &       &        \\
\midrule
\multicolumn{6}{l}{\textit{Training configuration}} \\
PH & $9.0H \times 3.036H$   & $40.6^{\circ}$ & $10{,}595$ & $15{,}600$ & streamwise periodic \\
\midrule
\multicolumn{6}{l}{\textit{Unseen test configurations}} \\
Parm-PH $\alpha=0.5$  & $10.1H \times 3.036H$  & $59.8^{\circ}$ & $5600$     & $15{,}600$ & streamwise periodic \\
Parm-PH $\alpha=1.0$  & $12.0H \times 3.036H$  & $40.7^{\circ}$ & $5600$     & $15{,}600$ & streamwise periodic \\
Parm-PH $\alpha=1.5$  & $10.9H \times 3.036H$  & $29.8^{\circ}$ & $5600$     & $15{,}600$ & streamwise periodic \\
CBFS                  & $22.7H \times 5.0H$    & $34.0^{\circ}$ & $13{,}700$ & $21{,}000$ & fixed profile \\
NASA hump             & $1.52c \times 0.91c$   & $40.4^{\circ}$ & $9.36\times10^{5}$ & $51{,}626$ & fixed, $U_{\infty} = 34.6$\,m\,s$^{-1}$ \\
\bottomrule
\end{tabular}

\label{tab:cfd_setup}
\end{table}

All RANS calculations were performed with the steady incompressible solver using SIMPLE pressure--velocity coupling. Spatial gradients
were evaluated with Gauss linear interpolation, convection of $U$, $k$, and $\omega$
with bounded Gauss linear-upwind schemes, and diffusion terms with Gauss linear corrected
discretization.

\subsection{Invariant features and tensor-basis representation}
\label{sec:features_basis}

To ensure physical consistency and generalizability across different flow configurations, the correction fields are expressed as functions of invariant quantities derived from the local mean velocity gradients. This guarantees that the learned models are independent of coordinate rotations and Galilean transformations.
The mean velocity gradient tensor is decomposed into its symmetric and antisymmetric parts, defined as the strain-rate tensor
$S_{ij} = \tfrac{1}{2}\left(\partial U_i/\partial x_j + \partial U_j/\partial x_i\right)$
and the rotation-rate tensor
$\Omega_{ij} = \tfrac{1}{2}\left(\partial U_i/\partial x_j - \partial U_j/\partial x_i\right)$.
From these tensors, a set of scalar invariant features is constructed:
\[
\mathbf{x}
=
\left[
S_{ij}S_{ji},\;
\Omega_{ij}\Omega_{ji},\;
S_{ij}S_{jk}S_{ki},\;
\Omega_{ij}\Omega_{jk}S_{ki},\;
\phi_{Re_t}= \frac{Re_t - Re_t^{\min}}{Re_t^{\max} - Re_t^{\min}}
\right]^T,
\qquad
Re_t = \frac{\nu_t}{\nu}.
\]

These features capture the local balance between strain, rotation, and turbulence activity, while remaining computable during a RANS simulation.
The velocity-gradient tensors used to construct the invariants are nondimensionalized
using a local mean-flow time scale,
\begin{equation}
\tilde{S}_{ij} = \tau_s S_{ij}, \qquad
\tilde{\Omega}_{ij} = \tau_s \Omega_{ij}, \qquad
\tau_s = \frac{1}{\lVert \nabla U \rVert_F},
\qquad
\lVert \nabla U \rVert_F =
\left(
\sum_{i,j}
\left(
\frac{\partial U_i}{\partial x_j}
\right)^2
\right)^{1/2},
\label{eq:meanflow_timescale}
\end{equation}
and the invariants above are evaluated using
$\tilde{S}_{ij}$ and $\tilde{\Omega}_{ij}$.
This differs from normalization with the turbulence time scale $1/\omega$ and avoids
introducing the state of the baseline turbulence model into the velocity-gradient
normalization. In the present data sets, $\lVert\nabla U\rVert_F$ remains nonzero within
the cells to which the correction model is applied.

After construction of the feature vector, each input feature is scaled by its standard
deviation over the PH training data without subtracting the corresponding mean,
\begin{equation}
\tilde{x}_m =
\frac{x_m}{\sigma_{x_m}^{\mathrm{PH}}},
\label{eq:input_scaling}
\end{equation}
where $\sigma_{x_m}^{\mathrm{PH}}$ is computed once from the PH training cells and retained
unchanged for all other flow configurations. No case-specific normalization statistics are
therefore introduced during inference.
The mean shift is omitted so that the physical zero of the invariant representation is
preserved across flow configurations. In contrast, centering by training-set statistics
would map $\mathbf{x}=\mathbf{0}$ to a location determined by the feature means of the
training flow. The quantitative effect of this choice, including comparison with the
$z$-score normalization, is examined in \S\ref{sec:results}.

Rather than learning the anisotropy correction $b_{ij}^{\Delta}$
directly, it is represented using a tensor-basis expansion following Pope’s framework \cite{pope2001turbulent}:
\begin{equation}
{b_{ij}^{\Delta}}
=
\sum_{n=1}^{N_T} g_n(\mathbf{x})\, T_{ij}^{(n)}.
\label{eq:pope_expansion}
\end{equation}

Here, $T_{ij}^{(n)}$ are nonlinear tensor basis functions constructed from $S_{ij}$ and $\Omega_{ij}$, while $g_n(\mathbf{x})$ are scalar coefficient functions of the invariant feature vector. This formulation preserves the tensorial structure of the Reynolds-stress correction while reducing the learning problem to predicting a small number of scalar fields.
For the two-dimensional separated flows considered in this work, the tensor-basis representation is restricted to the first four reduced basis tensors,
\[
T_{ij}^{(1)} = S_{ij}, \quad
T_{ij}^{(2)} = S_{ik}\Omega_{kj} - \Omega_{ik}S_{kj}, \quad
T_{ij}^{(3)} = S_{ik}S_{kj} - \tfrac{1}{3}\delta_{ij} S_{mn}S_{nm}, \quad
T_{ij}^{(4)} = \Omega_{ik}\Omega_{kj} - \tfrac{1}{3}\delta_{ij} \Omega_{mn}\Omega_{nm}.
\]

The tensor basis coefficients are obtained from the extracted correction field by projection onto this basis, transforming the tensor regression problem into a low-dimensional coefficient learning problem.

After introducing $b_{ij}^{\Delta}$, the scalar correction $k_{\text{deficit}}$ represents the residual model-form error in the TKE equation . Consistent with the structure of the production term, this correction can be interpreted as a production-like contribution and related to the local turbulence activity.
Instead of using a tensor expansion, the scalar correction $k_{\text{deficit}}$ is represented through a scalar basis model driven by the same invariant feature vector. The detailed form of this model is introduced in \S\ref{sec:correction_models}.
The invariant feature vector $\mathbf{x}$ therefore serves as the common input for both the anisotropy and scalar correction models, while their outputs are represented respectively through a tensor-basis expansion and a scalar basis formulation.

\subsection{Bayesian correction models}
\label{sec:correction_models}

The discrepancy fields introduced in the previous subsections ($k_{\text{deficit}}$ and $b_{ij}^{\Delta}$) are modeled using BNNs that map the invariant feature vector $\mathbf{x}$ to the corresponding correction quantities. Two distinct models are constructed: a scalar model for $k_{\text{deficit}}$ and a tensor-basis model for $b_{ij}^{\Delta}$.
The scalar correction $k_{\text{deficit}}$ is modeled using a
physics-guided basis formulation in which the prediction is expressed as a scaled function of the local dissipation rate. Specifically, the normalized output is written as
\begin{equation}
\tilde{y} = g(\mathbf{x})\, \varepsilon,
\label{eq:kdef_model}
\end{equation}
where $g(\mathbf{x})$ is a learned coefficient function and $\varepsilon$ is the
local dissipation rate, used in its raw dimensional form. The normalized target is
$\tilde{y} = k_{\text{deficit}} / y_s$ with the single fixed scalar
$y_s = \mathrm{std}\bigl(k_{\text{deficit}}^{\mathrm{PH}}\bigr)$ taken over the classified
training cells, so that the physical prediction is recovered as
$\hat{y} = y_s\, g(\mathbf{x})\, \varepsilon$ and the learned coefficient is the
dimensionless ratio $g \simeq k_{\text{deficit}} / \varepsilon$. No training mean is
subtracted from $k_{\text{deficit}}$, and the basis is used raw rather than divided by the
training peak $\varepsilon_{\max}^{\mathrm{PH}}$. Scaling the basis by a training-set
maximum would pin the correction amplitude to the dissipation level of the PH,
while the raw basis rescales itself with the local flow. The fixed scalar $y_s$ only sets
the units of $g$ and transfers unchanged.
This formulation introduces a physically meaningful scaling, ensuring that the correction magnitude increases in regions of strong turbulence activity while remaining bounded. The network therefore learns only the coefficient function $g(\mathbf{x})$, which modulates the dissipation-based basis.
To account for data-dependent noise, a second output head predicts a positive
input-dependent noise coefficient $c(\mathbf{x})$, yielding a heteroscedastic standard
deviation of the form
\begin{equation}
\sigma_y = c(\mathbf{x})\, \varepsilon,
\end{equation}
where positivity of $c(\mathbf{x})$ is enforced by the exponential transformation of the
noise-head output.

The anisotropy correction $b_{ij}^{\Delta}$ is modeled through the tensor-basis expansion introduced in \S\ref{sec:features_basis}. Rather than predicting the tensor field directly, the network learns the coefficient vector
\begin{equation}
\mathbf{g}(\mathbf{x}) = [g_1(\mathbf{x}), \dots, g_{N_T}(\mathbf{x})]^T,
\end{equation}
from which the correction tensor is reconstructed as \cref{eq:pope_expansion}.
By projecting the extracted anisotropy correction onto the chosen tensor basis the training targets for the coefficients $g_n$ are obtained, as described in \S\ref{sec:features_basis}.

\subsection{Bayesian neural network formulation}
\label{sec:bnn_formulation}

The correction models are implemented using BNNs, in which the network weights are treated as random variables rather than deterministic parameters. This probabilistic formulation enables the model to capture epistemic uncertainty arising from limited training data \cite{magris2023bayesian}.
Each linear layer is defined by a weight matrix $\mathbf{W}$ and bias vector $\mathbf{b}$, for which a variational posterior distribution is introduced. Assuming a mean-field Gaussian approximation \cite{kendall2017uncertainties}, the posterior
over weights is written as
\begin{equation}
q(\mathbf{W}) = \mathcal{N}(\boldsymbol{\mu}_W,\boldsymbol{\sigma}_W^2),
\end{equation}
with an analogous expression for $\mathbf{b}$.
To enable gradient-based optimization, samples from the posterior are drawn using the reparameterization trick,
\begin{equation}
\mathbf{W} = \boldsymbol{\mu}_W + \boldsymbol{\sigma}_W \odot \boldsymbol{\epsilon},
\quad
\boldsymbol{\epsilon} \sim \mathcal{N}(0,1),
\label{eq:reparam}
\end{equation}
which transforms stochastic sampling into a differentiable operation.
Its standard deviation is parameterized through an unconstrained variable $\rho_W$ using a \emph{softplus} transformation,
$\sigma_W = \log(1+\exp(\rho_W)) + 10^{-6}$, which guarantees positivity. In all layers, $\rho_W$ is initialized to $-5.0$, corresponding to a small initial posterior standard deviation.
A zero-mean Gaussian prior is placed over all weights,
\begin{equation}
p(\mathbf{W}) = \mathcal{N}(0,\alpha^{-1}\mathbf{I}),
\end{equation}
where $\alpha$ denotes the weight precision. Instead of fixing this parameter, it is treated as a learnable variable shared across all layers. To regularize its value, a Gamma prior is imposed,
\begin{equation}
\alpha \sim \mathrm{Gamma}(a_0, b_0),
\end{equation}
which encourages physically meaningful weight scales while allowing the model to adapt its complexity during training \cite{neal2012bayesian, tipping2001sparse}. This learnable precision provides an automatic mechanism for controlling the effective capacity of the network.

Both correction models employ fully connected Bayesian architectures with nonlinear activation functions. In both cases, the predictive mean and the input-dependent noise are represented separately. For the scalar correction model, the network predicts the coefficient function $g(\mathbf{x})$ together with the corresponding noise coefficient $c(\mathbf{x})$. For the anisotropy correction model, the network produces the tensor-basis coefficients along with their associated heteroscedastic uncertainty estimates. A preliminary version of this Bayesian correction formulation was presented in \cite{eidi2025physicsguidedbayesianneuralnetworks}; the architecture employed in the present study has since been revised, particularly in the representation and treatment of the predictive mean and heteroscedastic uncertainty.

All layers are treated within the Bayesian framework described above, such that each forward pass corresponds to a stochastic realization of the network weights. This enables the propagation of epistemic uncertainty through both correction models. For brevity, the overall architecture is illustrated only for the scalar correction model in \autoref{fig:architecture_combined}; the anisotropy model follows the same Bayesian formulation, with the output generalized to the tensor-basis coefficients and their corresponding uncertainty estimates.
The probabilistic formulation and training procedure are described in the following subsections.

\begin{figure}[htbp]
\centering

\begin{subfigure}[t]{0.95\textwidth}\label{fig:architecture_overall}
\centering
\resizebox{\textwidth}{!}{%
\begin{tikzpicture}[
    node distance=0.8cm and 1.2cm,
    box/.style={rectangle, draw, rounded corners, minimum height=0.9cm, minimum width=1.8cm, align=center, font=\small},
    bayesian/.style={box, fill=lightblue, draw=bayesblue, line width=1pt},
    deterministic/.style={box, fill=lightgreen, draw=detgreen, line width=1pt},
    basis/.style={box, fill=lightorange, draw=basisorange, line width=1pt},
    output/.style={box, fill=lightred, draw=outputred, line width=1pt},
    input/.style={box, fill=gray!10, draw=gray},
    arrow/.style={-{Stealth[length=2.5mm]}, thick},
    label/.style={font=\footnotesize\itshape},
]

\node[input] (input) {$\tilde{\bx} \in \mathbb{R}^5$\\{\small Invariants}};
\node[basis, below=2.5cm of input] (epsilon) {$\varepsilon$\\{\small Local dissipation}};

\node[bayesian, right=1.0cm of input] (h1) {VarLinear\\$5 \to 32$};
\node[bayesian, right=0.8cm of h1] (h2) {VarLinear\\$32 \to 32$};
\node[bayesian, right=0.8cm of h2] (out) {VarLinear\\$32 \to 1$};
\node[right=0.6cm of out] (g) {$g(\bx)$};

\node[bayesian, below=1.5cm of h1] (sh1) {VarLinear\\$5 \to 32$};
\node[bayesian, right=0.8cm of sh1] (sh2) {VarLinear\\$32 \to 32$};
\node[bayesian, right=0.8cm of sh2] (sout) {VarLinear\\$32 \to 1$};
\node[right=0.6cm of sout] (sg) {$c(\bx)$};

\node[output, right=0.5cm of g, yshift=-0.8cm] (mult1) {$\times$};
\node[output, right=0.5cm of sg] (mult2) {$\times$};

\node[output, right=0.5cm of mult1] (yhat) {$\hat{\tilde{y}} = g \cdot \varepsilon$};
\node[output, right=0.5cm of mult2] (sigmay) {$\sigma_y = c \cdot \varepsilon$};

\draw[arrow] (input) -- (h1);
\draw[arrow] (h1) -- node[above, label] {tanh} (h2);
\draw[arrow] (h2) -- node[above, label] {tanh} (out);
\draw[arrow] (out) -- (g);
\draw[arrow] (g) -- (mult1);

\draw[arrow] (input.south) -- (sh1.west);
\draw[arrow] (sh1) -- node[above, label] {tanh} (sh2);
\draw[arrow] (sh2) -- node[above, label] {tanh} (sout);
\draw[arrow] (sout) -- node[above, label] {$\exp$} (sg);
\draw[arrow] (sg) -- (mult2);

\draw[arrow] (epsilon.north) -- ++(0,2.1) |- (mult1.west);
\draw[arrow] (epsilon.east) -- ++(10.6,0) -| (mult2.south);

\draw[arrow] (mult1) -- (yhat);
\draw[arrow] (mult2) -- (sigmay);

\node[above=0.2cm of h2, font=\small\bfseries, text=bayesblue] {Coefficient Head (Bayesian)};
\node[above=0.2cm of sh2, font=\small\bfseries, text=bayesblue] {Noise Head (Bayesian)};

\begin{scope}[shift={(0,-5)}]
    \node[bayesian, minimum width=1.5cm, minimum height=0.6cm] at (0,0) {};
    \node[right, font=\small] at (0.9,0) {Variational layer (weight distributions)};
    \node[basis, minimum width=1.5cm, minimum height=0.6cm] at (8,0) {};
    \node[right, font=\small] at (8.75,0) {Basis function};
\end{scope}

\end{tikzpicture}}
\caption{Overall Bayesian basis-model architecture.}
\end{subfigure}

\vspace{0.6cm}

\begin{subfigure}[t]{0.75\textwidth} \label{fig:architecture_varlinear}
\centering
\resizebox{\textwidth}{!}{%
\begin{tikzpicture}[
    node distance=0.6cm and 1cm,
    param/.style={rectangle, draw, rounded corners, fill=bayesblue!20, minimum height=0.8cm, minimum width=1.6cm, align=center, font=\small},
    op/.style={circle, draw, fill=white, minimum size=0.8cm, font=\small},
    arrow/.style={-{Stealth[length=2mm]}, thick},
]

\node[font=\bfseries] at (4, 3) {Variational Linear Layer};

\node[param] (mu) at (-1.0, 1.5) {$\mu_W$\\{\small mean}};
\node[param] (rho) at (-1.0, 0) {$\rho_W$\\{\small init $= -5.0$}};

\node[op] (softplus) at (2, 0) {sp};
\node[param, fill=bayesblue!40] (sigma) at (4, 0) {$\sigma_W$\\{\small std dev}};

\node[param, fill=gray!20] (eps) at (4, -1.5) {$\epsilon \sim \Normal(0,1)$};

\node[op] (mult) at (6, 0) {$\times$};
\node[op] (add) at (8, 0.75) {$+$};

\node[param, fill=bayesblue!60] (W) at (10, 0.75) {$W$\\{\small sampled}};

\draw[arrow] (rho) -- node[above, font=\small] {softplus} (softplus);
\draw[arrow] (softplus) -- (sigma);
\draw[arrow] (sigma) -- (mult);
\draw[arrow] (eps) -- (mult);
\draw[arrow] (mult) -- (add);
\draw[arrow] (mu) -| (add);
\draw[arrow] (add) -- (W);

\node[draw, dashed, rounded corners, fill=yellow!10, align=center, font=\small] at (5, 2) {
    $W = \mu_W + \sigma_W \cdot \epsilon$\\[0.2em]
    {\small Reparameterization trick (\ref{eq:reparam})}
};

\node[font=\small, align=center] at (2, -0.8) {$\sigma = \log(1+e^\rho) + 10^{-6}$\\ensures $\sigma > 0$};

\end{tikzpicture}}
\caption{Internal structure of a \texttt{VarLinear} layer used in panel (a).}
\end{subfigure}

\caption{Fully Bayesian BNN architecture and variational layer detail. \textbf{(a)} The
scalar correction model. Variational linear layers predict the coefficient function
$g(\bx)$ and the noise coefficient $c(\bx)$, both scaled by the raw dissipation
rate $\varepsilon$} as defined in \S\ref{sec:correction_models}. Only the scalar model is
drawn; the anisotropy model shares this structure and predicts the tensor-basis
coefficients instead.
\textbf{(b)} Internal structure of one variational linear layer, showing the
reparameterization of \cref{eq:reparam}.
\label{fig:architecture_combined}
\end{figure}

\subsection{Heteroscedastic likelihood and ELBO training}
\label{sec:elbo_training}

The BNNs described in \S\ref{sec:bnn_formulation} are trained using variational inference by optimizing an evidence lower bound (ELBO)-type objective that combines data fit, posterior regularization, and additional priors on the learned uncertainty quantities.
For the scalar correction model, the predictive distribution is assumed to be Gaussian with input-dependent variance,
\begin{equation}
p(y \mid \mathbf{x}, \mathbf{w}, \mathbf{w}_{\sigma})
=
\mathcal{N}\big(y \mid \hat{y}(\mathbf{x}), \sigma_y^2(\mathbf{x})\big).
\end{equation}
An analogous formulation is adopted for the anisotropy correction model, where a multi-output Gaussian likelihood is used for the tensor-basis coefficients. In this case, each coefficient is modeled with its own mean and heteroscedastic variance, and the same ELBO objective is applied component-wise. For brevity, the loss formulation is presented here for the scalar model, as the extension to the multi-output case is
straightforward.
Following \S\ref{sec:correction_models}, the mean and standard deviation are modeled as
\begin{equation}
\hat{\tilde{y}}(\mathbf{x}) = g(\mathbf{x})\,\tilde{\varepsilon},
\qquad
\sigma_y(\mathbf{x}) = c(\mathbf{x})\,\tilde{\varepsilon},
\label{eq:noise_scaling}
\end{equation}
where $c(\mathbf{x})$ is predicted by the noise head.
For a dataset of size $N$, the negative log-likelihood (NLL) is
\begin{equation}
\mathcal{L}_{\text{NLL}}
=
\frac{1}{N}\sum_{i=1}^N
\left[
\log \sigma_{y,i}
+
\frac{(y_i-\hat{y}_i)^2}{2\sigma_{y,i}^2}
\right].
\end{equation}

Its variational posterior over the network weights is regularized through the Kullback--Leibler (KL) divergence with respect to the Gaussian prior,
\begin{equation}
\mathcal{L}_{\text{KL}}
=
\frac{1}{N}\,\mathrm{KL}\!\left(q(\mathbf{w})\,\|\,p(\mathbf{w})\right).
\end{equation}

As described in \S\ref{sec:bnn_formulation}, the weight precision $\alpha$ is treated as a learnable parameter with Gamma prior. This contributes the additional regularization term (the Alpha Term in \autoref{fig:loss}):
\begin{equation}
\mathcal{L}_{\alpha}
=
-(a_0-1)\log \alpha + b_0 \alpha.
\label{eq:alpha_loss}
\end{equation}

To prevent the heteroscedastic noise model from collapsing to unrealistically small values or inflating excessively, a Gamma prior is imposed on the predicted aleatoric standard deviation,
\begin{equation}
p(\sigma_y)=\mathrm{Gamma}(\sigma_y \mid a_{\sigma}, b_{\sigma}).
\end{equation}
This yields the regularization term
\begin{equation}
\mathcal{L}_{\sigma}
=
\frac{1}{N}\sum_{i=1}^{N}
\left[
-(a_{\sigma}-1)\log(\sigma_{y,i}+10^{-8})
+
b_{\sigma}\sigma_{y,i}
\right].
\label{eq:sigma_loss}
\end{equation}
A weighting factor $\lambda_{\sigma}$ is introduced to control the strength of this prior.
The final training objective is
\begin{equation}
\mathcal{L}
=
\mathcal{L}_{\text{NLL}}
+
\mathcal{L}_{\text{KL}}
+
\mathcal{L}_{\alpha}
+
\lambda_{\sigma}\mathcal{L}_{\sigma}.
\label{eq:loss}
\end{equation}

This is the objective minimized during Bayesian fine-tuning. A schematic overview of the loss construction is shown in \autoref{fig:loss}.

\begin{figure}[htbp]
\centering
\resizebox{0.8\linewidth}{!}{%
\begin{tikzpicture}[
    node distance=0.5cm,
    box/.style={rectangle, draw, rounded corners, minimum height=1cm, minimum width=3cm, align=center, font=\small},
    lossbox/.style={box, fill=outputred!20, draw=outputred},
    arrow/.style={-{Stealth[length=2mm]}, thick},
]

\node[box, fill=gray!10] (forward) at (0, 0) {\textbf{Forward Pass}\\Sample $\bW, \bW_\sigma \sim q_\phi$\\Compute $\hat{y}, \sigma_y$};

\node[lossbox] (nll) at (-6.5, -3) {
    \textbf{NLL Term}\\[0.3em]
    $\displaystyle\frac{(y-\hat{y})^2}{2\sigma_y^2} + \log\sigma_y$
};

\node[lossbox] (kl) at (5.5, -3) {
    \textbf{KL Term (Both Heads)}\\[0.3em]
    $\displaystyle\sum_{\text{coeff}} {\KL}_{l} + \sum_{\text{noise}} {\KL}_{l}$
};

\node[lossbox] (alphaprior) at (1.5, -3) {
    \textbf{Alpha Term}\\[0.3em]
    $\mathcal{L}_\alpha$
};

\node[lossbox] (sigprior) at (-2.5, -3) {
    \textbf{Sigma Term}\\[0.3em]
    $\lambda_\sigma \mathcal{L}_\sigma$ (\cref{eq:sigma_loss})
};

\node[box, fill=yellow!20, draw=orange, line width=1.5pt] (loss) at (0, -5.5) {
    \textbf{Total Loss}\\[0.3em]
$\mathcal{L} = \text{NLL} + \text{KL}/N + \mathcal{L}_\alpha + \lambda_\sigma\mathcal{L}_\sigma$
};

\draw[arrow] (forward) -- ++(-3.5, -1) -- (nll);
\draw[arrow] (forward) -- ++(3.5, -1) -- (kl);
\draw[arrow] (forward) -- (sigprior);
\draw[arrow] (forward) -- (alphaprior);
\draw[arrow] (nll) -- (loss);
\draw[arrow] (kl) -- (loss);
\draw[arrow] (sigprior) -- (loss);
\draw[arrow] (alphaprior) -- (loss);
\node[font=\small, anchor=north] at (0, -6.3) {Proper ELBO ($\beta = 1$), learnable $\alpha$ via Gamma hyperprior};

\end{tikzpicture}}
\caption{ELBO loss computation. Both the coefficient and noise head weights contribute to the KL divergence term (with learnable $\alpha$).}
\label{fig:loss}
\end{figure}

To improve convergence, training is performed in two phases. First, a deterministic network, equivalent to a TBNN \cite{ling2016reynolds}, with the same architecture is pretrained using a mean squared error objective, providing stable initialization for the posterior means of the Bayesian weights. Second, the full Bayesian model is optimized using \cref{eq:loss}, allowing both epistemic and aleatoric uncertainty to be learned jointly. Both phases are optimized with Adam \cite{kingma2014adam}: $10^{4}$ iterations at a learning rate of $10^{-2}$ for the deterministic phase and $4\times10^{4}$ at $10^{-3}$ for the Bayesian phase, full-batch in both cases. The classified region of the training flow contains $2378$ cells, split $80/20$ into training and held-out sets by a seeded permutation, and all inference uses $M = 100$ Monte Carlo weight samples.

\subsection{Monte Carlo inference and uncertainty decomposition}
\label{sec:mc_inference}

After training, predictions are obtained by performing MC sampling over the variational posterior of the network weights. Each forward pass corresponds to a stochastic realization of the model, enabling the propagation of epistemic uncertainty through the network.
Given an input $\mathbf{x}$, $M=100$ samples of the network weights are drawn from the variational posterior. This produces a set of predictions $\{y^{(m)}, \sigma_y^{(m)}\}_{m=1}^M$, from which the predictive mean is computed as
\begin{equation}
\bar{y}(\mathbf{x}) = \frac{1}{M}\sum_{m=1}^M y^{(m)}.
\end{equation}

The predictive uncertainty is decomposed into epistemic and aleatoric components.
Given $M$ MC samples, the epistemic variance is
\begin{equation}
\sigma_{\mathrm{epi}}^2
=
\frac{1}{M}\sum_{m=1}^M
\left(y^{(m)}-\bar{y}\right)^2,
\end{equation}
while the aleatoric variance is obtained as the mean predicted conditional variance,
\begin{equation}
\sigma_{\mathrm{alea}}^2
=
\frac{1}{M}\sum_{m=1}^M
\left(\sigma_y^{(m)}\right)^2.
\end{equation}

In addition, we monitor the posterior variability of the predicted aleatoric scale,
\begin{equation}
\sigma_{\mathrm{alea\text{-}epi}}^2
=
\frac{1}{M}\sum_{m=1}^M
\left(
\sigma_y^{(m)}-\bar{\sigma}_y
\right)^2,
\qquad
\bar{\sigma}_y
=
\frac{1}{M}\sum_{m=1}^M \sigma_y^{(m)}.
\end{equation}
This diagnostic measures how strongly the predicted aleatoric standard deviation changes
when the network weights are resampled from their posterior. It therefore quantifies
uncertainty in the estimated aleatoric scale itself. It is not a third additive component
of the predictive variance and is not included in
$\sigma_{\mathrm{total}}^2$.

The total predictive variance follows from the law of total variance as
\begin{equation}
\sigma_{\mathrm{total}}^2
=
\sigma_{\mathrm{epi}}^2
+
\sigma_{\mathrm{alea}}^2.
\end{equation}

The epistemic component reflects uncertainty in the learned model parameters and is
expected to increase in regions of the feature space that are poorly represented by the
training data. The aleatoric component represents input-dependent variability in the
feature-to-correction relationship that cannot be reduced solely by improving the estimate
of the network weights. The epistemic-on-aleatoric diagnostic quantifies posterior
variability in the predicted noise scale and therefore indicates how consistently the
heteroscedastic uncertainty itself is inferred across posterior samples. The overall inference procedure, including MC sampling and the
decomposition of predictive uncertainty into epistemic and aleatoric contributions, is
illustrated in \autoref{fig:inference}. A concise summary of the training and inference
procedures is provided in \ref{app:method_summary}.

In the present work, uncertainty propagation through the flow solver (OpenFOAM) is performed using a frozen-realization approach. For each posterior sample of the network weights, the predictive mean output defines a deterministic correction field, which is held fixed during the RANS simulation. Repeating this procedure for multiple samples yields an ensemble of flow solutions, from which the propagated uncertainty is quantified.
For propagation through the RANS solver, both components of the predictive uncertainty are
included in each realization. Consequently, the ensemble of propagated solutions represents
the total predictive uncertainty; epistemic and aleatoric contributions are not subsequently
decomposed at the level of the flow solution. An epistemic-only ensemble is considered
separately as a diagnostic control in \autoref{fig:corrlen_amplification}, whereas all
propagated uncertainty bands reported elsewhere include both contributions.

For realization $s$, the correction field supplied to the solver is written as
\begin{equation}
q^{(s)}(\mathbf{x})
=
\mu\!\left(\mathbf{x};\mathbf{w}_{\mu}^{(s)}\right)
+
\sigma\!\left(\mathbf{x};\mathbf{w}_{\sigma}^{(s)}\right)
\,\xi^{(s)}(\mathbf{x}),
\label{eq:aleatoric_sample}
\end{equation}
where $\mathbf{w}_{\mu}^{(s)}$ and $\mathbf{w}_{\sigma}^{(s)}$ denote posterior samples
associated with the predictive mean and heteroscedastic standard deviation, respectively.
The first term therefore carries uncertainty in the learned mean correction, while the
second represents input-dependent variability in the feature-to-correction relationship.

The random field $\xi^{(s)}$ has zero mean and unit pointwise variance. It is generated by
smoothing Gaussian white noise with a Gaussian kernel of bandwidth $h$ and subsequently
renormalizing the field to unit variance. The resulting autocorrelation is
\begin{equation}
C(r)
=
\exp\!\left(-\frac{r^2}{L^2}\right),
\qquad
L=2h,
\label{eq:aleatoric_acf}
\end{equation}
where $L$ denotes the $1/e$ correlation length, i.e.\ the spatial separation at which the
autocorrelation has decayed to $C(L)=e^{-1}$. Because $C(0)=1$, introducing spatial
correlation does not modify the local variance
$\sigma^2(\mathbf{x};\mathbf{w}_{\sigma}^{(s)})$ predicted by the Bayesian model; it changes
only the spatial covariance structure of the perturbation.

Spatial correlation was found to be important for retaining the influence of the
aleatoric contribution during propagation through the RANS solver. When statistically
independent perturbations of the same pointwise variance were applied cellwise, their
contribution to the propagated solution uncertainty was strongly attenuated by the
pressure--velocity coupling. Introducing spatial coherence increased the propagated
uncertainty band, with the dependence on correlation length quantified in
\autoref{fig:corrlen_amplification}.

The spatial correlation scale is calibrated once using the PH training case. The empirical
autocorrelation of the training residual gives
$(L_x,L_y)=(0.178,0.054)H$ in the streamwise and wall-normal directions, respectively,
corresponding to an anisotropy ratio of approximately $3.3$. Because the estimated
correlation length is sensitive to small-scale residual scatter and mesh resolution, the
effect of the correlation scale was examined on the PH configuration. Based on this
calibration, $(L_x,L_y)=(0.355,0.108)H$, preserving the same anisotropy ratio, is used for
the propagation studies. These values are subsequently retained unchanged for every test
case, so that no HF information from the test configurations enters the uncertainty
propagation. Each realization uses an independently generated field with a seed determined
by the sample index, ensuring reproducibility of the ensembles.

Since the training targets are obtained from a single deterministic inversion of one LES,
the aleatoric contribution should not be interpreted as measurement noise in the conventional
sense. Rather, it represents irreducible scatter in the learned feature-to-correction
relationship. This scatter can arise from several sources: non-uniqueness of the mapping,
since identical values of the selected invariants can correspond to different inverted
corrections; statistical convergence error inherited from the LES reference data; and
representation error associated with information that cannot be expressed by the selected
feature set. The predicted $\sigma(\mathbf{x})$ therefore characterizes conditional scatter
around the learned mean correction. Its magnitude is non-negligible: over the classified
PH training cells, the mean predicted standard deviation is approximately $0.70$ times the
RMSE of the corresponding mean prediction.

The principal computational cost of uncertainty propagation is associated with the ensemble
of RANS solutions. Each realization uses the same mesh, numerical discretization, and
convergence procedure as the corresponding baseline computation and is executed serially
on a single CPU core. Consequently, an $M$-member ensemble requires approximately $M$
RANS solves in terms of total computational work. The realizations are mutually
independent and can therefore be executed concurrently.

In the present calculations, up to ten realizations were run concurrently, with one CPU
core assigned to each realization. An isolated PH realization required approximately
$4.5$ min, while the median wall-clock time per PH realization during the concurrent
production runs was approximately $15$ min. Under the same computational setup, the
corresponding median times were approximately $17$ min for the CBFS and $25$ min for the
NASA hump, whose mesh contains approximately three times as many cells. With ten realizations executed concurrently, the $100$-member PH and CBFS ensembles
required approximately $2.5$ and $2.8$ h of elapsed wall-clock time, respectively. The
NASA-hump calculations used $50$ realizations, corresponding to approximately $2.1$ h at
the reported production timing.

Training of the Bayesian models constitutes a one-time computational cost and is not
repeated for individual test configurations. Application of the trained models without
uncertainty propagation requires only the baseline RANS solution needed to evaluate the
input features and classifier mask; the subsequent neural-network evaluation requires only
a few seconds. No HF data from the test configuration are required. The additional
computational expense of the proposed framework therefore arises primarily from propagation
of the predictive uncertainty rather than from evaluation of the learned correction itself.

\begin{figure}[htbp]
\centering
\resizebox{\linewidth}{!}{%
\begin{tikzpicture}[
    x=1cm,y=1cm,
    node distance=0.45cm,
    sample/.style={rectangle, draw, rounded corners, fill=bayesblue!20, minimum height=0.8cm, minimum width=2.8cm, align=center, font=\small},
    result/.style={rectangle, draw, rounded corners, fill=lightgreen, minimum height=0.8cm, minimum width=2.7cm, align=center, font=\small},
    aggregate/.style={rectangle, draw, rounded corners, fill=yellow!20, minimum height=2.3cm, minimum width=4.2cm, align=center, font=\small},
    aleabox/.style={rectangle, draw, rounded corners, fill=lightorange, minimum height=1.3cm, minimum width=4.2cm, align=center, font=\small},
    totalbox/.style={rectangle, draw, rounded corners, fill=outputred!20, line width=1.2pt, minimum height=1.0cm, minimum width=4.2cm, align=center, font=\small},
    arrow/.style={-{Stealth[length=2.2mm]}, thick},
    note/.style={font=\scriptsize\itshape, align=center},
]

\node[sample, fill=gray!20] (input) at (-4.3,0) {Input $\bx,\varepsilon$};

\node[sample] (s1) at (-0.4,2.2) {$\bW^{(1)},\bW_\sigma^{(1)} \sim q_\phi$};
\node[sample] (s2) at (-0.4,1.0) {$\bW^{(2)},\bW_\sigma^{(2)} \sim q_\phi$};
\node[font=\Large] at (-0.4,-0.2) {$\vdots$};
\node[sample] (sM) at (-0.4,-1.4) {$\bW^{(M)},\bW_\sigma^{(M)} \sim q_\phi$};

\node[result] (y1) at (3.5,2.2) {$\hat{y}^{(1)}, \sigma_y^{(1)}$};
\node[result] (y2) at (3.5,1.0) {$\hat{y}^{(2)}, \sigma_y^{(2)}$};
\node[font=\Large] at (3.5,-0.2) {$\vdots$};
\node[result] (yM) at (3.5,-1.4) {$\hat{y}^{(M)}, \sigma_y^{(M)}$};

\node[aggregate] (agg) at (8.5,1.0) {
    \textbf{Aggregate}\\[0.25em]
    $\bar{y} = \dfrac{1}{M}\sum_m \hat{y}^{(m)}$\\[0.25em]
    $\sigma_{\mathrm{epi}} = \mathrm{std}\!\left[\hat{y}^{(m)}\right]$\\[0.25em]
    $\sigma_{\mathrm{alea}} = \mathrm{mean}\!\left[\sigma_y^{(m)}\right]$
};

\node[aleabox] (alea_epi) at (8.5,-2.1) {
    \textbf{Posterior variability of aleatoric scale}\\[0.25em]
    $\sigma_{\mathrm{alea\mbox{-}epi}} = \mathrm{std}\!\left[\sigma_y^{(m)}\right]$
};

\node[totalbox] (total) at (8.5,-4.3) {
    $\sigma_{\mathrm{total}} = \sqrt{\sigma_{\mathrm{epi}}^2 + \sigma_{\mathrm{alea}}^2}$
};

\coordinate (collectorTop) at (5.6,2.2);
\coordinate (collectorBot) at (5.6,-1.4);

\draw[arrow] (input) -- (s1);
\draw[arrow] (input) -- (s2);
\draw[arrow] (input) -- (sM);

\draw[arrow] (s1) -- (y1);
\draw[arrow] (s2) -- (y2);
\draw[arrow] (sM) -- (yM);

\draw[arrow] (y1.east) -- (collectorTop);
\draw[arrow] (y2.east) -- (5.6,1.0);
\draw[arrow] (yM.east) -- (collectorBot);

\draw[thick] (collectorTop) -- (collectorBot);

\draw[arrow] (5.6,1.0) -- (agg.west);
\draw[arrow] (5.6,-1.0) |- (alea_epi.west);

\draw[arrow] (agg.south) -- (alea_epi.north);
\draw[arrow] (alea_epi.south) -- (total.north);

\node[note, font=\small] at (3.8,2.9) {both $\hat{y}$ and $\sigma_y$ vary across MC samples};

\end{tikzpicture}}
\caption{MC inference with the fully Bayesian model. Each posterior weight sample produces a different prediction $\hat{y}^{(m)}$ and a different noise estimate $\sigma_y^{(m)}$. The predictive mean $\bar{y}$ is obtained by averaging over samples. The spread of $\hat{y}^{(m)}$ gives epistemic uncertainty, the mean of $\sigma_y^{(m)}$ gives aleatoric uncertainty, and the spread of $\sigma_y^{(m)}$ measures posterior variability of the predicted aleatoric scale..}
\label{fig:inference}
\end{figure}

\section{ Correction models and propagation on the training flow}
\label{sec:results}

The model is trained on the PH dataset of Breuer et al.~\cite{breuer2009flow} at $Re_H = 10{,}595$, following the methodology described in \S\ref{sec:methodology}. The training and evaluation within this configuration are performed using a train/test split of the same flow case. 
To assess generalization beyond the training configuration, the trained model is applied
without retraining to five unseen flow cases: the CBFS of Bentaleb et
al.~\cite{bentaleb2012large}, three parameterized periodic-hill configurations, and the
NASA wall-mounted hump. The propagation results for all five unseen configurations are
presented in \S\ref{sec:generalization}.Additional implementation details and hyperparameter selection are provided in \ref{app:design}.

\begin{figure}[htbp]
\centering
\includegraphics[width=\textwidth]{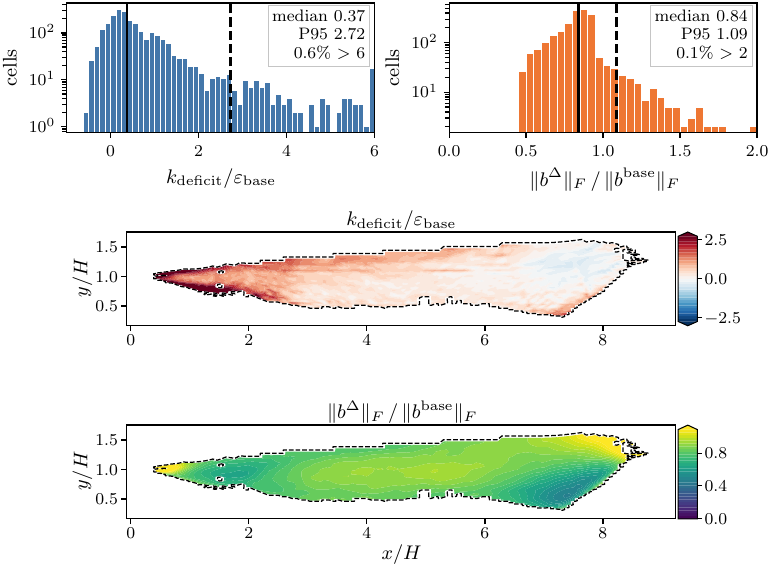}
\caption{Size of the two corrections relative to the baseline $k$--$\omega$ SST model on the PH, over the $2378$ classified cells. Top: distributions of the two dimensionless ratios, with medians and $95$th percentiles marked. Bottom: where in the flow the largest corrections sit. The scalar ratio uses the baseline dissipation $\varepsilon_{\mathrm{base}} = \beta^{*}k\omega$, since $k_{\text{deficit}}$ carries the dimensions of a TKE production rate; the anisotropy ratio compares Frobenius norms against the Boussinesq anisotropy $b^{\mathrm{base}} = -(\nu_t/k)S$. Dashed line: classifier boundary. Flood maps clipped at the 95th percentile.}
\label{fig:correction_magnitude_PH}
\end{figure}

To provide a scale for interpreting the correction errors discussed in the following
sections, \autoref{fig:correction_magnitude_PH} shows the magnitude of the two corrections
relative to corresponding terms in the baseline $k$--$\omega$ SST closure. Both ratios
are defined to be dimensionless. The scalar correction is normalized by the baseline
dissipation,
$\varepsilon_{\mathrm{base}} = \beta^{*} k \omega$, since
$k_{\text{deficit}}$ has the dimensions of a production rate. The anisotropy correction
is compared with the Boussinesq anisotropy,
$\mathbf{b}_{\mathrm{base}} = -(\nu_t/k)\mathbf{S}$, using the Frobenius norm.
On the classified cells, the median of
$k_{\text{deficit}}/\varepsilon_{\mathrm{base}}$ is $0.37$, with a $95$th percentile
of $2.72$. For
$\lVert b^{\Delta}\rVert/\lVert b_{\mathrm{base}}\rVert$, the median is $0.84$ and the
$95$th percentile is $1.09$. Relative to these baseline scales, the corrections are
therefore of the same order as the modeled terms they modify and, locally, cannot be
regarded as small perturbations to the baseline closure.

\subsection{\texorpdfstring{Scalar correction: $k_{\text{deficit}}$ model performance}{Scalar correction: k deficit model performance}}
\label{sec:kdef_results}

\begin{table}[htbp]
\centering
\small
\caption{Quantitative results for the final scalar $k_{\text{deficit}}$ BNN on
the PH case with an 80/20 train/test split, 40\,000 Bayesian
training epochs, and $M=100$ MC samples.}
\begin{tabular}{@{}lr@{}}
\toprule
\textbf{Metric} & \textbf{Value} \\
\midrule
Deterministic pretraining, test MSE (Phase 1) & $2.39 \times 10^{-5}$ \\
BNN test MSE, MC mean (Bayesian training) & $1.41 \times 10^{-5}$ \\
Deterministic pretraining, train MSE & $2.52 \times 10^{-6}$ \\
BNN train MSE (posterior-mean weights) & $1.19 \times 10^{-5}$ \\
Mean $|\bar{y} - y|$ & $2.00 \times 10^{-3}$ \\
\midrule
Mean aleatoric $\sigma_{\text{alea}}$ & $2.16 \times 10^{-3}$ \\
Mean epistemic $\sigma_{\text{epi}}$ & $1.19 \times 10^{-3}$ \\
Mean posterior variability of aleatoric scale
$\sigma_{\text{alea-epi}}$ & $2.23 \times 10^{-4}$ \\
\midrule
Calibration (aleatoric $1\sigma / 2\sigma$) & 68\% / 92\% \\
Calibration (total $1\sigma / 2\sigma$) & 71\% / 95\% \\
Ideal calibration ($1\sigma / 2\sigma$) & 68\% / 95\% \\
\midrule
Converged $\alpha$ & 1.12\\
Equivalent $\sigma_{\text{equiv}} = 1/\sqrt{\alpha}$ & 0.95 \\
Final NLL & $-0.22$ \\
Final KL$/N$ & 0.88 \\
\midrule
MC sample correlation $r(s_1, s_2)$ & 0.989\\
MC sample correlation $r(s_1, s_{50})$ & 0.951 \\
\bottomrule
\end{tabular}

\label{tab:results}
\end{table}

The quantitative performance of the learned correction model on the PH is
summarized in \autoref{tab:results}.
On the identical $476$ held-out cells and in the same physical units, the Bayesian
model achieves a test MSE of $1.41 \times 10^{-5}$, compared with
$2.39 \times 10^{-5}$ for the deterministic pretraining model, corresponding to a
reduction by a factor of approximately $1.7$. The held-out cells are drawn from the
$2378$ classified cells using a seeded permutation, with the training and test subsets
disjoint by construction.

The deterministic model fits the training cells substantially more closely than the
held-out cells, with an MSE of $2.52 \times 10^{-6}$ on the training set and
$2.39 \times 10^{-5}$ on the test set. For the Bayesian model, the posterior-mean
training MSE is $1.19 \times 10^{-5}$, while the MC-mean test MSE is
$1.41 \times 10^{-5}$. Although these Bayesian values correspond to slightly different
predictive summaries, their similar magnitude indicates a much smaller train--test gap
than in the deterministic pretraining phase. This behavior is consistent with reduced
overfitting during Bayesian fine-tuning rather than with improved interpolation accuracy
on the training data.

Another key outcome of the uncertainty decomposition is that aleatoric uncertainty exceeds
epistemic uncertainty, with mean values of $2.16 \times 10^{-3}$ and
$1.19 \times 10^{-3}$, respectively. In the present single-flow training setting, this
indicates that conditional variability in the learned correction contributes more strongly
to the predictive uncertainty than uncertainty in the learned mean model. The posterior variability of the aleatoric scale is considerably smaller, indicating that
the predicted noise magnitude changes relatively little across posterior weight samples.
The total predictive uncertainty is well calibrated, with empirical coverage of
71\% and 95\% at the $1\sigma$ and $2\sigma$ levels, respectively, close to the
corresponding Gaussian reference values of 68\% and 95\%. The aleatoric-only coverage
(68\%/92\%) shows similarly good agreement at the $1\sigma$ level, with modest
undercoverage at $2\sigma$.
The learned prior precision converges to $\alpha \approx 1.12$, corresponding to
an effective prior standard deviation of approximately $0.95$. This indicates that the
optimized prior retains substantial flexibility in the network weights.

\begin{figure}[htbp]
\centering
\includegraphics[width=\textwidth]{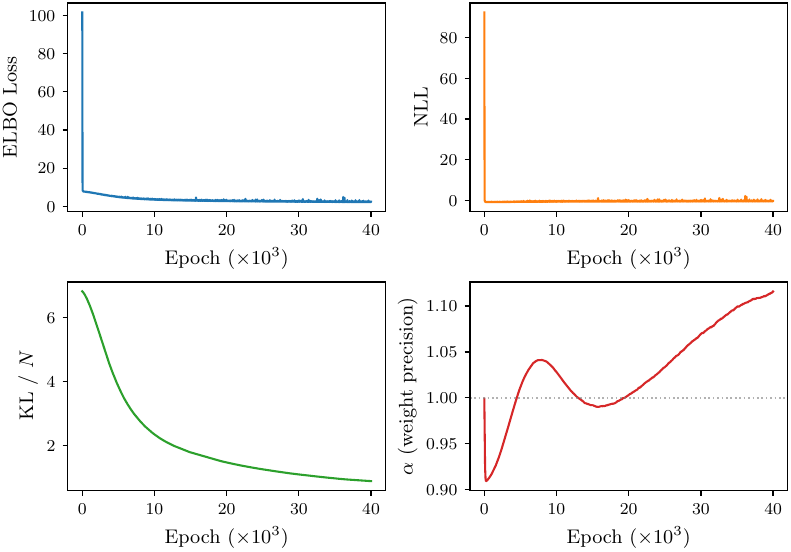}
\caption{Training convergence during the Bayesian training phase
(40\,000 epochs). Top-left: ELBO loss. Top-right: negative log-likelihood (NLL).
Bottom-left: per-sample KL divergence. Bottom-right: learnable weight
precision $\alpha$, with the initial value $\alpha = 1.0$ shown as a
dashed reference line.}
\label{fig:training_curves}
\end{figure}

The training process remains stable throughout the 40\,000-epoch Bayesian training phase, as illustrated in \autoref{fig:training_curves}. The ELBO decreases steadily, indicating consistent improvement of the variational posterior. At the same time, the NLL reaches negative values, showing that the heteroscedastic likelihood accurately captures the data distribution and assigns high confidence to the predictions.
The KL divergence decreases monotonically to 0.88, reflecting a progressive concentration of the posterior distributions as the
model gains confidence in the learned weights. The learnable precision parameter $\alpha$ exhibits a non-monotonic trajectory, initially decreasing to approximately 0.91 during early exploration and subsequently increasing to approximately 1.12 at convergence. This behavior confirms that the data favors a relatively broad prior while still allowing posterior concentration.

\begin{figure}[htbp]
\centering
\includegraphics[width=0.95\textwidth]{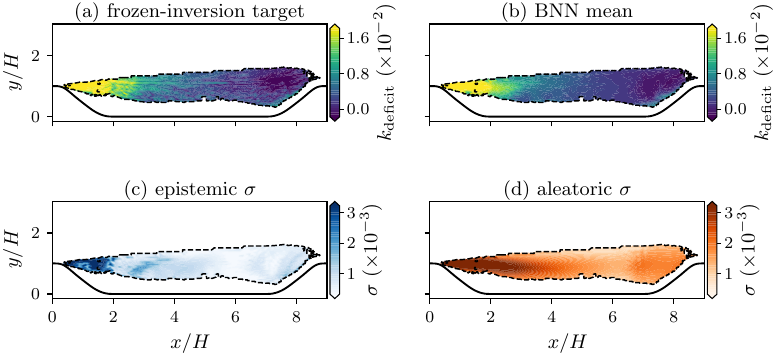}
\caption{\textit{A priori} scalar correction on the PH, classified cells. (a) Frozen-inversion target. (b) BNN mean over $M=100$ posterior samples. (c) Epistemic uncertainty. (d) Aleatoric uncertainty. Before propagation the two uncertainty components are separable, so both are shown; (a)--(b) share one colour scale and (c)--(d) another.Colour scales are clipped at the 90th percentile of the classified values, since the correction is strongly right-skewed.}
\label{fig:spatial_pred}
\end{figure}

The learned model also accurately captures the spatial structure of the
correction field. As shown in \autoref{fig:spatial_pred}, the high-deficit region downstream of the hill crest associated with flow separation is well resolved, while the freestream region remains close to zero. This demonstrates that the mapping from invariant features to correction magnitude generalizes consistently across different flow regimes.
The corresponding error field remains small and largely unstructured, with the largest deviations localized near the hill crest where gradients are steepest.
The absence of coherent spatial patterns indicates that the model does not exhibit systematic bias and maintains consistent accuracy across the domain.

The spatial distribution of uncertainty further reflects the underlying flow physics, in panels (c) and (d) of the same figure, which share a colour scale so that the two components can be read against each other. The epistemic component is compact, concentrated immediately downstream of the crest where the shear layer forms and where the training data constrain the model least. It falls to near zero over most of the recovery region. The aleatoric component is broader and follows the magnitude of the correction itself across the whole classified band. It is the larger of the two nearly everywhere, consistent with the mean values quoted above. This behavior is consistent with the heteroscedastic formulation, in which the predicted noise scales with the local turbulence activity.
The total uncertainty combines both contributions and remains localized in physically complex regions while staying low in the freestream, indicating reliable predictions in well-resolved flow regimes.

\subsection{\texorpdfstring{Anisotropy correction: $b_{ij}^{\Delta}$ model performance}{Anisotropy correction: bij Delta model performance}}
\label{sec:bijdelta_results}

In addition to the scalar correction, the anisotropy correction
$b_{ij}^{\Delta}$ is modeled to capture deviations of the Reynolds stress tensor from the Boussinesq hypothesis. The tensor-basis formulation introduced in \S\ref{sec:features_basis} is used to represent the correction field in terms of scalar coefficient functions.
The selection of an appropriate tensor basis is critical for accurately
representing the anisotropy correction. To this end, all combinations of the first four Pope basis tensors are evaluated using a pointwise
least-squares projection of $b_{ij}^{\Delta}$ onto the candidate bases
over the classified correction region (2{,}378 cells).

\begin{figure}[htbp]
\centering
\includegraphics[width=\textwidth]{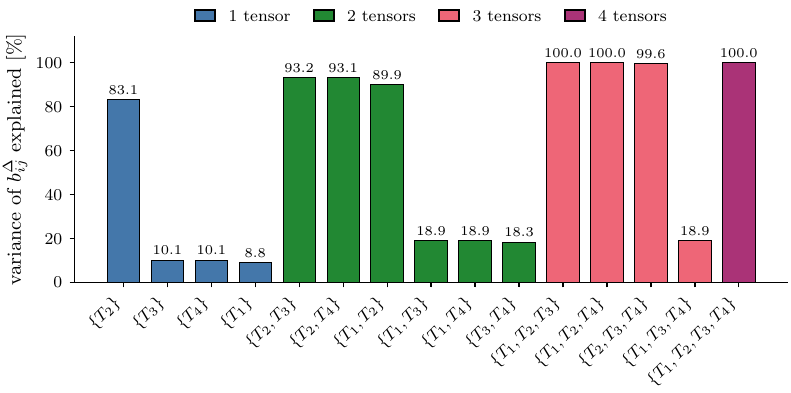}
\caption{Variance explained ($R^2$) by projection of
$b_{ij}^{\Delta}$ onto combinations of the first four Pope basis tensors.
The strain--rotation commutator ($T_2$) dominates among single bases,
while the combination ${T_1, T_2, T_3}$ achieves near-complete
representation of the anisotropy correction.}
\label{fig:basis_selection}
\end{figure}

The resulting variance-explained ($R^2$) values are shown in
\autoref{fig:basis_selection}. Among single-basis representations, the
strain--rotation commutator term $T_2$ dominates, achieving
$R^2 = 76.0\%$, while the remaining bases contribute significantly less
($T_3$: 17.2\%, $T_4$: 17.1\%, $T_1$: 9.4\%). This clearly reflects the
underlying flow physics of the PH case, where strong
strain--rotation interactions arise in the separation and reattachment
regions.
Combining basis functions leads to a substantial improvement in
representation accuracy. The two-basis combinations $\{T_2, T_3\}$ and $\{ T_2, T_4\}$ both reach approximately $93\%$ variance explained.
However, these combinations fail to adequately capture the
off-diagonal shear component $b_{xy}^{\Delta}$.
The inclusion of the strain-aligned tensor $T_1$ proves essential.
Although $T_1$ explains only a small fraction of the variance when used alone, it provides complementary information that is not represented by $T_2$ and $T_3$. The three-basis combination $\{T_1, T_2, T_3\}$ achieves $R^2 = 99.9\%$, representing a significant improvement over any two-basis configuration.
This gain is disproportionate to the standalone contribution of $T_1$, indicating that it captures shear-aligned components that are orthogonal to the span of $T_2$ and $T_3$. In contrast, adding the fourth basis $T_4$ does not provide any further improvement, confirming that it is redundant for this flow configuration.

Instead of predicting the full anisotropy tensor $b_{ij}^{\Delta}$ directly, the model learns the scalar coefficients $g_n$ associated with the selected tensor basis. These coefficients are obtained from the reference data through a pointwise least-squares projection onto the chosen basis (see ~\ref{app:g_projection}), resulting in
well-conditioned targets with magnitudes $\mathcal{O}(1\text{--}10)$ and approximately Gaussian distributions.
A key design choice is to perform direct coefficient regression rather than end-to-end prediction of the reconstructed tensor. In an end-to-end formulation, the loss is evaluated on the reconstructed tensor $b_{ij}^{\Delta}$, and the gradients with respect to the coefficients are scaled by the magnitude of the local basis tensors. As a result, regions where the basis tensors have small norm provide only weak supervision, even if the prediction error is large. This leads to gradient dilution effect, where learning becomes spatially inconsistent and dominated by regions with large basis magnitude.

Direct regression of the $g$-coefficients avoids this issue. Each coefficient is supervised independently, ensuring that all training points contribute uniformly to the learning signal. This leads to more stable optimization and better-conditioned training across the entire domain.
This formulation also provides important advantages for the Bayesian framework. First, uncertainty can be decomposed at the level of individual coefficients, allowing the model to learn which components of the anisotropy correction are intrinsically more uncertain. Second, calibration can be assessed both at the coefficient level and after reconstruction of the tensor, providing insight into how uncertainty propagates through the basis representation. Finally, the
projected coefficients form well-behaved targets that are naturally compatible with Gaussian likelihood assumptions, improving both training stability and uncertainty calibration.
Overall, direct $g$-coefficient regression provides a stable and interpretable learning framework, while preserving the physical structure imposed by the tensor basis representation.

The $b_{ij}^{\Delta}$ model extends the scalar formulation to a
multi-output setting, where the three coefficients associated with the
selected tensor basis are predicted simultaneously. The architecture
follows the same fully Bayesian design introduced for the
$k_{\text{deficit}}$ model, with shared hidden representations and
multiple outputs.
A key feature of this formulation is that each coefficient is assigned
an independent aleatoric uncertainty. This allows the model to learn
which components of the anisotropy correction are intrinsically more
uncertain, providing a more detailed uncertainty decomposition than in
the scalar case.
Unlike the $k_{\text{deficit}}$ model, no additional basis scaling is
applied during training. The $g$-coefficients are already well-conditioned quantities, and the tensor reconstruction is performed only at inference time. This separation ensures that the learning problem remains well-scaled while preserving the physical structure imposed by the tensor basis.
Further architectural and normalization details follow directly from the scalar model and are not repeated here.

The training procedure follows the same two-phase strategy used for the
$k_{\text{deficit}}$ model, consisting of deterministic pretraining followed by Bayesian fine-tuning with the full ELBO objective. All prior assumptions and hyperparameters are kept identical to ensure consistency between the scalar and tensor formulations.
A notable difference arises in the learned prior precision. The
$b_{ij}^{\Delta}$ model converges to $\alpha \approx 1.61$, corresponding to a tighter effective prior than in the scalar case
($\alpha \approx 1.12$). This reflects the increased output dimensionality and the need for stronger regularization when learning multiple correlated coefficients simultaneously.

At inference time, the model generates MC samples of the
$g$-coefficients, which are then mapped back to the physical tensor space through the selected basis. This reconstruction step produces spatially resolved realizations of the anisotropy correction field, enabling uncertainty propagation through the RANS solver.
Further implementation details of the likelihood, training procedure,
and inference strategy are provided in ~\ref{app:bij_training}.

The $b_{ij}^{\Delta}$ BNN is trained on the PH case using the $\{T_1, T_2, T_3\}$ basis. The quantitative performance is summarized in \autoref{tab:bijdelta_results}.
The reconstructed anisotropy correction achieves a global $R^2$ of $67.5\%$, indicating
that the model captures a substantial fraction of the variability in the target field,
although appreciable reconstruction error remains. Among the individual components,
$b_{zz}^{\Delta}$ exhibits the highest accuracy ($R^2=84.2\%$), followed by
$b_{xy}^{\Delta}$ ($68.1\%$), $b_{xx}^{\Delta}$ ($60.2\%$), and
$b_{yy}^{\Delta}$ ($48.3\%$). The components satisfy the tracelessness constraint
$b_{zz}^{\Delta}=-(b_{xx}^{\Delta}+b_{yy}^{\Delta})$ through the tensor-basis
representation.
The predicted tensor-basis coefficients are reasonably calibrated, with empirical
coverages of $64$--$78\%$ at $1\sigma$ and $91$--$94\%$ at $2\sigma$. After
reconstruction of the anisotropy tensor, the corresponding component-wise coverages are
lower, ranging from $37$--$49\%$ at $1\sigma$ and $58$--$75\%$ at $2\sigma$.
Thus, calibration of the coefficient distributions does not translate directly into
equivalent coverage of the reconstructed tensor components.

The revised shift-free normalization described in \S\ref{sec:features_basis} results in
lower in-distribution accuracy for the anisotropy model than the normalization used in
the originally submitted formulation. In the revised normalization, feature means from
the PH training flow are not subtracted, so no case-specific centering offset is carried
into unseen configurations. The consequences of this choice for out-of-distribution
prediction and uncertainty calibration are assessed in \S\ref{sec:generalization}.

A key observation is the importance of the $T_1$ basis for capturing the shear component $b_{xy}^{\Delta}$. Without this term, the anisotropy correction fails to reproduce the off-diagonal stress, leading to degraded downstream predictions.

\begin{table}[htbp]
\centering
\small
\caption{Performance of the $b_{ij}^{\Delta}$ BNN on the PH case.}
\begin{tabular}{@{}lr@{}}
\toprule
\textbf{Metric} & \textbf{Value} \\
\midrule
Global $R^2$ (reconstructed $b_{ij}^{\Delta}$) & 67.5\% \\
\midrule
$b_{xx}^{\Delta}$ & 60.2\% \\
$b_{xy}^{\Delta}$ & 68.1\% \\
$b_{yy}^{\Delta}$ & 48.3\% \\
$b_{zz}^{\Delta}$ & 4.2\% \\
\midrule
$g_1$ calibration ($1\sigma / 2\sigma$) & 72\% / 93\% \\
$g_2$ calibration ($1\sigma / 2\sigma$) & 78\% / 94\% \\
$g_3$ calibration ($1\sigma / 2\sigma$) & 64\% / 91\% \\
\midrule
$b_{xx}^{\Delta}$ coverage ($1\sigma / 2\sigma$) & 37\% / 62\% \\
$b_{xy}^{\Delta}$ coverage ($1\sigma / 2\sigma$) & 42\% / 67\% \\
$b_{yy}^{\Delta}$ coverage ($1\sigma / 2\sigma$) & 49\% / 75\% \\
$b_{zz}^{\Delta}$ coverage ($1\sigma / 2\sigma$) & 37\% / 58\% \\
\midrule
Converged $\alpha$ & 1.61 \\
\bottomrule
\end{tabular}

\label{tab:bijdelta_results}
\end{table}

\begin{figure}[htbp]
\centering
\includegraphics[width=\textwidth]{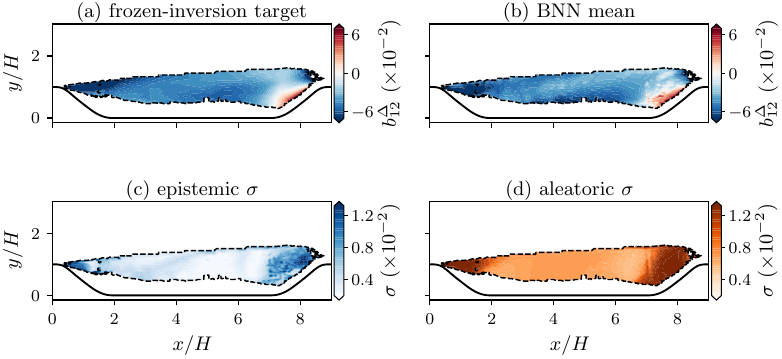}
\caption{\textit{A priori} in-plane shear component of the dimensionless anisotropy correction, $b^{\Delta}_{12}$, on the PH. Panels as in \autoref{fig:spatial_pred}.}
\label{fig:bijdelta_spatial}
\end{figure}

The spatial structure of the anisotropy correction is shown in
\autoref{fig:bijdelta_spatial}.
Only the in-plane shear component $b^{\Delta}_{12}$ is plotted, this being the component
that acts on the mean momentum balance in a two-dimensional shear flow. The other
components are predicted simultaneously and their accuracy is reported in
\autoref{tab:bijdelta_results}. The model reproduces
the magnitude, sign and spatial evolution of the plotted component, which is largely
negative and spatially smooth with its strongest values in the shear layer downstream of
separation. Panels (a) and (b) differ little over most of the domain, and most near the
trailing edge ($x/H > 8$), where the flow changes rapidly and the training data are
comparatively sparse.
The predicted field is smoother than the target, consistent with the residual reconstruction
error quantified in \autoref{tab:bijdelta_results}.

The epistemic uncertainty in panel~(c) is highest near the boundaries of the classified
region and in areas with strong spatial gradients, and remains low within the core shear
layer where the training data are dense. Its magnitude is approximately $10$--$20\%$ of
the signal amplitude, so the model provides informative but controlled uncertainty
estimates.

\begin{figure}[htbp]
\centering
\includegraphics[width=\textwidth]{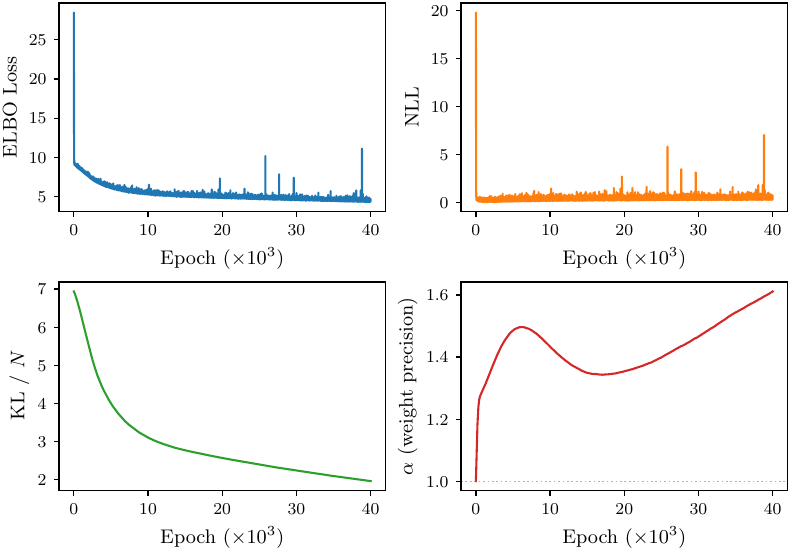}
\caption{Training convergence of the $b_{ij}^{\Delta}$ BNN.}
\label{fig:bijdelta_training}
\end{figure}

The training process remains stable throughout the Bayesian fine-tuning
phase, as shown in \autoref{fig:bijdelta_training}. The ELBO decreases
monotonically, while the NLL quickly converges to negative values,
indicating accurate modeling of heteroscedastic noise.
The KL divergence decreases smoothly, reflecting a controlled expansion of the variational posterior from its narrow initialization. The learned precision parameter $\alpha$ exhibits a characteristic overshoot followed by relaxation, converging to $\alpha \approx 1.91$. This value is significantly higher than in the scalar model, indicating that stronger regularization is required for multi-output learning with shared representations.

\subsection{Stochastic propagation results}
\label{sec:propagation_results}

The trained BNN corrections are propagated through the RANS solver as
stochastic ensembles. To isolate the physical role of each correction,
propagation is performed in two stages: (i)~using only
$k_{\text{deficit}}$, and (ii)~using the combined
$k_{\text{deficit}} + b_{ij}^{\Delta}$ model.
Each MC sample is propagated through the augmented
k--$\omega$ SST solver (OpenFOAM), producing an ensemble of steady-state flow
solutions. The learned corrections are applied only within the
identified correction region via the classifier mask, while outside this region the model reverts to the baseline RANS formulation.
The ensemble is evaluated against the frozen SpaRTA inversion
reference, obtained by propagating the exact inversion-derived corrections through the solver. This provides a deterministic target that isolates the learning error of the BNN from modeling limitations.
Uncertainty quality is assessed using a coverage metric. For a quantity of interest $\phi$, the coverage at confidence level $p$ is defined as
\begin{equation}
    C_p
    =
    \frac{1}{N_{\text{cells}}}
    \sum_{i=1}^{N_{\text{cells}}}
    \mathbb{1}\!\left[
        \phi_i^{\sigma}
        \in
        \left[
            \bar{\phi}_i^{\text{BNN}} - z_p\,\hat{\sigma}_i,
            \;
            \bar{\phi}_i^{\text{BNN}} + z_p\,\hat{\sigma}_i
        \right]
    \right],
\end{equation}
where $\bar{\phi}^{\text{BNN}}$ and $\hat{\sigma}$ denote the ensemble
mean and standard deviation, respectively, and $z_p$ is the Gaussian
quantile ($z_{0.68}=1$, $z_{0.95}=2$).

\begin{figure}[htbp]
\centering
\includegraphics[width=\textwidth]{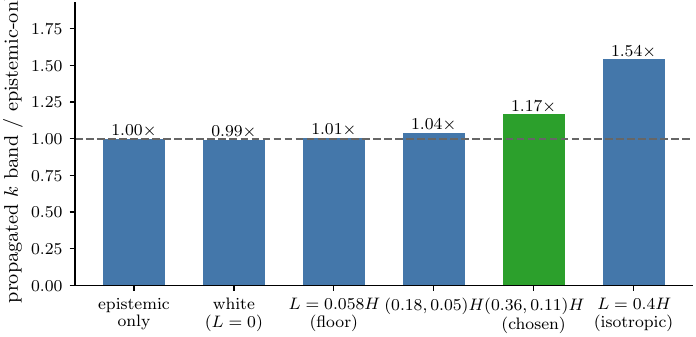}
\caption{Propagated TKE band relative to the epistemic-only ensemble, for aleatoric fields of increasing correlation length on the periodic hill (ten samples per configuration, sixty solves in total). Lengths are in units of the hill height $H$; where two values are given, the first is the streamwise ($x$) and the second the wall-normal ($y$) correlation length. White noise ($L = 0$) produces no measurable widening; the value used throughout this study is highlighted.}
\label{fig:corrlen_amplification}
\end{figure}

The screening study associated with \cref{eq:aleatoric_acf} is summarized in
\autoref{fig:corrlen_amplification}. Statistically independent cellwise perturbations
produce essentially no additional propagated spread relative to the epistemic-only
ensemble ($0.99\times$), despite having the same pointwise variance. This indicates that
the influence of the aleatoric perturbation is strongly attenuated when no spatial
coherence is imposed. As the correlation length is increased, a progressively larger
fraction of the imposed variability survives propagation through the RANS solver.

The correlation length used in the subsequent calculations is calibrated once on the PH
training case and is then retained unchanged for all test configurations. The empirical
residual correlation lengths are $(L_x,L_y)=(0.178,0.054)H$, while the values adopted for
propagation are $(0.355,0.108)H$, preserving the measured anisotropy ratio. The screening
study shows that the adopted correlation length increases the propagated band by
approximately $17\%$ relative to the epistemic-only ensemble, compared with approximately
$4\%$ when the empirical length is used directly. A larger isotropic correlation length
of $0.4H$ produces a substantially stronger amplification of approximately $54\%$.
Consequently, the absolute width of the propagated uncertainty band remains sensitive to
the assumed spatial correlation scale. The adopted value should therefore be interpreted
as a fixed calibration choice rather than as a uniquely determined correlation length.

\subsection*{Stage 1: $k_{\text{deficit}}$-only propagation (PH)}

The first propagation stage considers only the scalar $k_{\text{deficit}}$ correction, with $b_{ij}^{\Delta}=0$, in order to isolate the effect of the TKE correction before introducing anisotropy corrections. This stage therefore serves as a baseline for assessing how much of the flow reconstruction can already be achieved through scalar correction alone. Here, the term BP (best possible) refers to the reference solution obtained by directly propagating the correction fields computed from the frozen inversion framework through the RANS solver. In other words, BP represents the ideal outcome in which the correction model perfectly reproduces the target correction fields. Deviations between the propagated BNN results and BP therefore reflect the approximation error of the learned correction model, rather than limitations of the underlying propagation procedure.

\begin{figure}[htbp]
\centering
\includegraphics[width=\textwidth]{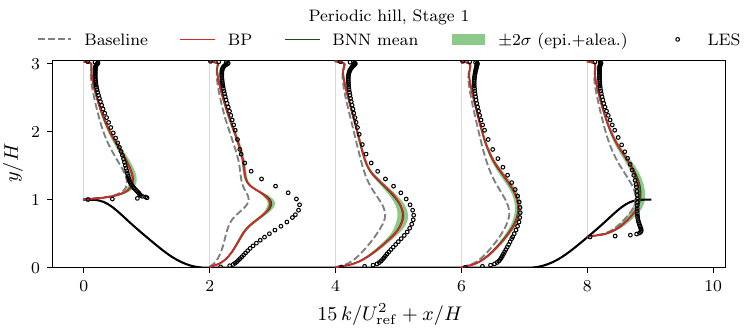}
\caption{Vertical profiles of TKE $k$ at five streamwise stations for $k_{\text{deficit}}$-only propagation. Baseline RANS, grey dashed; BP reference, red; propagated BNN mean, green, with the shaded band its propagated uncertainty; LES, black circles.}
\label{fig:konly_profiles_k}
\end{figure}

~\autoref{fig:konly_profiles_k} shows vertical profiles of TKE at five streamwise stations. As expected, baseline RANS systematically underpredicts the shear-layer TKE level throughout the separated region, with the largest discrepancies appearing around $x/H \approx 2$--$5$. In contrast, the propagated BNN mean closely follows the BP reference at all stations, reproducing both the magnitude and location of the TKE peak. The propagated uncertainty bands remain relatively narrow and are largest in the separated shear layer, where the learned scalar correction is strongest. Overall, the profile comparison confirms that the scalar BNN reproduces the BP reference in the recirculation region. BP itself remains below the LES through that region, so the residual gap to the HF data reflects the limits of the correction framework, which acts only within the classified region, rather than the accuracy of the network.

\begin{figure}[htbp]
\centering
\includegraphics[width=\textwidth]{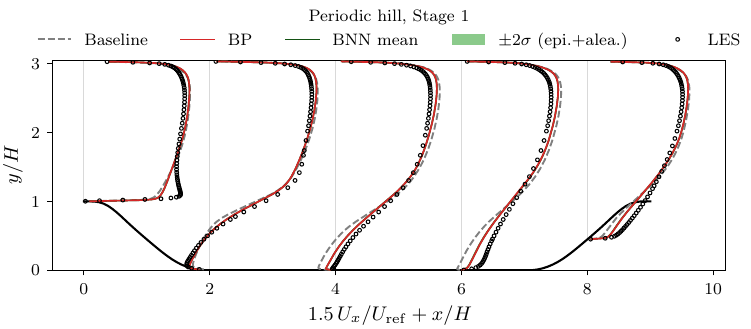}
\caption{Vertical profiles of streamwise velocity $U_x$ at five streamwise stations for $k_{\text{deficit}}$-only propagation. Symbols as in \autoref{fig:konly_profiles_k}.}
\label{fig:konly_profiles_Ux}
\end{figure}

The corresponding streamwise velocity profiles are shown in
\autoref{fig:konly_profiles_Ux}. Compared with the baseline solution, the scalar correction
produces a clear improvement in the mean velocity field, especially in the separated and
recovering regions. The propagated BNN mean remains very close to the BP reference across
all stations, indicating that the stochastic propagation preserves the deterministic
correction behavior.
The uncertainty bands remain narrow in absolute terms and are difficult to distinguish
visually at the scale of the velocity profiles. Their spatial variation is more clearly
resolved in the contour representation shown in \autoref{fig:konly_contours_Ux}. Although the improvement in $U_x$ is less dramatic than for $k$, these results show
that correcting the scalar turbulence level alone already induces a meaningful modification
of the momentum field.

\begin{figure}[htbp]
\centering
\includegraphics[width=0.78\textwidth]{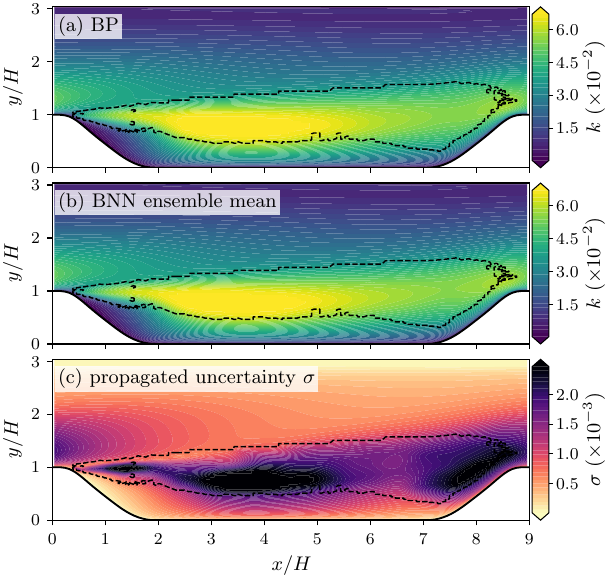}
\caption{Propagated turbulent kinetic energy on the PH, $k_{\text{deficit}}$-only propagation. (a) BP reference. (b) BNN ensemble mean. (c) Propagated uncertainty $\sigma(k)$, the standard deviation over the ensemble. The dashed contour marks the RITA-classified region where the correction is applied. Epistemic and aleatoric contributions cannot be separated after propagation, so (c) is the total.}
\label{fig:konly_contours_k}
\end{figure}

\begin{figure}[htbp]
\centering
\includegraphics[width=0.78\textwidth]{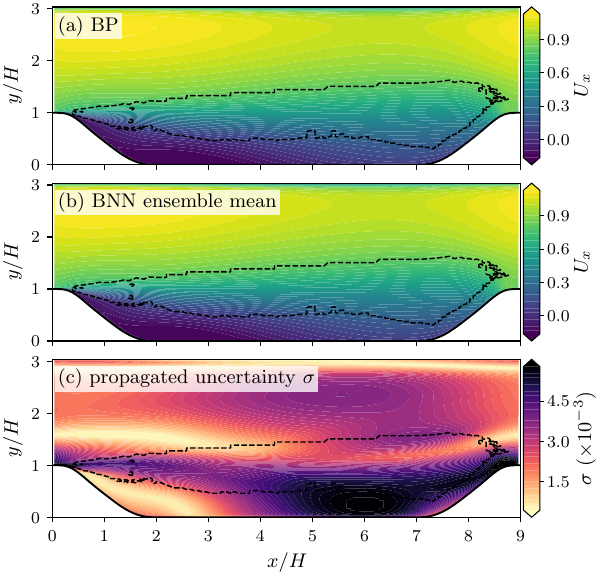}
\caption{Propagated streamwise velocity on the PH, $k_{\text{deficit}}$-only propagation. Panels as in \autoref{fig:konly_contours_k}.}
\label{fig:konly_contours_Ux}
\end{figure}

The spatial structure of the propagated fields is summarized in
\autoref{fig:konly_contours_k} for $k$ and
\autoref{fig:konly_contours_Ux} for $U_x$.
For the TKE, the propagated BNN mean reproduces the global structure of the BP reference well, with elevated turbulence levels through the separated shear layer and downstream toward reattachment.
Residual errors remain relatively small compared with the magnitude of the field and are primarily localized within the RITA-classified region, highlighted by the black dashed contour. This region corresponds to the dynamically relevant part of the flow where the correction is active.

The propagated uncertainty in $k$ is strongly concentrated along this same region, forming a coherent band that extends from the upstream shear layer through the recirculation zone and toward reattachment. This spatial alignment confirms that the largest propagated uncertainty appears where the scalar correction is most influential and where nonlinear transport effects amplify uncertainty.

The propagated $U_x$ field shows the same pattern as $k$. The BNN mean preserves the main
features of the corrected flow and improves on the baseline, the residual error relative
to BP is largest along the separated shear layer and near reattachment, and the
uncertainty is concentrated in the separation and recovery regions. The outer flow is
little affected in either field.

\begin{figure}[htbp]
\centering
\includegraphics[width=0.78\textwidth]{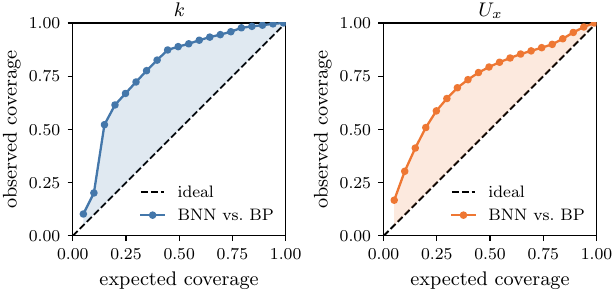}
\caption{Reliability diagrams for the $k_{\text{deficit}}$-only propagation stage on the PH, against BP. Left: turbulent kinetic energy. Right: streamwise velocity. The dashed diagonal is ideal calibration.}
\label{fig:konly_calibration}
\end{figure}

The calibration behavior is summarized in
\autoref{fig:konly_calibration}. Coverage at this stage is evaluated against BP, which
provides a reference for the error introduced by learning and propagating the
$k_{\text{deficit}}$ correction, independently of the discrepancy that remains between
the inverted correction and the HF solution. For $k$, the full-field empirical coverage
is $94.3\%$ within $\pm1\sigma$ and $99.8\%$ within $\pm2\sigma$, compared with nominal
Gaussian values of $68\%$ and $95\%$. The corresponding values for $U_x$ are
$86.6\%$ and $99.0\%$. Within the RITA-classified region, where the correction is active,
the coverage increases further to $99.6\%/100.0\%$ for $k$ and
$100.0\%/100.0\%$ for $U_x$. Accordingly, the reliability curves lie above the ideal
diagonal, showing that the Stage~1 propagated ensemble is conservative, or
over-dispersed, relative to the BP reference.

This over-coverage is consistent with the relatively small displacement of the Stage~1
mean solution from BP compared with the width of the propagated uncertainty band, which
contains both epistemic and aleatoric contributions. It should therefore be interpreted
as a limitation of the Stage~1 calibration rather than as evidence of under-confidence
in the correction model. The behavior changes when the complete correction is propagated
and when the model is applied to unseen configurations, for which the corresponding
coverage results are reported in \autoref{tab:summary_coverage} and
\S\ref{sec:generalization}.

Overall, the $k_{\text{deficit}}$-only propagation stage demonstrates that the scalar BNN correction reproduces the BP reference, which is the field it is trained to reproduce rather than the LES itself. It reconstructs the TKE accurately, propagates structured uncertainty through the separated region, and leads to noticeable improvement in the mean velocity field over the baseline. At the same time, the remaining discrepancies in $U_x$, especially around the separated shear layer and reattachment, suggest that scalar correction alone is not sufficient to fully recover the Reynolds-stress redistribution. This motivates the second stage, in which the anisotropy correction $b_{ij}^{\Delta}$ is introduced in addition to $k_{\text{deficit}}$.

\subsection*{Stage 2: Combined $k_{\text{deficit}} + b_{ij}^{\Delta}$ propagation (PH)}

The second propagation stage applies both the scalar TKE correction and the anisotropy
correction simultaneously. The tensorial correction $b_{ij}^{\Delta}$ is reconstructed
using the $\{T_1,T_2,T_3\}$ basis, enabling direct modification of the Reynolds-stress
structure. This stage reveals the dominant role of anisotropy in improving the mean-flow
prediction.

\begin{figure}[htbp]
\centering
\includegraphics[width=\textwidth]{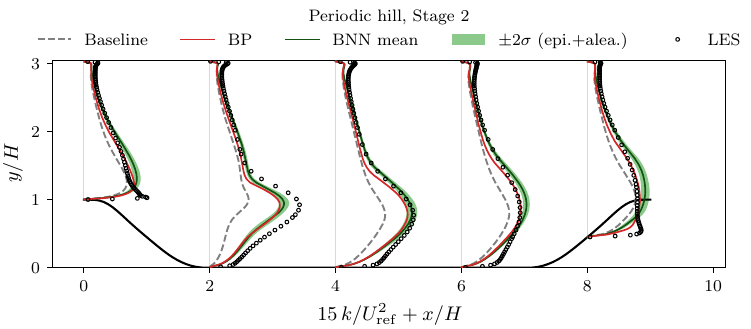}
\caption{Vertical profiles of TKE $k$ at five streamwise stations for combined propagation. Symbols as in \autoref{fig:konly_profiles_k}, with the shaded band showing $\pm 2\sigma$.}
\label{fig:combined_profiles_k}
\end{figure}

Vertical profiles of TKE at five streamwise stations are shown in
\autoref{fig:combined_profiles_k}. Compared with the $k_{\text{deficit}}$-only case,
the combined model produces higher TKE levels throughout the separated shear layer.
The BNN ensemble mean remains close to BP over the profiles, with BP contained within the
$\pm2\sigma$ interval on $87.8\%$ of the RITA-classified cells. Both the BNN and BP,
however, retain a systematic discrepancy relative to LES through much of the separated
region. Since a similar discrepancy is present when the exact inversion-derived
corrections are propagated, this difference reflects a limitation of the correction
framework in addition to the approximation introduced by the BNN.
Downstream of reattachment, the BNN and BP predictions continue to show an excess in TKE
relative to LES, whereas their corresponding velocity predictions remain much closer to
one another. This behavior is consistent with previous observations for corrected
$k$--$\omega$ SST closures \cite{buchanan2025data}, where improvement of the mean-flow
prediction does not necessarily imply an equally accurate reconstruction of the turbulence
energy level.

\begin{figure}[htbp]
\centering
\includegraphics[width=\textwidth]{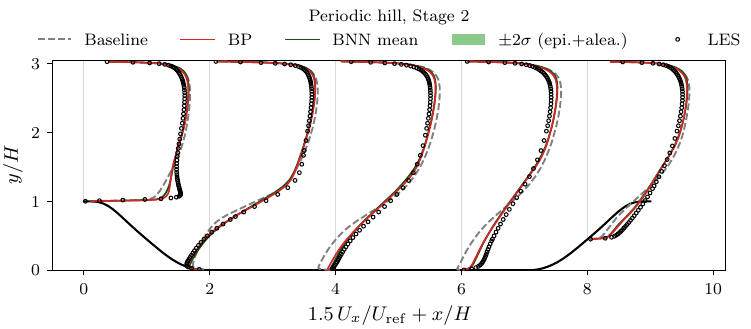}
\caption{Vertical profiles of streamwise velocity $U_x$ at five stations for combined propagation. Symbols as in \autoref{fig:konly_profiles_k}.}
\label{fig:combined_profiles_Ux}
\end{figure}

The effect on the velocity field is substantially more pronounced than in Stage~1, as shown
in \autoref{fig:combined_profiles_Ux}. The inclusion of $b_{ij}^{\Delta}$ enables accurate
prediction of separation, recirculation, and recovery. Negative velocities in the
recirculation region and the downstream evolution of the shear layer are well captured,
whereas the baseline RANS solution shows substantially larger discrepancies.
The BNN mean remains close to BP across the profiles, with BP contained within the
$\pm2\sigma$ interval on $97.6\%$ of the classified cells. A residual discrepancy relative
to LES remains, however, and is appreciably larger than the difference between the BNN
mean and BP.

\begin{figure}[htbp]
\centering
\includegraphics[width=0.78\textwidth]{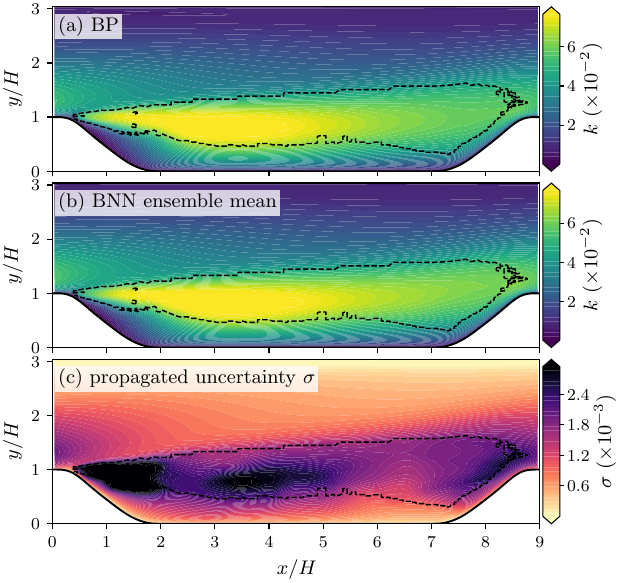}
\caption{Propagated turbulent kinetic energy on the PH, combined propagation. Panels as in \autoref{fig:konly_contours_k}.}
\label{fig:combined_contours_k}
\end{figure}

\begin{figure}[htbp]
\centering
\includegraphics[width=0.78\textwidth]{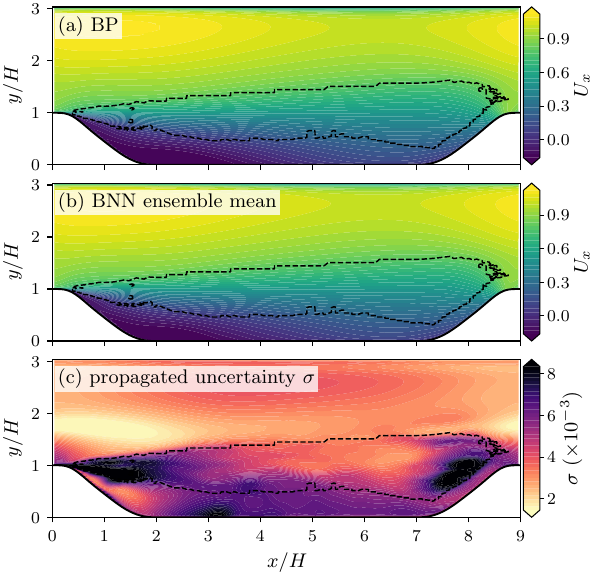}
\caption{Propagated streamwise velocity on the PH, combined propagation. Panels as in \autoref{fig:konly_contours_k}.}
\label{fig:combined_contours_Ux}
\end{figure}

The spatial structure of the propagated fields is shown in
\autoref{fig:combined_contours_k} for $k$ and
\autoref{fig:combined_contours_Ux} for $U_x$.
For TKE, the propagated BNN mean reproduces the BP structure well, including the elevated
turbulence levels within the separated region. The propagated uncertainty is spatially
structured and reaches its largest values in and around the separated shear layer and
recirculation region.

For the streamwise velocity, the improvement relative to Stage~1 is clear. The
ensemble-mean field closely follows the BP solution and captures the separation bubble and
subsequent recovery. The propagated uncertainty in $U_x$ is also spatially nonuniform,
with larger values near the lower wall and toward the recovery region, reflecting the
propagation of uncertainty in the corrected Reynolds stresses through the coupled RANS
equations.

\begin{figure}[htbp]
\centering
\includegraphics[width=0.8\textwidth]{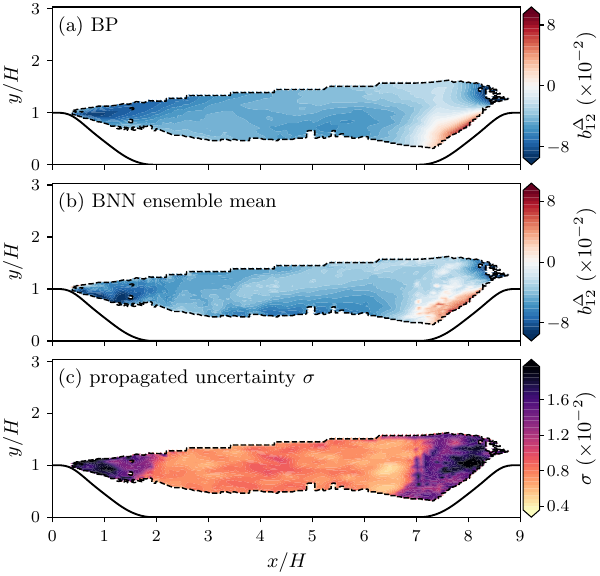}
\caption{Propagated anisotropy correction on the PH under combined propagation. Only the in-plane shear component $b^{\Delta}_{12}$ is shown.} (a) BP reference. (b) BNN ensemble mean. (c) Total propagated uncertainty, epistemic and aleatoric combined; the two are inseparable after propagation. Shown on the classified cells, where the correction is active.
\label{fig:combined_bijdelta_components}
\end{figure}

The reconstructed anisotropy field is shown in
\autoref{fig:combined_bijdelta_components}.
As in \autoref{fig:bijdelta_spatial}, only the in-plane shear component
$b^{\Delta}_{12}$ is shown; the normal components are corrected simultaneously but are not
plotted.
The BNN mean reproduces the principal spatial distribution and sign of the BP correction,
although some attenuation of the peak amplitudes remains. The largest visible differences
occur near the boundaries of the RITA-classified region.
The propagated uncertainty is concentrated primarily near these regions and toward
reattachment, where the predicted correction also varies most strongly.

\begin{figure}[htbp]
\centering
\includegraphics[width=0.78\textwidth]{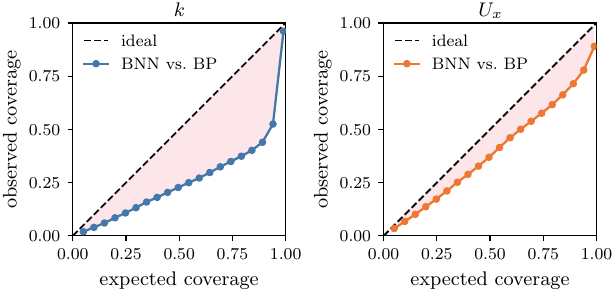}
\caption{Reliability diagrams for the combined propagation stage on the PH, against BP. Left: turbulent kinetic energy. Right: streamwise velocity.}
\label{fig:combined_calibration}
\end{figure}

Coverage statistics for the combined propagation are shown in
\autoref{fig:combined_calibration}, with the corresponding values for all configurations
collected in \autoref{tab:summary_coverage}.
In contrast to Stage~1, the combined correction produces a substantially larger displacement
of the flow solution and the propagated intervals become under-dispersed relative to BP.
Over the full field, the $1\sigma/2\sigma$ coverage is $31.9\%/68.5\%$ for $k$ and
$53.0\%/80.1\%$ for $U_x$, below the nominal Gaussian values of $68\%/95\%$.
Within the RITA-classified region, where the corrections are applied, coverage improves to
$56.9\%/87.8\%$ for $k$ and $68.6\%/97.6\%$ for $U_x$.

The difference between full-field and classified-region coverage reflects the spatial
localization of the correction and its subsequent influence on the flow. Uncertainty is
introduced within the classified region, whereas the resulting perturbations to the flow
can be transported beyond that region. Consequently, the full-field statistic includes
downstream locations where a residual flow error persists even though the locally generated
correction uncertainty is smaller. The classified-region results therefore provide a
complementary assessment of calibration in the region where the correction model acts
directly.

Taken together, these results show that the propagated uncertainty on the PH is not
uniformly calibrated: velocity coverage within the corrected region is close to nominal,
whereas TKE and the corresponding full-field statistics remain under-dispersed. The
streamwise velocity achieves higher coverage than $k$, consistent with the stronger
influence of the anisotropy correction on the momentum prediction. This distinction is
considered further in \S\ref{sec:conclusion}.

Overall, the combined propagation substantially improves the physical fidelity of the flow
prediction, particularly for the streamwise velocity. At the same time, the close agreement
between the BNN mean and BP, together with their similar discrepancies relative to LES,
indicates that a substantial part of the remaining prediction error is associated with
limitations of the correction framework rather than with the BNN approximation itself.

\section{Generalization to unseen configurations}
\label{sec:generalization}

The model developed in \S\ref{sec:results} is trained exclusively on the PH configuration
of Breuer et al.\ \cite{breuer2009flow} at $Re_H = 10{,}595$. Its generalization is
assessed here, without retraining, on five unseen configurations: the curved
backward-facing step (CBFS) of Bentaleb et al.\ \cite{bentaleb2012large} at
$Re_H = 13{,}700$; three parameterized periodic hills of Xiao et al.\
\cite{xiao2020flows}, with hill-width parameters $\alpha = 0.5$, $1.0$, and $1.5$, all
at $Re_H = 5600$; and the NASA wall-mounted hump of Uzun and Malik
\cite{uzun2017nasahump} at $Re_c = 9.36\times10^{5}$.

The parameterized hills provide a graded generalization test within the same geometry
family. Changing the hill width modifies the separation length and reattachment location,
while all three cases are evaluated at approximately half the Reynolds number of the
training flow ($Re_H=5600$ compared with $10{,}595$). They therefore assess the combined
effect of geometric and Reynolds-number variation while spanning flows from relatively
limited to extensive separation. The NASA hump provides a more substantial departure
from the training configuration: its geometry, coordinate plane, dimensional formulation,
and Reynolds number all differ. Because the hump is posed in the $x$--$z$ plane, its
in-plane anisotropy correction is $b_{13}^{\Delta}$ rather than
$b_{12}^{\Delta}$. Moreover, $29.7\%$ of its classified cells lie beyond three standard
deviations of the PH training-feature distribution. Each configuration is propagated
through the same Stage~1 and Stage~2 procedure used for the training flow, using
100-member ensembles for the CBFS and parameterized hills and a 50-member ensemble for
the NASA hump.

\subsection{Curved backward-facing step}
\label{sec:cbfs_results}

The CBFS provides the first test of the trained framework on an unseen flow configuration.
Its geometric and flow characteristics differ from those of the PH training case, while
the learned models, normalization parameters, classifier threshold, and uncertainty
settings are retained unchanged.

A notable difference from the PH is the spatial extent of the region in which the
corrections are active. The RITA-classified region comprises only $1.8\%$ of the CBFS
cells, compared with $15.2\%$ for the training flow. The local correction magnitudes,
however, remain comparable. The median
$k_{\text{deficit}}/\varepsilon_{\mathrm{base}}$ ratio is $0.38$ on the CBFS and
$0.37$ on the PH, while the corresponding median anisotropy ratios are $0.83$ and
$0.84$, respectively. Thus, the weaker overall influence of the correction on this flow
is associated more naturally with the smaller spatial extent of the corrected region
than with a reduction in the local correction amplitude.

\begin{figure}[htbp]
\centering
\includegraphics[width=\textwidth]{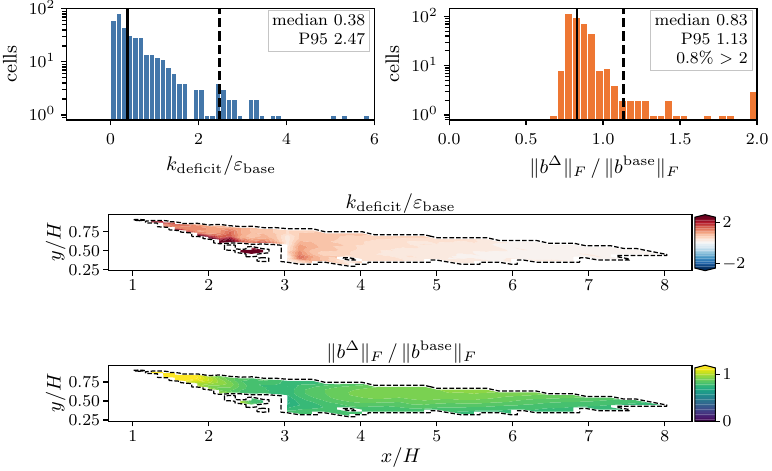}
\caption{Magnitude of the two corrections on the CBFS over the $379$ classified cells,
using the same definitions and axes as \autoref{fig:correction_magnitude_PH}. The median
values ($0.38$ and $0.83$) are close to those of the PH training case, whereas the
classified region comprises only $1.8\%$ of the CBFS cells compared with $15.2\%$ for
the PH.}
\label{fig:correction_magnitude_CBFS}
\end{figure}

\autoref{fig:correction_magnitude_CBFS} shows the distributions and spatial support of
the two correction ratios using the same definitions and axes as
\autoref{fig:correction_magnitude_PH}. The CBFS is therefore milder in terms of the
spatial extent of the corrected region, while the local correction amplitudes remain
comparable to those of the PH training case. The remaining test configurations extend the
generalization assessment to stronger geometric variation and substantially different
Reynolds numbers.
Given the smaller spatial extent of the corrected region, the resulting changes to the
global flow field are expected to be more localized. The propagation results are presented
in two stages, consistent with the PH analysis.

\subsubsection*{Stage 1: $k_{\text{deficit}}$-only propagation}

\begin{figure}[htbp]
\centering
\includegraphics[width=\textwidth]{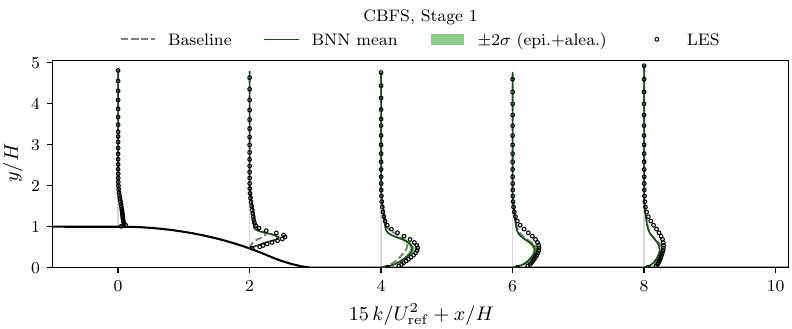}
\caption{Vertical profiles of turbulent kinetic energy $k$ for the CBFS case under $k_{\text{deficit}}$-only propagation. Baseline RANS, grey dashed; propagated BNN mean, green, with the shaded band showing $\pm2\sigma$; LES, black circles.}
\label{fig:cbfs_profiles_k}
\end{figure}

The first stage applies only the scalar turbulent kinetic energy correction, with
$b_{ij}^{\Delta}=0$, allowing the effect of the $k$-correction to be isolated in this
unseen configuration.
Vertical profiles of turbulent kinetic energy are shown in
\autoref{fig:cbfs_profiles_k}.
The scalar correction modifies the TKE primarily within the separated shear layer. Near
the step and in the early separated region, the propagated mean moves closer to the LES
reference than the baseline solution, whereas farther downstream the corrected solution
slightly overpredicts the LES level. The propagated uncertainty remains narrow relative
to the residual discrepancy from LES. Consistent with this behavior, the
$1\sigma/2\sigma$ coverage on the RITA-classified cells is only
$0.0\%/6.6\%$ (\autoref{tab:summary_coverage}). Thus, although the scalar correction
improves the local TKE prediction in part of the separated region, its propagated
uncertainty does not encompass the remaining discrepancy on the unseen CBFS configuration.

\begin{figure}[htbp]
\centering
\includegraphics[width=\textwidth]{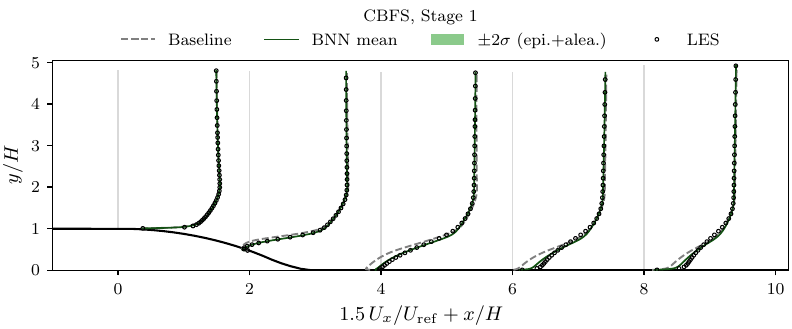}
\caption{Vertical profiles of streamwise velocity $U_x$ for the CBFS case under $k_{\text{deficit}}$-only propagation. Symbols as in \autoref{fig:cbfs_profiles_k}. Profiles are shown at the five stations at which high-fidelity data are available; the sampling lines further downstream carry no LES reference and are omitted from both stages.}
\label{fig:cbfs_profiles_Ux}
\end{figure}

The corresponding velocity profiles are presented in
\autoref{fig:cbfs_profiles_Ux}.
In contrast to the PH Stage~1 results, the scalar correction produces a noticeable change
in the mean velocity on the CBFS and generally moves the prediction toward the LES
reference. The improvement is concentrated in the near-wall and separated-flow regions,
where the baseline exhibits the largest discrepancies. Away from the wall, the propagated
mean approaches the baseline solution and the influence of the correction becomes small.
Nevertheless, the propagated intervals remain under-dispersed relative to LES, with
classified-region coverage of $20.8\%/39.6\%$ at the $1\sigma/2\sigma$ levels
(\autoref{tab:summary_coverage}).

\begin{figure}[htbp]
\centering
\includegraphics[width=0.8\textwidth]{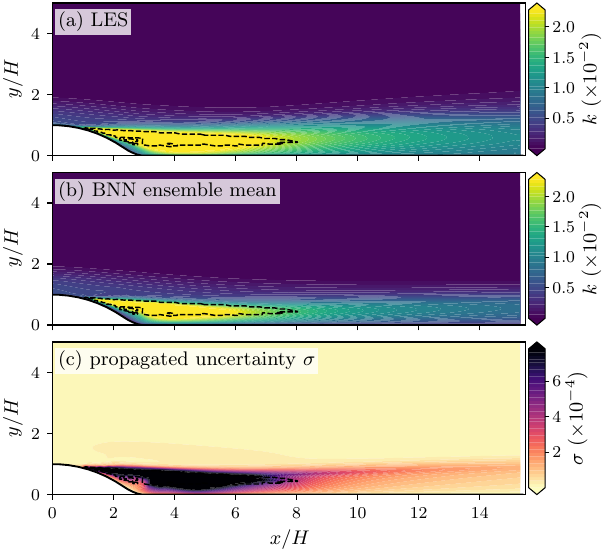}
\caption{Propagated turbulent kinetic energy on the CBFS, $k_{\text{deficit}}$-only propagation. (a) LES. (b) BNN ensemble mean. (c) Propagated uncertainty $\sigma(k)$.}
\label{fig:cbfs_contours_k}
\end{figure}

\begin{figure}[htbp]
\centering
\includegraphics[width=0.8\textwidth]{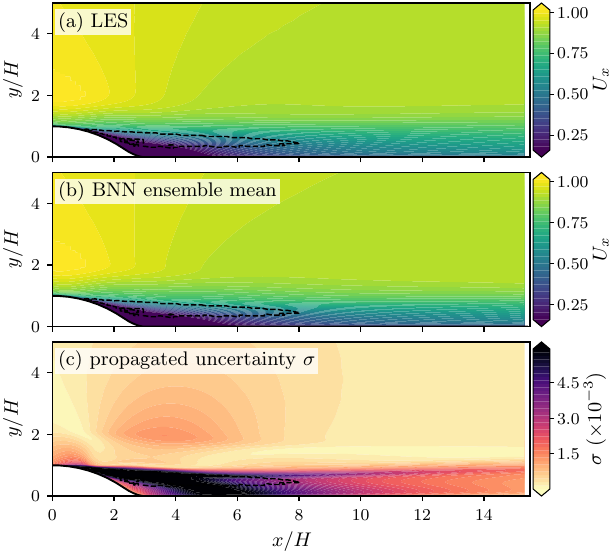}
\caption{Propagated streamwise velocity on the CBFS, $k_{\text{deficit}}$-only propagation. Panels as in \autoref{fig:cbfs_contours_k}.}
\label{fig:cbfs_contours_Ux}
\end{figure}

The spatial structure of the propagated fields is shown in
\autoref{fig:cbfs_contours_k} and \autoref{fig:cbfs_contours_Ux}.
For turbulent kinetic energy, the BNN mean modifies the solution primarily within and
downstream of the RITA-classified region, with the strongest changes occurring in the
separated shear layer.
The largest remaining differences from the LES reference are also concentrated in this
region, while the agreement improves away from the separated flow.
The propagated uncertainty is spatially localized around the corrected region
and remains comparatively small elsewhere.
For the streamwise velocity, the propagated mean follows the LES reference more closely
than the baseline within the separated and near-wall regions, while the differences among
the fields become small farther from the wall.
The propagated uncertainty exhibits a similarly localized structure, with its largest
values near the separation and recovery regions and much smaller values elsewhere.

For the unseen CBFS, propagated coverage is evaluated against the LES reference. On the
RITA-classified cells, the $1\sigma/2\sigma$ coverage is
$0.0\%/6.6\%$ for $k$ and $20.8\%/39.6\%$ for $U_x$. Over the full field, the
corresponding values are $0.1\%/0.3\%$ for $k$ and $7.1\%/14.5\%$ for $U_x$
(\autoref{tab:summary_coverage}). The propagated intervals are therefore substantially
under-dispersed relative to the HF solution at this stage.
These coverage values assess the complete predictive framework rather than the BNN
approximation alone. On the PH training case, where the exact inversion-derived
corrections can also be propagated, comparison with BP shows that a finite discrepancy
from the HF solution remains even when the learned-model error is removed. A CBFS-specific BP reference can be constructed retrospectively because HF data are
available for this benchmark, but it is not used as the primary generalization reference
because such information would not be available in a genuinely predictive application.
The CBFS coverage reported here consequently contains both transfer error in the learned
correction and the residual limitation of the correction framework.
Coverage results for all configurations are collected in
\autoref{tab:summary_coverage}.

\subsubsection*{Stage 2: Combined $k_{\text{deficit}} + b_{ij}^{\Delta}$ propagation}

\begin{figure}[htbp]
\centering
\includegraphics[width=\textwidth]{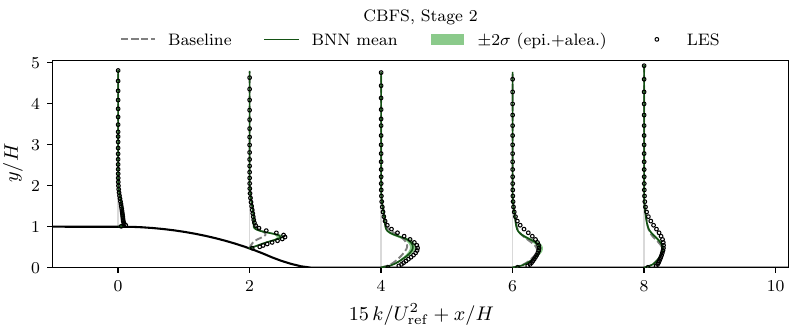}
\caption{Vertical profiles of turbulent kinetic energy $k$ for the CBFS under combined $k_{\text{deficit}} + b_{ij}^{\Delta}$ propagation.}
\label{fig:prop_combined_profiles_k_CBFS}
\end{figure}

In the second stage, the anisotropy correction $b_{ij}^{\Delta}$ is introduced in addition
to the scalar $k_{\text{deficit}}$ correction.
This modifies the Reynolds-stress tensor directly. In Stage~1, the momentum balance is
affected only indirectly through changes in the turbulence variables, whereas the
anisotropy correction permits a direct modification of the stress structure.
Vertical profiles of turbulent kinetic energy are shown in
\autoref{fig:prop_combined_profiles_k_CBFS}. Compared with the
$k_{\text{deficit}}$-only case, the combined model produces a stronger modification of
the TKE field within the separated shear layer.
The BNN mean moves closer to the LES reference than the baseline over much of the
classified region, particularly within the recirculation zone and toward reattachment.
A residual discrepancy nevertheless remains, and the propagated uncertainty bands do not
fully span the LES reference, as reflected in the coverage statistics reported below and
in \autoref{tab:summary_coverage}.

\begin{figure}[htbp]
\centering
\includegraphics[width=\textwidth]{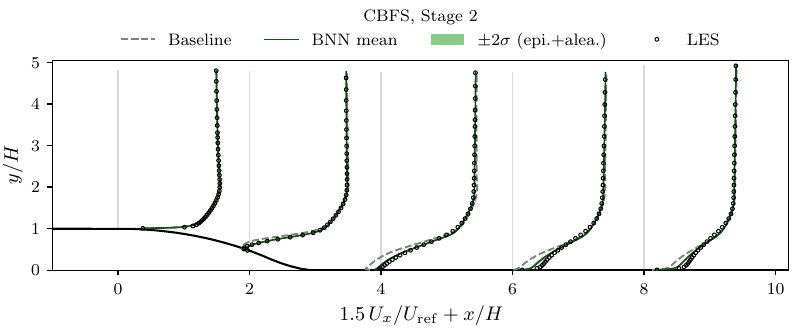}
\caption{Vertical profiles of streamwise velocity $U_x$ for the CBFS under combined propagation. Profiles are shown at the five stations with high-fidelity data, as in \autoref{fig:cbfs_profiles_Ux}.}
\label{fig:prop_combined_profiles_Ux_CBFS}
\end{figure}

The impact of the anisotropy correction on the mean flow is shown in
\autoref{fig:prop_combined_profiles_Ux_CBFS}. Relative to Stage~1, the combined correction
produces additional changes within the recirculation region and separated shear layer,
consistent with the direct modification of the Reynolds-stress anisotropy.
These changes do not, however, improve the velocity prediction relative to LES when
measured over the classified cells. The velocity RMSE increases slightly from
$2.03\times10^{-2}$ in Stage~1 to $2.26\times10^{-2}$ in Stage~2, while the
$1\sigma$ coverage increases from $20.8\%$ to $31.4\%$
(\autoref{tab:summary_coverage}). Thus, on the unseen CBFS configuration, introducing the
anisotropy correction modifies the momentum field and increases the propagated coverage,
but does not reduce the mean velocity error relative to LES.

\begin{figure}[htbp]
\centering
\includegraphics[width=0.8\textwidth]{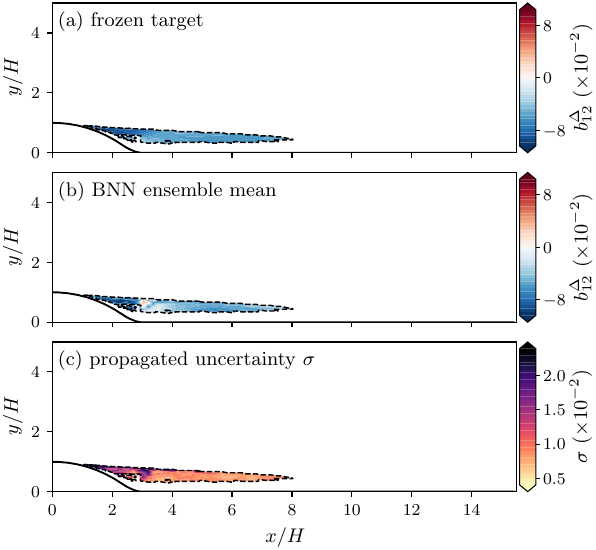}
\caption{In-plane shear correction $b_{12}^{\Delta}$ on the CBFS under combined
propagation. (a) Frozen-inversion target. (b) BNN ensemble mean. (c) Propagated
uncertainty. Results are shown on the RITA-classified cells.}
\label{fig:prop_combined_bij_CBFS}
\end{figure}

The learned anisotropy correction is shown in
\autoref{fig:prop_combined_bij_CBFS}.
Only the in-plane shear component $b_{12}^{\Delta}$ is shown. The BNN ensemble mean
reproduces the overall sign and spatial organization of the frozen-inversion target,
with the correction concentrated within the RITA-classified region.
The strongest correction occurs in the separated shear layer downstream of the step and
decays farther downstream. The associated propagated uncertainty is likewise
localized within the active region, with larger values near the upstream part of the
separated shear layer and toward the boundaries of the classified region.

\begin{figure}[htbp]
\centering
\includegraphics[width=0.8\textwidth]{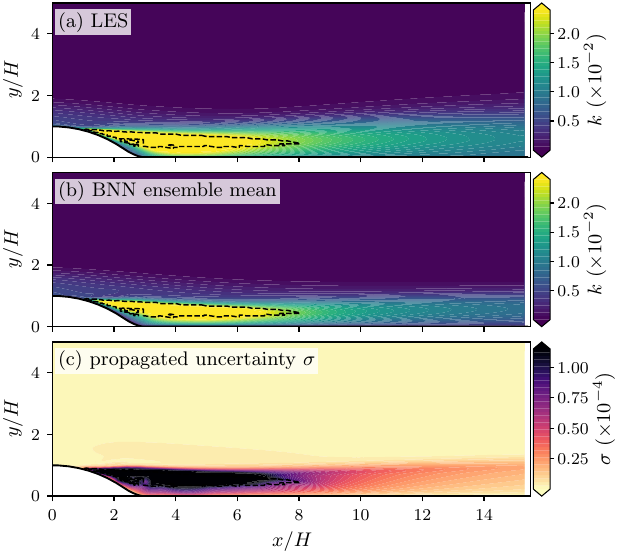}
\caption{Propagated turbulent kinetic energy on the CBFS, combined propagation. Panels as in \autoref{fig:cbfs_contours_k}.}
\label{fig:prop_combined_contours_k_CBFS}
\end{figure}

\begin{figure}[htbp]
\centering
\includegraphics[width=0.8\textwidth]{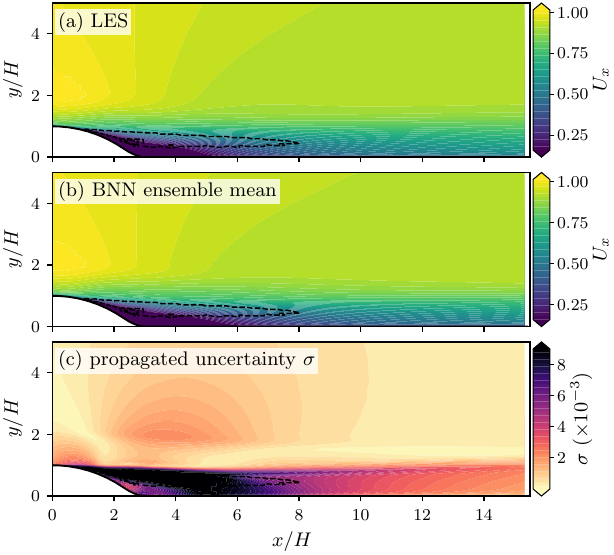}
\caption{Propagated streamwise velocity on the CBFS, combined propagation. Panels as in \autoref{fig:cbfs_contours_k}.}
\label{fig:prop_combined_contours_Ux_CBFS}
\end{figure}

The spatial structure of the propagated fields is shown in
\autoref{fig:prop_combined_contours_k_CBFS} and
\autoref{fig:prop_combined_contours_Ux_CBFS}. For turbulent kinetic energy, the BNN mean
shows a stronger response than in Stage~1 within the separated region and generally
follows the LES structure more closely there, although differences remain in the shear
layer and downstream recovery region. The propagated uncertainty is concentrated
primarily within and downstream of the RITA-classified region.

For the streamwise velocity, the anisotropy correction produces visible changes in the
recirculation and recovery regions. The overall BNN mean remains close to the LES field,
although these changes do not reduce the classified-region RMSE relative to Stage~1. The
propagated uncertainty is spatially structured, with its largest values concentrated near
the lower wall, within the separated region, and downstream toward recovery.

Adding the anisotropy correction also changes the calibration of the propagated ensemble.
Against LES, the combined propagation covers $35.4\%/64.1\%$ of the RITA-classified
cells for $k$ and $31.4\%/55.9\%$ for $U_x$ at the $1\sigma/2\sigma$ levels
(\autoref{tab:summary_coverage}). Compared with Stage~1, this represents a substantial
increase in coverage, particularly for $k$, although the intervals remain below the
nominal Gaussian levels of $68\%/95\%$.

As in Stage~1, these LES-referenced coverage values assess the complete predictive
framework on an unseen configuration rather than the BNN approximation alone. They
therefore contain both transfer error in the learned correction and residual limitations
of the correction framework. Separating these contributions would require constructing a
CBFS-specific inversion using its own HF data, which is outside the intended predictive
setting. Coverage results for all configurations are collected in
\autoref{tab:summary_coverage}.

With the normalization defined in \S\ref{sec:features_basis}, the out-of-distribution
anisotropy predictions on the CBFS remain bounded. The maximum predicted correction
magnitude is $\lVert b^{\Delta}\rVert_F = 0.62$, below the realizability bound
$\sqrt{2/3}$, while the mean magnitude over the classified cells is $0.19$. This indicates
that the bounded response is achieved without uniformly suppressing the anisotropy
correction throughout the active region.

\subsection{The parameterized hill family}

\begin{figure}[htbp]
\centering
\includegraphics[width=\textwidth]{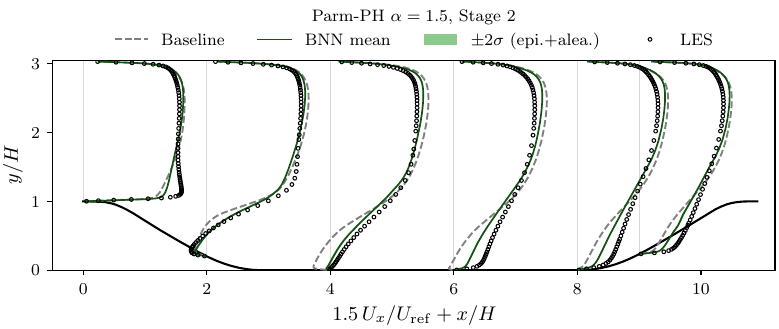}
\caption{Streamwise velocity profiles for the Parm-PH at $\alpha = 1.5$, Stage-2 propagation. Green: ensemble mean with the total $\pm 2\sigma$ band; red: LES; gray dashed: baseline RANS; circles: LES.}
\label{fig:parm_profiles_Ux_15}
\end{figure}

\begin{figure}[htbp]
\centering
\includegraphics[width=\textwidth]{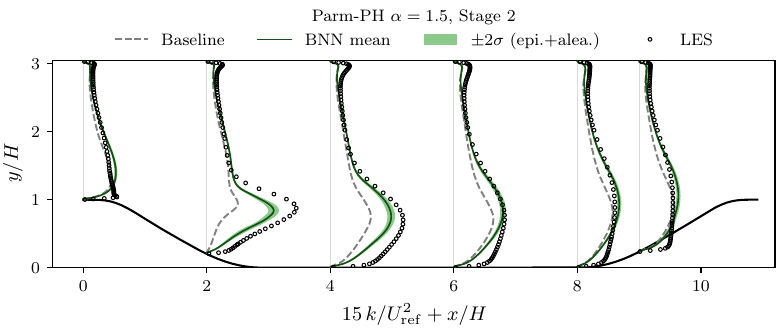}
\caption{Turbulent kinetic energy profiles for the Parm-PH at $\alpha = 1.5$, Stage-2 propagation. Symbols as in \autoref{fig:parm_profiles_Ux_15}.}
\label{fig:parm_profiles_k_15}
\end{figure}

\autoref{fig:parm_profiles_Ux_15} and \autoref{fig:parm_profiles_k_15} show the widened
hill, $\alpha=1.5$ at $Re_H=5600$, which is geometrically the closest of the
parameterized hills to the training configuration. It nevertheless represents an unseen
flow, with a wider hill and a Reynolds number approximately half that of the PH training
case ($Re_H=10{,}595$). The propagated mean generally improves upon the baseline
prediction, particularly within the separated and recovering regions, although substantial
discrepancies from LES remain and are not fully encompassed by the propagated uncertainty.
On the RITA-classified cells, the Stage~2 $1\sigma/2\sigma$ coverage against LES is
$15.2\%/28.2\%$ for $k$ and $1.9\%/3.8\%$ for $U_x$
(\autoref{tab:summary_coverage}). Thus, despite reasonable transfer of the mean correction,
the propagated uncertainty remains strongly under-dispersed on this unseen configuration.

\begin{figure}[htbp]
\centering
\includegraphics[width=\textwidth]{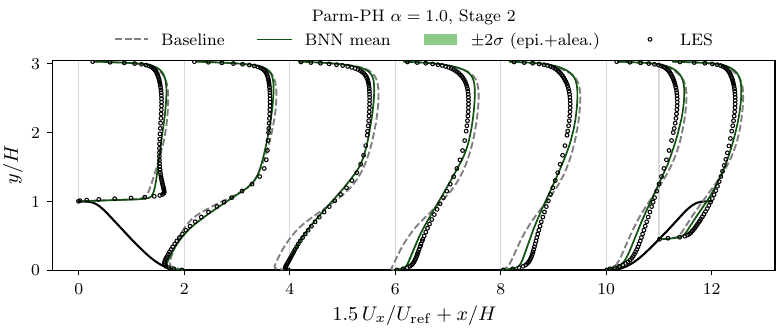}
\caption{Streamwise velocity profiles for the Parm-PH at $\alpha = 1.0$, Stage-2 propagation. Symbols as in \autoref{fig:parm_profiles_Ux_15}.}
\label{fig:parm_profiles_Ux_10}
\end{figure}

At $\alpha=1.0$ (\autoref{fig:parm_profiles_Ux_10}), the propagated mean again modifies
the baseline prediction toward the LES reference over much of the separated-flow region,
but the uncertainty bands remain insufficient to encompass the residual discrepancy.
The corresponding classified-region coverage is $10.1\%/21.4\%$ for $k$ and
$7.5\%/14.3\%$ for $U_x$ at the $1\sigma/2\sigma$ levels
(\autoref{tab:summary_coverage}).

\begin{figure}[htbp]
\centering
\includegraphics[width=\textwidth]{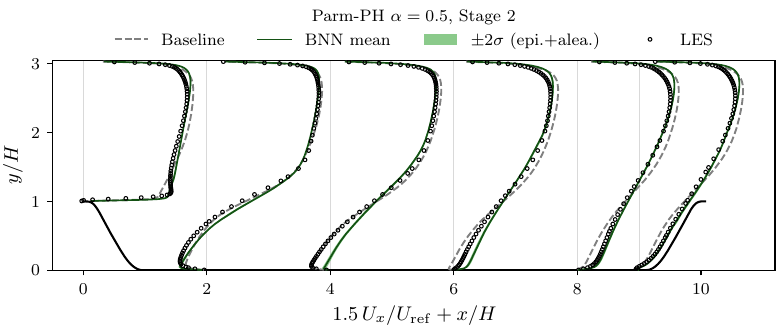}
\caption{Streamwise velocity profiles for the Parm-PH at $\alpha = 0.5$, Stage-2 propagation: the regime boundary of the model. Symbols as in \autoref{fig:parm_profiles_Ux_15}.}
\label{fig:parm_profiles_Ux_05}
\end{figure}

The contracted hill, $\alpha=0.5$, exhibits the weakest generalization of the three
parameterized configurations (\autoref{fig:parm_profiles_Ux_05}). On the classified cells,
the Stage~2 coverage falls to $3.8\%/8.4\%$ for $k$ and
$5.2\%/11.7\%$ for $U_x$ at the $1\sigma/2\sigma$ levels
(\autoref{tab:summary_coverage}). It is also the only configuration considered for which
the propagated corrected prediction becomes worse than the uncorrected baseline for at
least one of the evaluated quantities. The $\alpha=0.5$ case therefore exposes a clear
limitation of the transferability of the present feature-to-correction mapping and defines
the most challenging regime within the parameterized-hill family considered here.

\subsection{The NASA wall-mounted hump}

\begin{figure}[htbp]
\centering
\includegraphics[width=\textwidth]{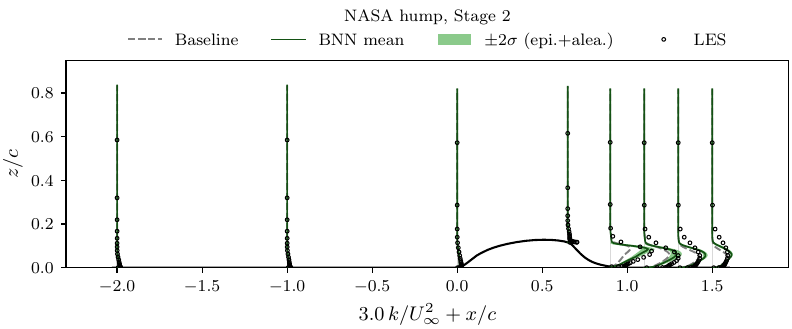}
\caption{Turbulent kinetic energy profiles for the NASA wall-mounted hump, Stage-2
propagation, at the CFDVAL2004 stations. The case is in the $x$--$z$ plane and is
normalized with the chord $c$ and $U_\infty$. Symbols as in
\autoref{fig:parm_profiles_Ux_15}.}
\label{fig:hump_profiles_k}
\end{figure}

\begin{figure}[htbp]
\centering
\includegraphics[width=\textwidth]{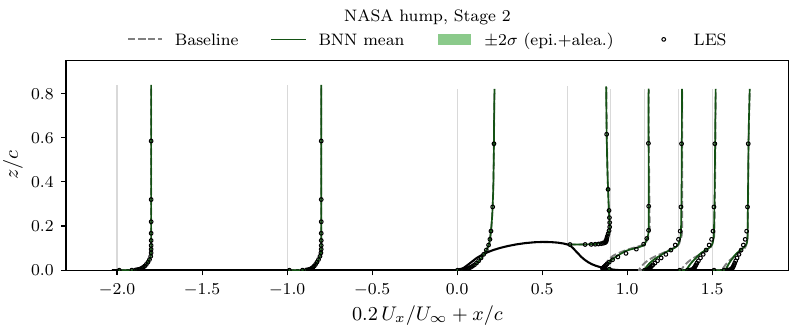}
\caption{Streamwise velocity profiles for the NASA wall-mounted hump, Stage-2 propagation. Symbols as in \autoref{fig:hump_profiles_k}.}
\label{fig:hump_profiles_Ux}
\end{figure}

The NASA hump provides the most substantial departure from the training configuration,
combining changes in geometry, coordinate plane, dimensional formulation, and Reynolds
number, with a chord-based $Re_c = 9.36 \times 10^{5}$
(\autoref{fig:hump_profiles_k} and \autoref{fig:hump_profiles_Ux}). The scalar correction
retains partial transferability: \textit{a priori}, it achieves $R^2 = 0.66$ against the inverted
target, while the propagated $k$ attains $31.8\%/44.3\%$ coverage against LES in the
RITA-classified region at the $1\sigma/2\sigma$ levels
(\autoref{tab:summary_coverage}), the highest LES-referenced Stage~2 $k$ coverage among
the unseen configurations considered here. The propagated mean generally improves upon
the baseline prediction in the separated and recovering regions, although residual
differences from LES remain.

The streamwise velocity shows substantially poorer uncertainty coverage, with only
$2.4\%/6.3\%$ of the classified cells covered at the $1\sigma/2\sigma$ levels
(\autoref{tab:summary_coverage}). Although the ensemble mean remains close to the LES
profiles at several stations and generally improves upon the baseline solution in the
separated-flow region, the propagated bands remain too narrow to encompass the residual
discrepancy consistently. This behavior is consistent with the weaker transfer of the
anisotropy correction on this configuration. The hump therefore illustrates that the
scalar correction retains useful predictive information farther from the training
distribution, whereas transfer of the anisotropy correction remains the more restrictive
component of the present framework.

\subsection{Where the calibration is lost}
\label{sec:coverage_chain}

Coverage can be examined at different stages of the prediction chain. Before CFD
propagation, the predicted correction fields can be compared directly with the
frozen-inversion targets. After propagation, the resulting flow solution can be compared
with either the BP solution obtained by propagating the exact inversion-derived
corrections or with the HF reference. These comparisons isolate different sources of
error: \textit{a priori} correction-field coverage reflects the learned correction model itself;
coverage against BP additionally includes the effect of propagation through the RANS
solver; and coverage against the HF reference further includes the residual limitation of
the correction framework.
For the benchmark cases considered here, BP can be constructed retrospectively because
HF data are available for the inversion. It is used in this subsection only as a
diagnostic reference to separate these contributions. Such a reference would not be
available in a predictive application to a genuinely new flow, since its construction
requires that flow's own HF data.

\begin{figure}[htbp]
\centering
\includegraphics[width=\textwidth]{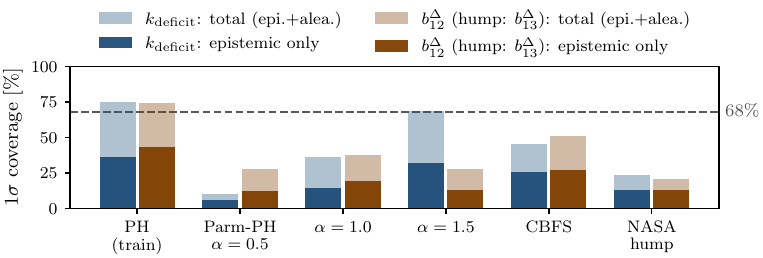}
\caption{\textit{A priori} $1\sigma$ coverage of the correction fields against the frozen-inversion targets, on the classified cells, for the scalar correction $k_{\text{deficit}}$ and the in-plane shear component of $b_{ij}^{\Delta}$ ($b_{13}^{\Delta}$ on the hump). Dark bars: epistemic-only band. Light bars: total band including the correlated aleatoric contribution. The dashed line marks the nominal $68\%$ level. No CFD is involved in this comparison.}
\label{fig:apriori_coverage_M3}
\end{figure}

\autoref{fig:apriori_coverage_M3} examines calibration before CFD propagation. On the PH
training flow, the total $1\sigma$ band covers $74.9\%$ of the
$k_{\text{deficit}}$ targets and $74.2\%$ of the in-plane anisotropy-correction targets,
compared with $36.5\%$ and $43.3\%$, respectively, for the epistemic-only intervals.
Including the correlated aleatoric contribution therefore substantially improves
correction-field coverage on the training configuration.
This improvement does not eliminate the loss of calibration under transfer. For
$k_{\text{deficit}}$, the total $1\sigma$ coverage decreases from $74.9\%$ on the PH to
$68.4\%$ for $\alpha=1.5$, $45.6\%$ on the CBFS, $36.2\%$ for $\alpha=1.0$,
$23.3\%$ on the NASA hump, and $10.1\%$ for $\alpha=0.5$. Thus, for the more distant
configurations, the predictive uncertainty of the correction model does not increase
sufficiently to encompass the corresponding transfer error.

\begin{figure}[htbp]
\centering
\includegraphics[width=\textwidth]{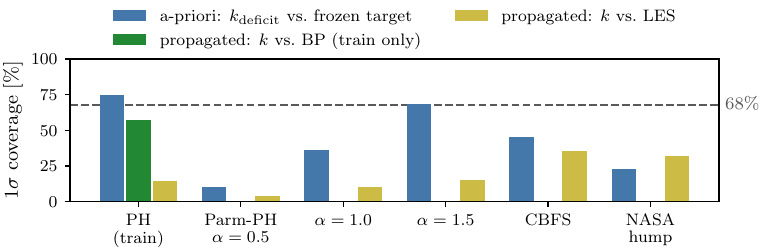}
\caption{Propagation of $1\sigma$ coverage through the prediction chain on the
RITA-classified cells. Blue: \textit{a priori} coverage of $k_{\text{deficit}}$ against the
frozen-inversion target; green: propagated $k$ coverage against BP for the PH training
case; yellow: propagated $k$ coverage against LES. BP can also be constructed
retrospectively for the benchmark test cases using their HF data, but it is not used as
the primary generalization reference because such information would not be available in a
predictive application. The dashed line denotes the nominal $68\%$ coverage level.}
\label{fig:coverage_chain_M3}
\end{figure}

\autoref{fig:coverage_chain_M3} relates the \textit{a priori} correction-field coverage to the
coverage obtained after propagation. The deterioration observed out of distribution is
already present before the correction fields enter the RANS solver. For example,
$k_{\text{deficit}}$ coverage falls from $74.9\%$ on the PH to $23.3\%$ on the NASA
hump and $10.1\%$ for $\alpha=0.5$ before any CFD propagation is performed. The
propagation step therefore cannot be regarded as the primary origin of the loss of
calibration under transfer.

On the PH training flow, where BP is shown explicitly in
\autoref{fig:coverage_chain_M3}, propagation reduces the $1\sigma$ coverage from
$74.9\%$ at the correction-field level to $56.9\%$ against BP. Propagation therefore
modifies the absolute calibration and is not strictly neutral. Nevertheless, the much
larger deterioration observed across the unseen configurations is already evident at the
correction-field level, indicating that the principal limitation lies in transfer of the
learned correction rather than in the CFD propagation procedure itself. Improving
out-of-distribution reliability must therefore primarily address the learned correction,
for example through broader training data, more informative invariant features, or a
posterior that better represents extrapolation uncertainty.

Coverage against LES is lower still for several configurations. For example, the
classified-region Stage~2 $1\sigma$ coverage of $k$ is $15.2\%$ for
$\alpha=1.5$ and $10.1\%$ for $\alpha=1.0$
(\autoref{tab:summary_coverage}). This additional reduction reflects the fact that the
correction framework itself retains a finite discrepancy from the HF solution even when
the exact inversion-derived corrections are propagated. We refer to this residual
discrepancy as the \emph{correction ceiling} of the present additive formulation: it is
the error that remains even when the exact inversion-derived corrections are supplied.

The BP reference provides a diagnostic means of making this distinction. Coverage against
BP isolates the combined effects of learning and propagation, whereas coverage against LES
additionally contains the residual limitation of the correction ansatz. Although BP can
be constructed retrospectively for the benchmark test cases using their HF data, the
generalization results in the preceding subsections are reported primarily against LES
because such a reference would not be available in a genuinely predictive application.
The LES-referenced coverage should therefore be interpreted as a measure of the
reliability of the complete predictive framework rather than of the BNN approximation
alone.

\subsection{Correction contributions}
\label{sec:stage_comparison}

Running Stage~1 on all six configurations, alongside the Stage~2 ensembles already
discussed, isolates the contribution of the anisotropy correction beyond that of
$k_{\text{deficit}}$ alone. The two stages are compared against a common LES reference,
because each stage has a different BP solution: Stage~1 propagates the inverted
$k_{\text{deficit}}$ alone, whereas Stage~2 propagates both inverted corrections.
BP-referenced errors would therefore not provide a common basis for comparing the two
stages. Although BP can be constructed retrospectively for all benchmark cases using
their HF data, \autoref{tab:stage_comparison} reports both stages against LES on the
classified cells.

\begin{table}[htbp]
\caption{Effect of adding the anisotropy correction, classified cells, both stages measured against LES. Negative RMSE changes are improvements. Coverage entries are $k$ / $U_x$.}
\centering
\small
\begin{tabular}{lrrrr}
\toprule
& \multicolumn{2}{c}{RMSE vs.\ LES, change $1 \rightarrow 2$} & \multicolumn{2}{c}{$1\sigma$ coverage vs.\ LES [\%]} \\
\cmidrule(lr){2-3}\cmidrule(lr){4-5}
Case & $k$ & $U_x$ & Stage 1 & Stage 2 \\
\midrule
PH (train)   & $-36\%$ & $-17\%$ & 15.4 / 1.2  & 14.2 / 4.6 \\
CBFS                    & $-34\%$ & $+12\%$ & 0.0 / 20.8  & 35.4 / 31.4 \\
Parm-PH $\alpha=0.5$    & $+16\%$ & $-4\%$  & 2.9 / 1.0   & 3.8 / 5.2 \\
Parm-PH $\alpha=1.0$    & $-17\%$ & $+2\%$  & 5.6 / 3.8   & 10.1 / 7.5 \\
Parm-PH $\alpha=1.5$    & $-24\%$ & $-2\%$  & 8.5 / 1.3   & 15.2 / 1.9 \\
NASA hump               & $+20\%$ & $+45\%$ & 18.8 / 1.0  & 31.8 / 2.4 \\
\bottomrule
\end{tabular}

\label{tab:stage_comparison}
\end{table}

Three observations follow. First, adding the anisotropy correction improves the
turbulent kinetic energy prediction on four of the six configurations, reducing the
RMSE by $17$--$36\%$, and increases its $1\sigma$ LES coverage on five of the six
cases. The largest change occurs on the CBFS, where the classified-region coverage of
$k$ rises from $0.0\%$ in Stage~1 to $35.4\%$ in Stage~2.

Second, the effect on the streamwise velocity is considerably less transferable. The
largest improvement occurs on the PH training flow, where the RMSE decreases by $17\%$.
Across the parameterized hill family the change is small, ranging from a $4\%$
improvement to a $2\%$ deterioration, while the CBFS and NASA hump deteriorate by
$12\%$ and $45\%$, respectively. Thus, the strong momentum field benefit observed on
the training flow does not persist consistently across the unseen configurations.

Third, the \textit{a priori} results in \autoref{tab:summary_apriori} show that the two learned
corrections do not lose transferability in the same way. The scalar correction retains
moderate predictive accuracy for $\alpha=1.0$, $\alpha=1.5$, and the NASA hump, with
$R^2=0.63$, $0.67$, and $0.66$, respectively, but performs poorly on the CBFS
($R^2=-0.15$) and $\alpha=0.5$ ($R^2=0.00$). The anisotropy correction is more
sensitive to the configuration: its shear-component $R^2$ becomes negative on the CBFS
($-1.84$), $\alpha=0.5$ ($-3.31$), and the NASA hump ($-0.57$), while remaining
positive for $\alpha=1.0$ and $\alpha=1.5$. The weak transfer of the anisotropy model
on several unseen configurations is consistent with the limited Stage~2 improvement of
the velocity field observed in \autoref{tab:stage_comparison}. Overall, the anisotropy correction provides a more
consistent benefit to the predicted $k$ field than to the streamwise velocity, for which
its strong training-flow improvement does not transfer reliably.

The distributions of the inversion-derived targets provide additional context for this
difference. Across the configurations considered here, the marginal distributions of the
anisotropy corrections remain relatively similar despite substantial changes in geometry
and Reynolds number, whereas the scalar-correction distribution varies more strongly
between cases. Similar target distributions, however, do not imply an invariant mapping
from the input features to those targets. The negative \textit{a priori} $R^2$ values obtained for
the anisotropy correction on several unseen configurations indicate that this mapping does
not transfer reliably with the present feature set.

The field-level comparisons are consistent with these aggregate metrics. On the
parameterized hills and the CBFS, the predicted fields retain parts of the spatial
structure of the inversion-derived targets but generally smooth their extrema. On the
NASA hump, the scalar correction retains recognizable spatial organization, whereas the
predicted anisotropy correction differs substantially from its target. The per-case
accuracy and coverage of both corrections are summarized in
\autoref{tab:summary_apriori}.

\begin{table}[H]
\caption{\textit{A priori}, before any CFD: the predicted correction fields against the frozen-inversion targets on the classified cells. Coverage is the percentage of target values inside the total ($\pm1\sigma$/$\pm2\sigma$) band; nominal values are $68/95$. $^{\dagger}$ marks a flow that is in the training set of that model. Stage~1 and Stage~2 share the same $k_{\text{deficit}}$ model, so the left half applies to both.}
\centering
\small
\setlength{\tabcolsep}{4.5pt}
\begin{tabular}{l cc cc cc cc}
\toprule
& \multicolumn{4}{c}{$k_{\text{deficit}}$} & \multicolumn{4}{c}{$b_{ij}^{\Delta}$ (shear component)} \\
\cmidrule(lr){2-5}\cmidrule(lr){6-9}
& \multicolumn{2}{c}{single-case training} & \multicolumn{2}{c}{multi-case training} & \multicolumn{2}{c}{single-case training} & \multicolumn{2}{c}{multi-case training} \\
\cmidrule(lr){2-3}\cmidrule(lr){4-5}\cmidrule(lr){6-7}\cmidrule(lr){8-9}
Case & $R^2$ & cov.\ & $R^2$ & cov.\ & $R^2$ & cov.\ & $R^2$ & cov.\ \\
\midrule
Periodic hill & 0.94$^{\dagger}$ & 75/96 & 0.93$^{\dagger}$ & 68/95 & 0.83$^{\dagger}$ & 74/95 & 0.76$^{\dagger}$ & 69/92 \\
CBFS & -0.15 & 46/79 & 0.83$^{\dagger}$ & 82/99 & -1.84 & 51/74 & 0.78$^{\dagger}$ & 90/99 \\
Parm-PH $\alpha=0.5$ & 0.00 & 10/25 & 0.09 & 19/38 & -3.31 & 28/51 & -3.47 & 19/43 \\
Parm-PH $\alpha=1.0$ & 0.63 & 36/76 & 0.63 & 38/75 & 0.41 & 38/74 & 0.48 & 41/83 \\
Parm-PH $\alpha=1.5$ & 0.67 & 68/91 & 0.63 & 61/85 & 0.48 & 28/51 & 0.43 & 29/47 \\
NASA hump & 0.66 & 23/52 & 0.70 & 27/57 & -0.57 & 21/44 & 0.41 & 33/63 \\
\bottomrule
\end{tabular}

\label{tab:summary_apriori}
\end{table}

\subsection{Multi-case extension}
\label{sec:m4}

The corrections considered above are learned from a single flow. A natural question is
whether the same formulation can be trained jointly on multiple configurations. To examine
this, both correction models are retrained using the PH and CBFS together, while retaining
the same architecture, normalization procedure, and training schedule, and are propagated
through the same Stage~2 pipeline.

Adding the CBFS to the training set substantially improves its \textit{a priori} prediction. For
$k_{\text{deficit}}$, the CBFS $R^2$ increases from $-0.15$ to $0.83$
and its $1\sigma$ coverage from $46\%$ to $82\%$
(\autoref{tab:summary_apriori}). On the four configurations that remain unseen by both
models, the mean $k_{\text{deficit}}$ $R^2$ changes only modestly, from $0.49$ with
single-case training to $0.51$ with multi-case training, with the individual cases
showing mixed changes. The NASA hump, for example, improves from $0.66$ to $0.70$,
while the PH training-flow accuracy is essentially retained
($R^2: 0.94 \rightarrow 0.93$). Thus, the additional training configuration strongly
improves the newly included case without producing a comparable systematic improvement
across the remaining unseen flows.

The effect on the anisotropy correction and the propagated velocity is less systematic.
On the PH, the \textit{a priori} shear-component $R^2$ decreases from $0.83$ to $0.76$, while
including the CBFS in training changes its $R^2$ from $-1.84$ to $0.78$
(\autoref{tab:summary_apriori}). Among the configurations that remain unseen, however,
the response is mixed: the NASA hump improves from $R^2=-0.57$ to $0.41$, whereas
$\alpha=0.5$ remains poorly predicted ($-3.31$ to $-3.47$), and only small changes occur
for $\alpha=1.0$ and $\alpha=1.5$.

This mixed behavior is also reflected after propagation. The classified-region
$1\sigma$ coverage of $U_x$ increases from $5.2\%$ to $11.2\%$ for $\alpha=0.5$,
changes from $7.5\%$ to $6.9\%$ for $\alpha=1.0$ and from $1.9\%$ to $1.2\%$
for $\alpha=1.5$, and decreases from $2.4\%$ to $0.0\%$ on the NASA hump
(\autoref{tab:summary_coverage}). Joint training therefore does not produce a consistent
improvement in propagated velocity on the configurations that remain unseen.

The multi-case model is therefore presented as an extension rather than as a replacement
for the single-case model used in the preceding generalization study. The experiment shows
that the formulation can incorporate data from multiple configurations and that adding a
case to the training set can substantially improve its correction-field prediction.
However, this improvement does not translate uniformly to configurations that remain
unseen, particularly for the anisotropy correction and the propagated velocity field.
Together with the single-case results of \S\ref{sec:stage_comparison}, this shows that
adding a second training configuration is not by itself sufficient to resolve the
remaining transfer limitation, and suggests that the representation of
$b_{ij}^{\Delta}$ and the associated input features are important targets for further
improvement.

\subsection{Summary across flows and models}
\label{sec:summary_tables}

The preceding sections report one configuration at a time. Tables
\ref{tab:summary_apriori}--\ref{tab:summary_rmse} collect the comparison across all
flows. Three model/propagation settings are considered: Stage~1 with the scalar correction
alone, Stage~2 with both corrections, and Stage~2 with the multi-case model. Each is
reported across all six flows, before and after CFD. The tables summarize three aspects of
performance: accuracy of the correction fields, coverage of the propagated solution, and
error of the propagated fields.

\begin{table}[H]
\caption{Propagated coverage, $\pm1\sigma$/$\pm2\sigma$ [\%], nominal $68/95$, on the RITA-classified cells (class.) and over the full field. BP is the propagation of the inverted corrections for that stage, the best the framework can do; LES is the high-fidelity reference. Each stage has its own BP, the inverted $k_{\text{deficit}}$ alone for Stage~1 and the full corrections for Stage~2, so BP-referenced entries are not comparable across stages. BP is tabulated for the training flow only, for the reason given in \S\ref{sec:coverage_chain}.}
\centering
\small
\setlength{\tabcolsep}{3.5pt}
\begin{tabular}{l l l cc cc cc}
\toprule
& & & \multicolumn{2}{c}{Stage 1, single-case} & \multicolumn{2}{c}{Stage 2, single-case} & \multicolumn{2}{c}{Stage 2, multi-case} \\
\cmidrule(lr){4-5}\cmidrule(lr){6-7}\cmidrule(lr){8-9}
Case & Ref. & Region & $k$ & $U_x$ & $k$ & $U_x$ & $k$ & $U_x$ \\
\midrule
Periodic hill & BP & class. & 99.6/100.0 & 100.0/100.0 & 56.9/87.8 & 68.6/97.6 & 64.3/88.2 & 62.7/94.7 \\
 & BP & full & 94.3/99.8 & 86.6/99.0 & 31.9/68.5 & 53.0/80.1 & 26.7/44.3 & 42.0/70.5 \\
 & LES & class. & 15.4/28.8 & 1.2/4.4 & 14.2/30.6 & 4.6/8.7 & 14.2/28.9 & 5.1/8.7 \\
 & LES & full & 4.5/9.1 & 0.6/1.6 & 10.7/20.6 & 2.4/4.7 & 10.4/19.9 & 2.6/5.2 \\
\addlinespace[1pt]
CBFS & LES & class. & 0.0/6.6 & 20.8/39.6 & 35.4/64.1 & 31.4/55.9 & 40.6/66.0 & 14.8/31.7 \\
 & LES & full & 0.1/0.3 & 7.1/14.5 & 1.2/2.2 & 4.6/11.0 & 1.1/1.9 & 2.6/6.7 \\
\addlinespace[1pt]
Parm-PH $\alpha=0.5$ & LES & class. & 2.9/6.3 & 1.0/1.9 & 3.8/8.4 & 5.2/11.7 & 3.8/7.8 & 11.2/20.3 \\
 & LES & full & 2.4/4.9 & 2.3/4.4 & 4.3/8.7 & 4.7/9.9 & 5.1/10.3 & 5.7/11.6 \\
\addlinespace[1pt]
Parm-PH $\alpha=1.0$ & LES & class. & 5.6/11.9 & 3.8/7.4 & 10.1/21.4 & 7.5/14.3 & 13.2/24.0 & 6.9/13.1 \\
 & LES & full & 2.7/5.5 & 0.9/1.8 & 7.3/13.5 & 2.9/5.6 & 5.6/10.1 & 2.7/5.3 \\
\addlinespace[1pt]
Parm-PH $\alpha=1.5$ & LES & class. & 8.5/16.3 & 1.3/3.2 & 15.2/28.2 & 1.9/3.8 & 11.2/22.2 & 1.2/3.0 \\
 & LES & full & 2.4/4.7 & 1.1/2.3 & 4.8/9.5 & 1.2/2.7 & 4.0/8.0 & 1.2/2.5 \\
\addlinespace[1pt]
NASA hump & LES & class. & 18.8/34.6 & 1.0/5.1 & 31.8/44.2 & 2.4/6.3 & 20.7/40.3 & 0.0/0.0 \\
 & LES & full & 0.9/1.8 & 2.2/3.9 & 2.3/3.8 & 4.0/7.4 & 1.3/2.6 & 3.0/5.2 \\
\addlinespace[1pt]
\bottomrule
\end{tabular}

\label{tab:summary_coverage}
\end{table}

\begin{table}[H]
\caption{Propagated field RMSE against LES on the classified cells, in solver units, with the uncorrected $k$--$\omega$ SST solution for scale. LES is used throughout because each stage has a different BP, so BP-referenced errors are not comparable across stages.}
\centering
\small
\setlength{\tabcolsep}{4pt}
\begin{tabular}{l cc cc cc cc}
\toprule
& \multicolumn{2}{c}{baseline $k$--$\omega$ SST} & \multicolumn{2}{c}{Stage 1, single} & \multicolumn{2}{c}{Stage 2, single} & \multicolumn{2}{c}{Stage 2, multi} \\
\cmidrule(lr){2-3}\cmidrule(lr){4-5}\cmidrule(lr){6-7}\cmidrule(lr){8-9}
Case & $k$ & $U_x$ & $k$ & $U_x$ & $k$ & $U_x$ & $k$ & $U_x$ \\
\midrule
Periodic hill & -- & -- & 1.63e-02 & 6.11e-02 & 1.04e-02 & 5.07e-02 & 1.06e-02 & 5.06e-02 \\
CBFS & 1.21e-02 & 1.25e-01 & 5.69e-03 & 2.03e-02 & 3.77e-03 & 2.26e-02 & 3.45e-03 & 2.75e-02 \\
Parm-PH $\alpha=0.5$ & 2.50e-02 & 5.97e-02 & 1.89e-02 & 7.60e-02 & 2.18e-02 & 7.30e-02 & 1.85e-02 & 6.65e-02 \\
Parm-PH $\alpha=1.0$ & 3.30e-02 & 9.02e-02 & 1.78e-02 & 4.71e-02 & 1.48e-02 & 4.82e-02 & 1.47e-02 & 4.80e-02 \\
Parm-PH $\alpha=1.5$ & 3.18e-02 & 1.41e-01 & 2.04e-02 & 8.37e-02 & 1.54e-02 & 8.18e-02 & 1.67e-02 & 8.52e-02 \\
NASA hump & 3.35e+01 & 7.94e+00 & 1.34e+01 & 2.86e+00 & 1.60e+01 & 4.13e+00 & 1.08e+01 & 4.34e+00 \\
\bottomrule
\end{tabular}

\label{tab:summary_rmse}
\end{table}

The \textit{a priori} results already reveal a substantial part of the generalization problem.
In \autoref{tab:summary_apriori}, correction-field coverage deteriorates markedly on
several unseen configurations, most strongly for $\alpha=0.5$ and, for the anisotropy
correction, the CBFS and NASA hump. This loss of calibration is therefore present before
the corrections enter the CFD solver. Propagation changes the absolute coverage and can
alter the ordering between cases, but it does not remove the underlying
out-of-distribution deterioration identified at the correction-field level.

Multi-case training has its clearest effect on the configuration added to the training
set. For the CBFS, $R^2$ increases from $-0.15$ to $0.83$ for
$k_{\text{deficit}}$ and from $-1.84$ to $0.78$ for the anisotropy correction
(\autoref{tab:summary_apriori}). Changes on the configurations that remain unseen are
less systematic. The NASA-hump anisotropy prediction improves from $R^2=-0.57$ to
$0.41$, whereas $\alpha=0.5$ remains poorly predicted under both models. Thus, adding
the CBFS substantially improves the newly represented configuration but does not produce
a uniform improvement in transfer across the remaining flows.

The PH results also illustrate why coverage depends strongly on the chosen reference.
Against BP, Stage~1 gives $99.6/100.0\%$ coverage for $k$ and
$100.0/100.0\%$ for $U_x$ on the classified cells
(\autoref{tab:summary_coverage}), indicating strongly conservative uncertainty relative
to the Stage~1 best-possible solution. Against LES, however, the same ensemble covers
only $15.4/28.8\%$ for $k$ and $1.2/4.4\%$ for $U_x$. The difference reflects the
residual discrepancy between the correction framework and the HF solution and shows why
BP- and LES-referenced coverage answer different questions.

For the benchmark test cases, BP can also be constructed retrospectively because HF data
are available for inversion. It is not used as the primary generalization reference,
however, because such information would not be available for a genuinely predictive
application to a new flow. The unseen-case entries in
\autoref{tab:summary_coverage} are therefore reported against LES and should be
interpreted as measures of the complete predictive framework: they combine transfer of
the learned corrections, their propagation through RANS, and the residual limitation of
the correction ansatz.

The propagated RMSE provides a complementary picture. Normalized by the uncorrected
baseline error, the $U_x$ RMSE at $\alpha=1.0$ is
$0.52/0.53/0.53$ for Stage~1, single-case Stage~2, and multi-case Stage~2,
respectively, and $0.59/0.58/0.60$ at $\alpha=1.5$
(\autoref{tab:summary_rmse}). The three settings therefore give very similar mean
velocity accuracy on these two configurations. Larger differences occur on the CBFS,
where the corresponding ratios are $0.16/0.18/0.22$, and on the NASA hump, where
Stage~1 is the most accurate of the three for $U_x$ ($0.36/0.52/0.55$).
The $\alpha=0.5$ case is the only configuration for which the corrected velocity is
worse than the baseline under all three settings, with ratios
$1.27/1.22/1.11$.

Taken together, the tables show that differences in mean-field accuracy and uncertainty
calibration do not necessarily follow one another. Several model settings produce similar
RMSE while exhibiting substantially different coverage, emphasizing that predictive
accuracy alone is insufficient for assessing the probabilistic correction framework.

\section{Conclusions}
\label{sec:conclusion}

This work presented a Bayesian framework for data-driven correction of RANS turbulence
models, combining a scalar turbulent kinetic energy correction
($k_{\text{deficit}}$) and a tensorial anisotropy correction
($b_{ij}^{\Delta}$) within a unified stochastic formulation.
The framework was trained on a periodic-hill configuration and evaluated without
retraining on five unseen flows: the CBFS, three parameterized periodic hills, and the
NASA wall-mounted hump, enabling a systematic assessment of correction accuracy,
uncertainty calibration, and transfer across geometries and Reynolds numbers.

The two corrections play distinct roles. The scalar correction primarily modifies the
turbulence field, while direct correction of the Reynolds-stress anisotropy has the
stronger influence on the momentum equation.
In contrast, the anisotropy correction $b_{ij}^{\Delta}$ leads to substantial
improvements in mean-flow predictions on the training flow. Combined with the scalar
correction, the model captures separation, recirculation and reattachment more accurately
there, highlighting the importance of the Reynolds-stress structure in governing the mean
flow.
This momentum-field benefit, however, does not transfer consistently to the unseen
configurations. The anisotropy correction remains beneficial for turbulent kinetic energy
in several cases, but its strong velocity improvement on the training flow is not
recovered systematically under distribution shift. The \textit{a priori} results identify
the anisotropy correction as the more restrictive component for transferring this
momentum-field benefit.

The uncertainty analysis shows that the loss of calibration under transfer is already
present at the correction-field level, before CFD propagation. Propagation changes the
absolute coverage but is not the primary source of the out-of-distribution
miscalibration. The propagated bands should therefore not be regarded as reliably
calibrated outside the training regime on the evidence considered here.

Coverage against BP and against the HF reference also answer different questions. BP,
which can be constructed retrospectively for the benchmark cases, separates the combined
effects of learning and propagation from the residual limitation of the correction
formulation, here referred to as the correction ceiling. HF-referenced coverage
instead evaluates the complete predictive framework. In a genuinely predictive
application BP would not be available, because its construction requires HF data from the
target flow.

The broader generalization study further shows that geometric similarity alone does not
ensure reliable transfer. The strongly contracted periodic hill is the most challenging
configuration considered, while the multi-case experiment shows that adding the CBFS to
the training set substantially improves that case but does not systematically improve the
configurations that remain unseen. Additional training data are therefore beneficial, but
are not by themselves sufficient to resolve the remaining transfer limitation.

Two methodological choices are important for the propagated uncertainty: the aleatoric
contribution is represented through spatially correlated stochastic fields, and the input
normalization preserves the physical zero of the invariant features rather than exporting
a training-set mean offset to unseen flows. These choices improve the consistency of the
probabilistic formulation, although they do not eliminate the remaining
out-of-distribution under-coverage.

Overall, the results show that mean prediction accuracy and uncertainty calibration do
not necessarily improve together. Bayesian turbulence-model corrections can provide
useful uncertainty information within and near the training regime, but reliable
application beyond that regime requires improved transfer of the learned corrections,
particularly the anisotropy model, together with posterior formulations that better
represent extrapolation uncertainty.

Future work should therefore focus on broader and more diverse training data, improved
representations and invariant features for $b_{ij}^{\Delta}$, and uncertainty models that
respond more appropriately under distribution shift.
Further work should also examine online evaluation of the corrections during the RANS
iteration and extension of the framework to fully three-dimensional flows.

\section*{Acknowledgments}
A.E. acknowledges the support from the GO-VIKING project (Grant NO. 101060826) funded by the Euratom Research and Training program of the European Union. 

\section*{CRediT authorship contribution statement}
T. B.: Conceptualization (equal); Methodology (equal); Software \& code (lead); Formal analysis (lead); Writing – original draft (supporting).
A. E.: Conceptualization (equal); Methodology (equal); Software \& code (supporting); Formal analysis (supporting); Writing – original draft (lead). 
R. P. D.: Conceptualization (equal); Methodology (equal); Supervision (lead); Writing – review and editing (equal).

\section*{Data availability}

The data and software required to reproduce the results of this study are openly
available through the BayesTBNNLite Zenodo archive. The repository contains the raw
final-time flow fields and frozen-inversion correction fields for all six configurations,
the classifier fields, trained network checkpoints, training code, and OpenFOAM cases
with their corresponding meshes. The archived release is available at
\url{https://doi.org/10.5281/zenodo.22021141}, with development history maintained at
\url{https://github.com/tsb-buchanan/BayesTBNNLite}.

The archive is self-contained and does not require the private feature-extraction library
used during development. The required feature calculations are reimplemented in the
public repository, and the raw stored fields allow alternative feature definitions to be
recomputed directly from the supplied data.

\section*{Declaration of competing interest}
The authors declare that they have no known competing financial interests or personal relationships that could have appeared to influence the work reported in this paper.

\appendix
\section{RITA-based shear-layer classifier}
\label{app:rita}

The correction region used in this work is identified using the \emph{Relative Importance Term Analysis} (RITA) framework of Buchanan et al.~\cite{buchanan2025data}, a physics-based classifier that isolates separated shear layers by comparing the magnitudes of terms in the turbulent kinetic energy equation of the baseline $k$--$\omega$ SST model. The method builds on the balance-regime analysis of Callaham
et al.~\cite{callaham2021learning}, who classified flow regimes from term balances in the RANS momentum equations; RITA instead targets the $k$-equation because the balance between its production, destruction, and transport terms varies systematically across boundary layers, shear layers, and free-stream regions.

\subsection*{Term ratios in the $k$-equation}

The transport equation for $k$ contains four dominant contributions:
convection ($C_k$), production ($P_k$), destruction ($D_k$), and diffusion ($d_k$). While their absolute magnitudes vary strongly with flow conditions, the \emph{ratios} between them remain characteristic
of particular flow regions. The most informative ratio for identifying shear layers is the regularised destruction-to-production ratio
\begin{equation}
  \varphi_{D_k/P_k} \;=\; \frac{|D_k|}{|P_k| + |D_k|} \;\in\; [0,1].
  \label{eq:phi_dkpk}
\end{equation}
The absolute values and the sum in the denominator ensure the ratio is bounded, avoiding the singularities that arise when $P_k \to 0$. In separated shear layers, strong mean shear drives $P_k \gg D_k$ and
$\varphi_{D_k/P_k} < 0.55$. In attached boundary layers destruction dominates and $\varphi_{D_k/P_k} > 0.55$, while in free-stream regions production is negligible and $\varphi_{D_k/P_k} \to 1$.

\subsection*{Additional flow indicators}

The ratio $\varphi_{D_k/P_k}$ alone cannot distinguish shear layers from the outer part of an attached boundary layer, where production also exceeds destruction. Two further indicators are therefore required: one sensitive to turbulence intensity, the other to local rotation. The first is the turbulent-to-total kinetic energy ratio,
\begin{equation}
  \varphi_k \;=\; \frac{k}{k + \tfrac{1}{2}|\mathbf{U}|^{2}},
  \label{eq:phi_k}
\end{equation}
which is large in shear layers, where fluctuations are strong and the mean velocity reduced, and small in attached boundary layers where mean kinetic energy dominates. Because $\varphi_k$ uses the mean velocity magnitude it is not Galilean invariant, so comparisons must be made in a common reference frame. The second is the vorticity Reynolds number,
\begin{equation}
  Re_{\Omega} \;=\; \frac{d_w^{2}\,|\Omega|}{\nu},
  \label{eq:reomega_raw}
\end{equation}
with $d_w$ the wall distance, $|\Omega|$ the vorticity magnitude, and $\nu$ the kinematic viscosity. It marks regions of strong rotation and shifts the classified layer a few cells from the wall, improving robustness in 3D flows. Since $Re_{\Omega}$ is unbounded, it is rescaled case-wise,
\begin{equation}
  \varphi_{Re_{\Omega}}
  \;=\;
  \frac{Re_{\Omega} - Re_{\Omega,\min}}
       {Re_{\Omega,\max} - Re_{\Omega,\min}}
  \;\in\; [0,1],
  \label{eq:phi_reomega}
\end{equation}
with the extrema taken over the full domain. A single threshold on $\varphi_{Re_{\Omega}}$ then transfers across configurations, at the cost of making the quantity a per-case global rather than strictly local one.

\subsection*{Binary classifier}

The three indicators are combined into the binary mask
\begin{equation}
  \sigma(\mathbf{x}) \;=\;
  \begin{cases}
    1, & \text{if } \varphi_{D_k/P_k} \le 0.55 \;\land\;
                    \varphi_k \ge 0.12 \;\land\;
                    \varphi_{Re_{\Omega}} \ge 0.02, \\[2pt]
    0, & \text{otherwise,}
  \end{cases}
  \label{eq:sigma}
\end{equation}
with thresholds tuned empirically by Buchanan et al.~\cite{buchanan2025data} on the PH, NASA-Hump, and CBFS training cases and shown there to generalise to 2D and 3D test cases. Cells with $\sigma = 1$ are labelled as lying in a separated shear layer.

\subsection*{Role in the present framework}

In this work, $\sigma$ defines the support of the data-driven corrections: training data are drawn only from cells with $\sigma = 1$, and at inference (propagation) time the learned corrections are multiplied by $\sigma$ before entering the transport equations, so the baseline $k$--$\omega$ SST model is recovered wherever $\sigma = 0$. The mask itself is never differentiated in the transport or momentum equations; its discrete jump at the boundary between classified and unclassified cells does not introduce numerical instability \cite{buchanan2025data}. Because the classifier is built from local term-balance ratios, it adapts to the flow state and identifies separated shear layers consistently across configurations. Further details on the RITA framework can be found in \cite{callaham2021learning} and \cite{buchanan2025data}.

\section{Summary of Methodology and Implementation Details}
\label{app:method_summary}
This appendix summarizes the key implementation steps of the proposed framework in a concise algorithmic form. While the main text provides the mathematical formulation and modeling rationale, the following pseudocode outlines the practical training and inference procedures used in this work. The presentation is intended to improve clarity and reproducibility by explicitly documenting the sequence of operations, including deterministic pretraining, Bayesian fine-tuning via variational inference, and MC prediction for uncertainty estimation.
The algorithms shown here are for the scalar $k_{\text{deficit}}$ model. The anisotropy correction model for $b_{ij}^{\Delta}$ follows an identical training and inference procedure, including deterministic pretraining, variational Bayesian fine-tuning, and MC inference. The key difference lies in the network output, which is vector-valued and corresponds to the tensor-basis coefficients. The ELBO loss formulation and UQ strategy remain the same.
\subsection*{Training procedure}

\begin{algorithm}[!htbp]
\caption{Training procedure for the scalar $k_{\text{deficit}}$ BNN}
\label{alg:training_bnn}
\begin{algorithmic}[1]
\Require Training inputs $\{\bx_i,\tilde{\varepsilon}_i,y_i\}_{i=1}^N$, learning rates $\eta_{\mathrm{det}}=0.01$, $\eta_{\mathrm{bnn}}=0.001$
\Require Pretraining epochs $N_{\mathrm{det}}=10{,}000$, Bayesian epochs $N_{\mathrm{BNN}}=40{,}000$
\Require Initial prior standard deviation $\sigma_{\mathrm{prior}}=1.0$, sigma-prior weight $\lambda_\sigma$
\Ensure Variational parameters $(\mu,\rho)$ and learned precision $\alpha$

\Statex
\State \textbf{Phase 1: Deterministic pretraining}
\State Initialize deterministic network with coefficient and noise heads
\For{$e = 1$ to $N_{\mathrm{det}}$}
    \State Compute deterministic prediction $\hat{\tilde{y}}_i$
    \State Compute mean-squared-error loss
    \[
    \mathcal{L}_{\mathrm{MSE}} = \frac{1}{N}\sum_{i=1}^N (\hat{\tilde{y}}_i - y_i)^2
    \]
    \State Update deterministic network weights using Adam with learning rate $\eta_{\mathrm{det}}$
\EndFor

\Statex
\State \textbf{Phase 2: Bayesian fine-tuning}
\State Initialize posterior means $\mu$ from pretrained deterministic weights
\State Initialize posterior log-scale parameters $\rho=-5.0$
\State Initialize $\log\alpha$ from $\sigma_{\mathrm{prior}}=1.0$

\For{$e = 1$ to $N_{\mathrm{bnn}}$}
    \ForAll{variational layers}
        \State Sample $\epsilon \sim \mathcal{N}(0,1)$
        \State Compute $\bW = \mu + \mathrm{softplus}(\rho)\cdot \epsilon$
    \EndFor
    \State Forward pass through coefficient and noise heads
    \State Compute normalized prediction
    \[
    \hat{\tilde{y}}_i = g(\bx_i;\bW)\,\tilde{\varepsilon}_i
    \]
    \State Compute predicted aleatoric standard deviation
    \[
    \sigma_{y,i} = c(\bx_i;\bW)\,|\tilde{\varepsilon}_i|
    \]
    \State Compute negative log-likelihood $\mathcal{L}_{\mathrm{NLL}}$
    \State Compute KL term $\mathcal{L}_{\mathrm{KL}}$
    \State Compute $\alpha$-prior regularization $\mathcal{L}_{\alpha}$
    \State Compute sigma-prior regularization $\mathcal{L}_{\sigma}$
    \State Form total loss
    \[
\mathcal{L}
=
\mathcal{L}_{\mathrm{NLL}}
+
\frac{1}{N}\mathcal{L}_{\mathrm{KL}}
+
\mathcal{L}_{\alpha}
+
\lambda_\sigma \mathcal{L}_{\sigma}
    \]
    \State Update $(\mu,\rho,\log\alpha)$ using Adam with learning rate $\eta_{\mathrm{bnn}}$
\EndFor
\end{algorithmic}
\end{algorithm}
\FloatBarrier

\subsection*{Inference and uncertainty estimation}
\begin{algorithm}[!htbp]
\caption{MC inference for the scalar $k_{\text{deficit}}$ BNN}
\label{alg:inference_bnn}
\begin{algorithmic}[1]
\Require Test inputs $\{\bx_i,\tilde{\varepsilon}_i\}_{i=1}^{N_{\mathrm{test}}}$, training normalization statistics $(\mu_{\mathrm{train}},\sigma_{\mathrm{train}})$
\Require Number of MC samples $M=100$
\Ensure Predictive mean, epistemic uncertainty, aleatoric uncertainty, and epistemic-on-aleatoric uncertainty

\For{$m = 1$ to $M$}
    \ForAll{variational layers}
        \State Sample $\epsilon^{(m)} \sim \mathcal{N}(0,1)$
        \State Compute sampled weights
        \[
        \bW^{(m)} = \mu + \mathrm{softplus}(\rho)\cdot \epsilon^{(m)}
        \]
    \EndFor
    \State Compute normalized prediction
    \[
    \hat{\tilde{y}}_i^{(m)} = g(\bx_i;\bW^{(m)})\,\tilde{\varepsilon}_i
    \]
    \State Compute aleatoric standard deviation
    \[
    \sigma_{y,i}^{(m)} = c(\bx_i;\bW^{(m)})\,\tilde{\varepsilon}_i
    \]
    \State Denormalize prediction
    \[
    \hat{y}_i^{(m)} = \hat{\tilde{y}}_i^{(m)}\,\sigma_{\mathrm{train}} + \mu_{\mathrm{train}}
    \]
    \State Denormalize aleatoric standard deviation
    \[
    \sigma_{\mathrm{alea},i}^{(m)} = \sigma_{y,i}^{(m)}\,\sigma_{\mathrm{train}}
    \]
\EndFor

\State Compute predictive mean
\[
\bar{y}_i = \frac{1}{M}\sum_{m=1}^M \hat{y}_i^{(m)}
\]

\State Compute epistemic uncertainty
\[
\sigma_{\mathrm{epi},i} = \mathrm{std}_m\!\left(\hat{y}_i^{(m)}\right)
\]

\State Compute aleatoric uncertainty
\[
\sigma_{\mathrm{alea},i} = \mathrm{mean}_m\!\left(\sigma_{\mathrm{alea},i}^{(m)}\right)
\]

\State Compute epistemic-on-aleatoric uncertainty
\[
\sigma_{\mathrm{alea\mbox{-}epi},i} = \mathrm{std}_m\!\left(\sigma_{\mathrm{alea},i}^{(m)}\right)
\]
\end{algorithmic}
\end{algorithm}

\FloatBarrier

\section{Model Design and Hyperparameter Selection}
\label{app:design}

This appendix summarizes the main architectural choices and hyperparameter selection procedure used for the proposed BNNs. Since the
scalar $k_{\text{deficit}}$ model and the anisotropy-correction
$b_{ij}^{\Delta}$ model share the same Bayesian training framework, the focus here is on the key design decisions, the hyperparameter sweep used to select the final configuration, and a compact side-by-side comparison of the two models.

\subsection*{Hyperparameter selection}

The final configuration was selected from a 48-configuration sweep over the hyperparameters \emph{prior\_std}, \emph{weight\_decay}, and $\lambda_\sigma$,
using candidate values
$\{0.05,\,0.1,\,0.5,\,1.0\}$,
$\{0,\,10^{-5},\,10^{-4}\}$, and
$\{0.05,\,0.1,\,0.5,\,1.0\}$, respectively.
The basis form (raw $\varepsilon$) and sigma-prior shape
$Gamma(2,10)$ were fixed from a prior 6-configuration comparison.

The main findings of this sweep are summarized as follows:
\begin{itemize}[noitemsep]
    \item $\beta = 1.0$ (proper ELBO) yields approximately $4\times$ lower MSE than $\beta = 0.1$ when combined with learnable $\alpha$.
    \item Larger values of \texttt{prior\_std} ($0.5$--$1.0$), together with zero weight decay, allow $\alpha$ to converge freely to
    $\sim 0.8$--$1.0$, resulting in the best predictive performance.
    \item The Gamma$(2,10)$ sigma prior with $\lambda_\sigma = 0.5$
    provides the best balance between prediction accuracy and uncertainty calibration.
    \item The selected configuration
    (MSE $= 9.24 \times 10^{-6}$, calibration $74\%/96\%$)
    was retrained with a fixed seed to produce the final reported model.
\end{itemize}

\subsection*{Model architecture and implementation}

The architectural design and training configuration of the proposed BNNs are summarized in \autoref{tab:model_comparison}. Both models share the same Bayesian training strategy, prior specification, and optimizer settings. The main differences lie in the output dimensionality and in how the correction terms are represented: the scalar model predicts a single dissipation-scaled correction, whereas the anisotropy model predicts the tensor-basis coefficients directly.

\begin{table}[htbp]
\centering
\small
\caption{Side-by-side comparison of the $k_{\text{deficit}}$ and
$b_{ij}^{\Delta}$ BNNs. Both models share the same prior choices and training schedule, but differ in output dimensionality and in the representation of the correction terms.}
\label{tab:model_comparison}
\begin{tabular}{@{}p{3.2cm}p{5.0cm}p{5.0cm}@{}}
\toprule
\textbf{Component} & \textbf{$k_{\text{deficit}}$ BNN} & \textbf{$b_{ij}^{\Delta}$ BNN} \\
\midrule

\multicolumn{3}{@{}l}{\textit{Architecture}} \\[2pt]
Hidden layers & $2 \times [32, 32]$, tanh & Same \\
Output dim (coeff.) & 1 & 3 ($g_1, g_2, g_3$) \\
Output dim (noise) & 1 & 3 \\
Mean prediction & $\hat{y} = g \cdot \varepsilon$ & $\hat{g}_n$ (direct, no basis) \\
Noise model & $\sigma = c \cdot \varepsilon$ & $\sigma_n = \exp(\log c_n) + 10^{-6}$ \\

\midrule
\multicolumn{3}{@{}l}{\textit{Inputs and targets}} \\[2pt]
Input features & \multicolumn{2}{l}{$S^2, \Omega^2, S^3, \Omega^2S, q_{\text{Re}_k}$ (5 invariants, scaled by $\sigma_x^{\mathrm{PH}}$, no mean shift)} \\
Time scale & \multicolumn{2}{l}{ $\tilde{S} = \tau_s S$, $\tilde{\Omega} = \tau_s \Omega$ with $\tau_s = 1/\lVert \nabla U \rVert_F$ (mean flow)} \\
Target & $k_{\text{deficit}} / y_s$, $y_s = \mathrm{std}(k_{\text{deficit}}^{\mathrm{PH}})$ & $g_1, g_2, g_3$ (fixed scalar scale each) \\
Basis normalization & $\varepsilon$ (raw) & N/A \\

\midrule
\multicolumn{3}{@{}l}{\textit{Priors (shared by both models)}} \\[2pt]
Weight precision & \multicolumn{2}{l}{Learnable $\alpha$ via Gamma$(1, 0.025)$; initialized with $\alpha_0 = 1.0$} \\
$\alpha$ scope & \multicolumn{2}{l}{Single $\alpha$ shared by coefficient and noise heads} \\
Sigma prior & \multicolumn{2}{l}{Gamma$(2, 10)$ with $\lambda_\sigma = 0.5$ (applied per output)} \\
KL weight $\beta$ & \multicolumn{2}{l}{1.0 (proper ELBO)} \\

\midrule
\multicolumn{3}{@{}l}{\textit{Initialization (shared)}} \\[2pt]
$\mu_W$ & \multicolumn{2}{l}{Kaiming normal (fan-in)} \\
$\mu_b$ & \multicolumn{2}{l}{0} \\
$\rho$ & \multicolumn{2}{l}{$-5.0$ ($\sigma_W \approx 0.007$)} \\
Log-noise bias & \multicolumn{2}{l}{$-3.0$} \\

\midrule
\multicolumn{3}{@{}l}{\textit{Training schedule (shared)}} \\[2pt]
Phase 1 (det.) & \multicolumn{2}{l}{10{,}000 epochs, lr $= 0.01$, Adam, MSE loss} \\
Phase 2 (BNN) & \multicolumn{2}{l}{40{,}000 epochs, lr $= 0.001$, Adam, ELBO loss} \\
Batch size & \multicolumn{2}{l}{Full batch (no mini-batching)} \\
Training cells & \multicolumn{2}{l}{2{,}378 (classified region, 80/20 split)} \\

\midrule
\multicolumn{3}{@{}l}{\textit{Converged values}} \\[2pt]
$\alpha$ & 0.97 ($\sigma_{\text{equiv}} = 1.01$) & 1.91 ($\sigma_{\text{equiv}} = 0.72$) \\
Final NLL & $-0.60$ & $-0.14$ \\
Final KL$/N$ & 0.93 & 2.12 \\

\midrule
\multicolumn{3}{@{}l}{\textit{Inference}} \\[2pt]
MC samples & \multicolumn{2}{l}{$M = 100$} \\
Inference domain & \multicolumn{2}{l}{Full field (15{,}600 cells)} \\

\bottomrule
\end{tabular}
\end{table}
\section{Coefficient Projection and Training Formulation}
\label{app:g_projection}

\subsection*{\texorpdfstring{Least-Squares Projection of $g$-Coefficients}{Least-Squares Projection of g coefficients}}

The target coefficients $g_n^{*}$ are obtained by projecting the reference anisotropy correction tensor $b_{ij}^{\Delta}$ onto the selected tensor basis. This is formulated as a pointwise least-squares problem:
\begin{equation}
    \bm{g}^{*} = \argmin_{\bm{g}} \left\| b_{ij}^{\Delta} - \sum_{n=1}^{3} g_n\, T_{ij}^{(n)} \right\|_F^2
    \label{eq:g_projection}
\end{equation}

Using Voigt notation, the system can be written in matrix form:
\begin{equation}
    \bm{g}^{*} = (A^T A)^{-1} A^T \bm{b}^{\Delta}
    \label{eq:g_lstsq}
\end{equation}
where $A_{k n} = T_k^{(n)}$ is the design matrix, $k$ indexes the six
independent tensor components, and $n$ indexes the basis functions.
To ensure numerical stability at degenerate points, a small Tikhonov
regularization is applied:
\begin{equation}
    \bm{g}^{*} = (A^T A + \epsilon I)^{-1} A^T \bm{b}^{\Delta},
    \quad \epsilon = 10^{-12}
    \label{eq:g_tikhonov}
\end{equation}

For the selected $T_1 + T_2 + T_3$ basis, the system is well-conditioned (condition number $\sim O(10)$), and the regularization term has negligible impact on the solution.

\subsection*{End-to-End vs.\ Direct Regression Formulations}

An alternative approach is to train the network in an end-to-end manner, where the loss is defined on the reconstructed tensor:
\begin{equation}
    \mathcal{L}_{\text{e2e}} =
    \left\| b_{ij}^{\Delta} - \sum_{n=1}^{N_T} \hat{g}_n(\bx)\, T_{ij}^{(n)} \right\|_F^2
    \label{eq:loss_e2e}
\end{equation}

The gradient of this loss with respect to each coefficient is:
\begin{equation}
    \frac{\partial \mathcal{L}_{\text{e2e}}}{\partial \hat{g}_n}
    =
    -2\left(b_{ij}^{\Delta} - \sum_m \hat{g}_m\, T_{ij}^{(m)}\right)
    T_{ij}^{(n)}
    \label{eq:grad_e2e}
\end{equation}

This expression shows that the gradient is scaled by the magnitude of the basis tensor $T_{ij}^{(n)}$, leading to spatially varying supervision strength across the domain.
In contrast, the direct regression formulation minimizes the error on the coefficients themselves:
\begin{equation}
    \mathcal{L}_{\text{direct}} =
    \sum_{n=1}^{3} \left(\hat{g}_n - g_n^{*}\right)^2
    \label{eq:loss_direct}
\end{equation}
with gradients:
\begin{equation}
    \frac{\partial \mathcal{L}_{\text{direct}}}{\partial \hat{g}_n}
    =
    2\left(\hat{g}_n - g_n^{*}\right)
    \label{eq:grad_direct}
\end{equation}

This formulation provides uniform, basis-independent gradients for all
training points.

\subsection*{Properties of the Projection}

The least-squares system is overdetermined (six equations, three unknowns), ensuring a unique solution for $\bm{g}^{*}$ without overfitting. The resulting coefficients have magnitudes $\mathcal{O}(1\text{--}10)$ and exhibit approximately Gaussian distributions, making them well-suited for z-score normalization and Gaussian likelihood-based training.
No additional regularization (e.g., ridge or elastic net) is applied during projection, as such methods would bias the coefficients toward zero and reduce the reconstruction accuracy. Instead, regularization is introduced at the learning stage through the BNN via weight priors and noise modeling.

\section{\texorpdfstring{Likelihood, Training, and Inference Details for $b_{ij}^{\Delta}$}{Likelihood, Training, and Inference Details for bij Delta}}
\label{app:bij_training}

\subsection*{Multi-Output Likelihood}

For the anisotropy correction model, the three coefficient outputs are treated as conditionally independent given the inputs and network weights. The likelihood is therefore written as
\begin{equation}
    p(\bm{g} \mid \bx, \bW, \bW_\sigma)
    =
    \prod_{n=1}^{3}
    \Normal\!\left(
        g_n \;\big|\;
        \hat{g}_n(\bx, \bW),\;
        \sigma_n^2(\bx, \bW_\sigma)
    \right),
\end{equation}
where each coefficient is assigned an independent aleatoric uncertainty
$\sigma_n(\bx)$. The corresponding NLL becomes
\begin{equation}
    \text{NLL}
    =
    \frac{1}{N}\sum_{i=1}^{N}\sum_{n=1}^{3}
    \left[
        \log\sigma_{n,i}
        +
        \frac{(g_{n,i} - \hat{g}_{n,i})^2}{2\sigma_{n,i}^2}
    \right].
\end{equation}

This extends the scalar heteroscedastic formulation in the main text to a three-output setting while preserving the same probabilistic structure.

\subsection*{Coefficient Projection}

The target coefficients are obtained by projecting the reference anisotropy correction tensor $b_{ij}^{\Delta}$ onto the selected tensor basis using a pointwise least-squares problem,
\begin{equation}
    \bm{g}^{*}
    =
    \argmin_{\bm{g}}
    \left\|
        b_{ij}^{\Delta}
        -
        \sum_{n=1}^{3} g_n T_{ij}^{(n)}
    \right\|_F^2.
\end{equation}

A small Tikhonov regularization is added for numerical stability in degenerate
configurations,
\begin{equation}
    \bm{g}^{*}
    =
    (A^T A + \epsilon I)^{-1} A^T \bm{b}^{\Delta},
    \qquad \epsilon = 10^{-12}.
\end{equation}

The resulting coefficients are well-conditioned and approximately
Gaussian-distributed, which makes them suitable targets for regression with z-score normalization.

\subsection*{Training Specifics}

Training follows the same two-phase strategy described in the main text and is applied jointly to all three coefficients. The main difference from the scalar model is the shared multi-output representation, which increases coupling between outputs and creates a stronger tendency toward overfitting. This is reflected in the learned prior precision, which converges to a higher value ($\alpha \approx 1.91$) than in the scalar case, indicating the need for stronger effective regularization.
All other prior assumptions and optimization settings are unchanged from the scalar model and are therefore not repeated here.

\subsection*{Inference and Tensor Reconstruction}

At inference time, MC sampling is performed in coefficient space. For each sampled set of network weights, a realization of the coefficient vector $\hat{\bm{g}}^{(m)}(\bx)$ is obtained and mapped back to the physical tensor space through the selected basis,
\begin{equation}
    b_{ij}^{\Delta,(m)}
    =
    \sum_{n=1}^{3}
    \hat{g}_n^{(m)}(\bx)\, T_{ij}^{(n)}.
    \label{eq:bij_reconstruct}
\end{equation}

Repeating this process yields an ensemble of anisotropy correction fields, from which predictive statistics and uncertainty measures are computed. These realizations are written as OpenFOAM \texttt{symmTensor} fields for downstream propagation through the RANS solver.

\section*{Declaration of generative AI and AI-assisted technologies in the manuscript preparation process}

During the preparation of this work, the authors used ChatGPT (OpenAI) and Claude (Anthropic) for language editing, wording refinement, and code syntax checking. The authors reviewed and edited the output as needed and take full responsibility for the content of the published article.
\bibliographystyle{elsarticle-num-names}
\bibliography{bibliography,MR}

\end{document}